\documentclass[aps,prc,nofootinbib,preprintnumbers,12pt,tightenlines,superscriptaddress]{revtex4}

\usepackage{graphicx}
\usepackage{rotating}
\usepackage{latexsym}

\newcommand{\beq}{\begin{equation}}
\newcommand{\eeq}{\end{equation}}
\newcommand{\bqa}{\begin{eqnarray}}
\newcommand{\eqa}{\end{eqnarray}}

\begin{document}

\title{New Developments in Relativistic Viscous Hydrodynamics}

\preprint{INT-PUB-09-010}

\author{Paul Romatschke}
\affiliation{
Institute for Nuclear Theory, University of Washington, 
Box 351550, Seattle, WA, 98195}

\begin{abstract}
Starting with a brief introduction into the basics of relativistic fluid dynamics, 
I discuss our current knowledge of a relativistic theory of fluid dynamics
in the presence of (mostly shear) viscosity. Derivations based on
the generalized second law of thermodynamics, kinetic theory, and 
a complete second-order gradient expansion are reviewed. 
The resulting fluid dynamic equations are shown to be consistent 
for all these derivations, when properly accounting for the respective region of 
applicability, and can be applied to both weakly and strongly coupled systems.
In its modern formulation, relativistic viscous hydrodynamics can directly
be solved numerically. This has been useful for the problem of 
ultrarelativistic heavy-ion collisions, and I will review the setup and results of
a hydrodynamic description of experimental data for this case.
\end{abstract}

\maketitle

\newpage

\tableofcontents
\newpage

\section{Introduction}

\subsection{Non-relativistic fluid dynamics}

Fluid dynamics is one of the oldest and most successful theories
in modern physics. In its non-relativistic form, 
it is intuitively understandable due to our everyday experience with 
hydrodynamics, or the dynamics of water\footnote{In some fields
it has been the tradition to use the term hydrodynamics synonymous
with fluid dynamics of other substances, and I will adopt this 
somewhat sloppy terminology.}. The degrees of freedom for an ideal, neutral,
uncharged, one-component fluid are the fluid velocity ${\vec v}(t,{\vec x})$,
the pressure $p(t,{\vec x})$, and the fluid mass density $\rho(t,{\vec x})$,
which are linked by the fluid dynamic equations \cite{Euler},\cite{LL}\S2,
\bqa
\partial_t {\vec v}+\left({\vec v}\cdot {\vec \partial}\right) {\vec v}&=&-\frac{1}{\rho} {\vec \partial} p\,,
 \label{Eulereq}\\
\partial_t \rho + \rho\, {\vec \partial}\cdot {\vec v}+{\vec v}\cdot {\vec \partial} \rho &=&0\ . \label{Continuityeq}
\eqa
These equations are referred to as ``Euler equation'' (\ref{Eulereq}) and ``Continuity equation'' (\ref{Continuityeq}),
respectively, and typically have to be supplemented by an equation of state $p=p(\rho)$ to 
close the system. For non-ideal fluids, where dissipation can occur, the Euler equation generalizes
to the ``Navier-Stokes equation'' \cite{Navier,Stokes},\cite{LL}\S15,
\bqa
\frac{\partial v^i}{\partial t}+v^k \frac{\partial v^i}{\partial x^k}&=&-\frac{1}{\rho} \frac{\partial p}{\partial x^i}
-\frac{1}{\rho}\frac{\partial \Pi^{k i} }{\partial x^k}\,,
\label{NS}\\
\Pi^{ki}&=&-\eta \left(\frac{\partial v^i}{\partial x^k}+\frac{\partial v^k}{\partial x^i}-\frac{2}{3}
\delta^{k i} \frac{\partial v^l}{\partial x^l}\right)-\zeta\, \delta^{i k} \frac{\partial v^l}{\partial x^l}\ ,
\label{visctens}
\eqa
where Latin indices denote the three space directions, e.g. $i=1,2,3$.
The viscous stress tensor $\Pi^{ki}$ contains the coefficients for 
shear viscosity, $\eta$, and bulk viscosity, $\zeta$, which are independent of velocity. 
The non-relativistic Navier-Stokes equation is well tested and found to be reliable in many applications,
so any successful theory of relativistic viscous hydrodynamics should reduce to it
in the appropriate limit. 

\subsection{Relativistic ideal fluid dynamics}
\label{sec:1b}

For a relativistic system, the mass density $\rho(t,{\vec x})$ is not a good degree of freedom because
it does not account for kinetic energy that may become sizable for 
motions close to the speed of light. Instead, it is useful to replace it by the total
energy density $\epsilon(t,{\vec x})$, which reduces to $\rho$ in the non-relativistic
limit. Similarly, $\vec{v}(t,{\vec x})$ is not a good degree of freedom because
it does not transform appropriately under Lorentz transforms. Therefore, it should be
replaced by the Lorentz 4-vector for the velocity,
\beq
u^\mu \equiv \frac{d x^\mu}{d {\cal T}},
\eeq
where Greek indices denote Minkowski 4-space, e.g. $\mu=0,1,2,3$ with metric
$g_{\mu \nu}={\rm diag}(+,-,-,-)$ 
(the same symbol for the metric will also be used for curved spacetimes).
The proper time increment $d{\cal T}$ is given by the line element,
\bqa
(d{\cal T})^2&=&g_{\mu \nu}dx^\mu dx^\nu 
= (dt)^2-(d{\vec x})^2\,,\nonumber\\
&=&(dt)^2\left[1-\left(\frac{d \vec{x}}{d t}\right)^2\right]=(dt)^2\left[1-(\vec{v})^2\right]\ ,
\nonumber
\eqa
where here and in the following, natural units $\hbar=c=k_B=1$ will be used. This implies that
\beq
u^\mu=\frac{d t}{d{\cal T}} \frac{d x^\mu}{d t}=\frac{1}{\sqrt{1-{\vec v}^{\,2}}}
\left(\begin{array}{c}1\\{\vec v}\end{array} \right)=\gamma(\vec{v})\left(\begin{array}{c}1\\{\vec v}\end{array}\right)\ ,
\label{umunr}
\eeq
which reduces to $u^\mu=(1,{\vec v})$ in the non-relativistic limit. In particular,
one has $u^\mu=(1,{\vec 0})$ if the fluid is locally at rest (``local rest frame'').
Note that the 4-vector $u^\mu$ only contains three independent components since it obeys
the relation 
\beq
u^2\equiv u^\mu g_{\mu \nu} u^\nu=\gamma^2 (\vec{v})\left(1-{\vec{v}}^{\,2}\right)=1,
\eeq
so one does not need additional equations when trading ${\vec v}$ for the 
fluid 4-velocity $u^\mu$. 

To obtain the relativistic fluid dynamic equations, it is sufficient to 
derive the energy-momentum tensor $T^{\mu \nu}$ for a relativistic fluid,
as will be shown below. The energy-momentum tensor of an
ideal relativistic fluid (denoted as $T^{\mu \nu}_{(0)}$) has to be built
out of the hydrodynamic degrees of freedom, namely two Lorentz scalars ($\epsilon,p$)
and one vector $u^\mu$, as well as the metric tensor $g_{\mu \nu}$. 
Since $T^{\mu \nu}$ should be symmetric and transform as a tensor under Lorentz 
transformations, the most general form allowed by symmetry is therefore
\beq
T^{\mu \nu}_{(0)}=\epsilon \left(c_0 g^{\mu \nu}+c_1 u^\mu u^\nu\right)
+p \left(c_2 g^{\mu \nu}+c_3 u^\mu u^\nu\right)\ .
\label{genform}
\eeq
In the local restframe, one requires the $T_{(0)}^{00}$ component to represent the 
energy density $\epsilon$ of the fluid. Similarly, in the local rest frame, 
the momentum density should be vanishing
$T_{(0)}^{0 i}=0$, and the space-like components should be proportional to the 
pressure, $T_{(0)}^{ij}=p\, \delta^{ij}$ \cite{LL} \S133. Imposing these conditions
onto the general form (\ref{genform}) leads to the equations 
\beq
(c_0+c_1)\epsilon+(c_2+c_3)p=\epsilon,\qquad
-c_0 \epsilon-c_2 p = p,
\eeq
which imply $c_0=0,c_1=1,c_2=-1,c_3=1$, or
$T^{\mu \nu}_{(0)}=\epsilon\, u^\mu u^\nu- p\,\left(g^{\mu \nu}-u^\mu u^\nu\right)$.
For later convenience, it is useful to introduce the tensor
\beq
\Delta^{\mu \nu}=g^{\mu \nu}-u^\mu u^\nu.
\eeq
It has the properties $\Delta^{\mu \nu}u_\mu = \Delta^{\mu \nu}u_\nu=0$
and $\Delta^{\mu \nu} \Delta_\nu^\alpha=\Delta^{\mu \alpha}$ and serves
as a projection operator on the space orthogonal to the fluid velocity $u^\mu$.
In this notation, the energy-momentum tensor of an ideal relativistic fluid
becomes
\beq
T^{\mu \nu}_{(0)}=\epsilon\ u^\mu u^\nu- p\ \Delta^{\mu \nu}.
\label{idTmunu}
\eeq

If there are no external sources, the energy-momentum tensor is conserved,
\beq
\partial_\mu T^{\mu \nu}_{(0)}=0\ .
\label{EMT0cons}
\eeq
It is useful project these equations in the direction parallel 
($u_\nu \partial_\mu T^{\mu \nu}_{(0)}$) and perpendicular 
($\Delta^\alpha_\nu \partial_\mu T^{\mu \nu}_{(0)}$) 
to the fluid velocity.
For the first projection, one finds
\bqa
u_\nu \partial_\mu T^{\mu \nu}_{(0)} &=& u^\mu \partial_\mu \epsilon
+ \epsilon\, (\partial_\mu u^\mu) + \epsilon\, u_\nu u^\mu \partial_\mu u^\nu
-p\, u_\nu \partial_\mu \Delta^{\mu \nu}\,,\nonumber\\
&&=(\epsilon + p)\partial_\mu u^\mu+u^\mu \partial_\mu \epsilon =0 \ ,\label{pe1}
\eqa
where the identity 
$u_\nu \partial_\mu u^\nu =\frac{1}{2}\partial_\mu (u_\nu u^\nu)=\frac{1}{2}\partial_\mu 1 = 0$ 
was used. For the other projection one finds
\bqa
\Delta^\alpha_\nu \partial_\mu T^{\mu \nu}_{(0)}&=&
\epsilon\, u^\mu \Delta^\alpha_\nu \partial_\mu u^\nu-\Delta^{\mu \alpha} (\partial_\mu p)
+p\, u^\mu \Delta^\alpha_\nu \partial_\mu u^\nu\,,\nonumber\\
&=&(\epsilon+p)\, u^\mu \partial_\mu u^\alpha- \Delta^{\mu \alpha} \partial_\mu p=0\ . \label{pe2}
\eqa
Introducing the shorthand notations
\beq
D\equiv u^\mu \partial_\mu, \qquad \nabla^\alpha = \Delta^{\mu \alpha} \partial_\mu
\eeq
for the projection of derivatives parallel and perpendicular to $u^\mu$,
equations (\ref{pe1}),(\ref{pe2}) can be written as
\bqa
D\epsilon + (\epsilon+p)\partial_\mu u^\mu &=&0 \label{relcont}\\
(\epsilon+p)D u^\alpha-\nabla^\alpha p &=&0\ \label{releuler}.
\eqa
These are the fundamental equations for a relativistic ideal fluid.
Their meaning becomes transparent in the non-relativistic limit:
for small velocities $|{\vec{v}}|\ll1$ one finds 
\beq
D=u^\mu \partial_\mu \simeq \partial_t + \vec{v}\cdot\vec{\partial} + {\cal O}(|\vec{v}|^2),\qquad
\nabla^i=\Delta^{i \mu}\partial_\mu \simeq \partial^i+ {\cal O}(|\vec{v}|),
\eeq
so $D$ and $\nabla^i$ essentially reduce to time and space derivatives, respectively.
Imposing further a non-relativistic equation of state where $p\ll\epsilon$,
and that energy density is dominated by mass density $\epsilon\simeq \rho$, 
Eq.~(\ref{relcont}) becomes the continuity equation (\ref{Continuityeq}), and 
Eq.~(\ref{releuler}) the non-relativistic Euler equation (\ref{Eulereq}).

One thus recognizes the fluid dynamic equations (both relativistic and 
non-relativistic) to be identical to the conservation equations 
for the fluid's energy-momentum tensor.

\section{Relativistic Viscous Hydrodynamics}

\subsection{The relativistic Navier-Stokes equation}
\label{sec:relns}

In the ideal fluid picture, all dissipative (viscous) effects are by definition
neglected. If one is interested in a fluid description that
includes for instance the effects of viscosity, one has to go beyond
the ideal fluid limit, and in particular the fluid's energy momentum
tensor will no longer have the form Eq.~(\ref{idTmunu}).
Instead, one writes 
\beq
T^{\mu \nu}=T^{\mu \nu}_{(0)}+\Pi^{\mu \nu},
\label{EMTdecomp}
\eeq
where $T^{\mu \nu}_{(0)}$ is the familiar ideal fluid part given by
Eq.~(\ref{idTmunu}) and $\Pi^{\mu \nu}$ is the viscous stress tensor
that includes the contributions to $T^{\mu \nu}$ stemming from 
dissipation. Considering for simplicity a system without conserved charges
(or at zero chemical potential), all momentum density is due
to the flow of energy density
\beq
u_\mu T^{\mu \nu}=\epsilon\, u^\nu \longrightarrow \quad u_\mu \Pi^{\mu \nu}=0.
\eeq
While here this is the only possibility, for a more general system 
with conserved charges one can view this as a 
choice of frame for the definition of the fluid 4-velocity, 
sometimes referred to as Landau-Lifshitz frame. This can be easily
understood by recognizing that in a system with a conserved charge
there will be an associated charge current $n^\mu$ that 
can be used alternatively to define the fluid velocity, 
e.g. the Eckart frame $u_\mu n^\mu = n$. These choices reflect
the freedom of defining the local rest frame either as
the frame where the energy density (Landau-Lifshitz)
or the charge density (Eckart) is at rest. Since the physics 
must be the same in either of these frames, one can show
that charge diffusion in one frame is related to heat flow in
the other frame, as done e.g. in the appendix of \cite{Son:2006em}. 
For other recent discussions of relativistic viscous hydrodynamics in 
the presence of conserved charges, see e.g. \cite{Rischke:1998fq,Muronga:2003ta}.

Similar to the case of ideal fluid dynamics studied in section \ref{sec:1b},
the fundamental equations of viscous fluid dynamics are found by
taking the appropriate projections of the conservation equations of
the energy momentum tensor,
\bqa
u_\nu \partial_\mu T^{\mu \nu}&=&D\epsilon +(\epsilon+p)\partial_\mu u^\mu + u_\nu \partial_\mu \Pi^{\mu \nu}=0\,,
\nonumber\\
\Delta_\nu^\alpha \partial_\mu T^{\mu \nu}&=&(\epsilon+p)Du^\alpha-\nabla^\alpha p + \Delta^\alpha_\nu \partial_\mu 
\Pi^{\mu \nu}=0\ .
\eqa
The first equation can be further simplified by rewriting
$u_\nu \partial_\mu \Pi^{\mu \nu}=\partial_\mu \left(u_\nu \Pi^{\mu \nu}\right)-\Pi^{\mu \nu} \partial_{(\mu} u_{\nu)}$,
and using the identity
\beq
\partial_\mu = u_\mu D + \nabla_\mu
\eeq
as well as the choice of frame, $u_\mu \Pi^{\mu \nu}=0$. Here and in the following
the $(\ldots)$ denote symmetrization, e.g.
$$
A_{(\mu} B_{\nu)}=\frac{1}{2}\left(A_\mu B_\nu+A_\nu B_\mu\right)\ .
$$
Hence, the fundamental equations for relativistic viscous fluid dynamics are
\begin{center}
\fbox{\parbox{10cm}
{\bqa
D\epsilon +(\epsilon+p)\partial_\mu u^\mu -\Pi^{\mu \nu}\nabla_{(\mu} u_{\nu)}&=&0\,,\nonumber\\
(\epsilon+p)Du^\alpha-\nabla^\alpha p + \Delta^\alpha_\nu \partial_\mu \Pi^{\mu \nu} &=&0\ .
\label{visceq}
\eqa}}
\end{center}

At this point, however, the viscous stress tensor has not been specified. Indeed, much 
of the remainder of this work will deal with deriving expressions for $\Pi^{\mu \nu}$,
which together with (\ref{visceq}) will give different theories of viscous hydrodynamics.

An elegant way of obtaining $\Pi^{\mu \nu}$ builds upon the second law of 
thermodynamics, which states that entropy must always increase locally.
The entropy density $s$ is connected to  energy density, pressure and
temperature $T$ by the basic equilibrium thermodynamic relations for a system without 
conserved charges (or zero chemical potential),
\beq
\epsilon+p=T s,\qquad T ds=d\epsilon\label{thermodyn}\ .
\eeq
The second law of thermodynamics can be recast in the covariant form 
\beq
\partial_\mu s^\mu \ge 0 \label{2ndlaw}
\eeq
using the entropy 4-current $s^\mu$ which in equilibrium is given by
\beq
s^\mu=s u^\mu\label{smueq}\ .
\eeq
The thermodynamic relations (\ref{thermodyn}) allow to rewrite 
the second law (\ref{2ndlaw}) as 
\beq
\partial_\mu s^\mu = D s + s \partial_\mu u^\mu =
\frac{1}{T} D\epsilon+\frac{\epsilon+p}{T} \partial_\mu u^\mu
=\frac{1}{T} \Pi^{\mu \nu}\nabla_{(\mu} u_{\nu)}\ge 0\ ,
\label{expli2ndlaw}
\eeq
where (\ref{visceq}) was used to rewrite $D\epsilon$.
It is customary to split 
$\Pi^{\mu \nu}$ into a part $\pi^{\mu \nu}$ that is traceless, $\pi^\mu_\mu=0$,
and a remainder with non-vanishing trace,
\beq
\Pi^{\mu \nu}=\pi^{\mu \nu}+\Delta^{\mu \nu} \Pi\ .
\label{Pidecomp}
\eeq
Similarly one introduces a new notation for the traceless part of $\nabla_{(\mu} u_{\nu)}$,
\beq
\nabla_{<\mu} u_{\nu>}\equiv2 \nabla_{(\mu} u_{\nu)}-\frac{2}{3} \Delta_{\mu \nu} \nabla_\alpha u^\alpha\,,
\label{pointbrackdef}
\eeq
so that the the  second law becomes
\beq
\partial_\mu s^\mu=\frac{1}{2 T}\pi^{\mu \nu} \nabla_{<\mu} u_{\nu>}+\frac{1}{T} \Pi \nabla_\alpha u^\alpha \ge 0\ .
\eeq
One recognizes that this inequality is guaranteed to be fulfilled if 
\beq
\pi^{\mu \nu}=\eta \nabla^{<\mu} u^{\nu>}\,,\quad \Pi =\zeta \nabla_\alpha u^\alpha\,,\quad
\eta\ge0\,,\quad \zeta\ge0\,,
\label{NSrelations}
\eeq
because then $\partial_\mu s^\mu$ is a positive sum of squares.

In the non-relativistic limit, the viscous stress tensor becomes that of the
Navier-Stokes equations (\ref{visctens}), which leads one to equate $\eta,\zeta$ with
the shear and bulk viscosity coefficient, respectively. Also, for this reason
we refer to the system of equations (\ref{visceq}),(\ref{Pidecomp}),(\ref{NSrelations}) as
the relativistic Navier-Stokes equation. 
While beautifully simple, it turns out that the relativistic Navier-Stokes equation
-- unlike its non-relativistic counterpart -- 
exhibits pathologies for all but the simplest flow profiles, as will be shown
below.

\subsection{Acausality problem of the relativistic Navier-Stokes equation}
\label{sec:acprob}

Let us consider small perturbations of the energy density and fluid velocity 
in a system that is initially in equilibrium and at rest,
\beq
\epsilon=\epsilon_0+\delta \epsilon (t,x),\qquad u^\mu=(1,{\vec{0}})+\delta u^\mu(t,x),
\label{smallpert}
\eeq
where for simplicity the perturbation was assumed to be dependent
on one space coordinate only. The relativistic Navier-Stokes equation then 
specifies the space-time evolution of the perturbations. For 
the particular direction $\alpha=y$, Eq.~(\ref{visceq}) gives
\bqa
(\epsilon+p) D u^y-\nabla^yp+\Delta^y_\nu \partial_\mu \Pi^{\mu\nu}&=&(\epsilon_0+p_0)\partial_t \delta u^y
+\partial_x \Pi^{xy}+{\cal O}(\delta^2)\,,\nonumber\\
\Pi^{xy}=\eta \left(\nabla^x u^y+\nabla^y u^x\right)+\left(\zeta-\frac{2}{3} \eta\right)
\Delta^{xy} \nabla_\alpha u^\alpha&=&-\eta_0\ \partial_x \delta u^y+{\cal O}(\delta^2)\ .\nonumber
\eqa
This implies a diffusion-type evolution equation for the perturbation $\delta u^y(t,x)$:
\beq
\partial_t \delta u^y-\frac{\eta_0}{\epsilon_0+p_0} \partial_x^2 \delta u^y={\cal O}(\delta^2)\ .
\label{diffusioneq}
\eeq
To investigate the individual modes of this diffusion process, one can insert a mixed Laplace-Fourier 
wave ansatz
$$
\delta u^y(t,x)=e^{-\omega t+ i k x} f_{\omega,k}
$$
into Eq.~(\ref{diffusioneq}). 
This gives the ``dispersion-relation'' of the diffusion equation,
\beq
\omega=\frac{\eta_0}{\epsilon_0+p_0} k^2\ , \label{diffeqdisprel}
\eeq
which one can use to estimate the speed of diffusion of a mode with wavenumber $k$,
\beq
v_T(k)=\frac{d \omega}{dk}=2 \frac{\eta_0}{\epsilon_0+p_0} k\ .
\eeq
One finds that $v_T$ is linearly dependent on the wavenumber, which implies that 
as $k$ becomes larger and larger, the diffusion speed will grow without bound.
In particular, at some sufficiently large value of $k$, $v_T(k)$ 
will exceed the speed of light, which violates causality\footnote{
One should caution that the diffusion speed exceeding the speed of light is
a hint -- but no proof -- of causality violation. The proof is given
in the appendix.}. Therefore the relativistic Navier-Stokes 
equation does not constitute a causal theory.

The obvious conclusion to draw from this argument is that the relativistic
Navier-Stokes equation exhibits unphysical behavior for the short wavelength ($k\gg1$)
modes and hence can only be valid in the description of the long wavelength modes. 
This is not a principal problem, as one can regard hydrodynamics simply as 
an effective theory of matter in the long wavelength, $k\rightarrow 0$ limit. However, having
a finite range of validity in $k$ typically is a practical problem when dealing
with more complicated flow profiles that do not lend themselves to analytic
solutions and have to be solved numerically. In this case, it turns out that
the high $k$ modes are associated with instabilities \cite{Hiscock:1985zz} that 
make it necessary to regulate the theory by other means. A simple argument
to understand the practical problem can be given as follows: modes
that travel faster than the speed of light in one Lorentz frame correspond
to modes traveling backwards in time in a different frame. 
Hydrodynamics is an initial value problem which requires a well defined
set of initial conditions. However, if there are modes present
in the equations that travel backwards in time, the initial conditions
cannot be given freely \cite{Kostadt:2000ty}, and as a consequence one cannot solve
the relativistic Navier-Stokes equation numerically.

One possible way to regulate the theory is provided by considering
the ``Maxwell-Cattaneo law'' \cite{Maxwell,Cattaneo}
\beq
\tau_\pi\partial_t \Pi^{xy}+\Pi^{xy}=-\eta_0 \partial_x \delta u^y \, \label{MC}
\eeq
instead of the Navier-Stokes equation. Here $\tau_\pi$ is a new transport 
coefficient sometimes referred to as relaxation time. The effect of 
this modification becomes apparent when recalculating the dispersion relation
for the perturbation $\delta u^y$ using Eq.~(\ref{MC}). One finds
\beq
\omega=\frac{\eta_0}{\epsilon_0+p_0} \frac{k^2}{1-\omega \tau_\pi}\ , \label{moddisp}
\eeq
which coincides with the dispersion relation of the diffusion equation Eq.~(\ref{diffeqdisprel})
in the hydrodynamic ($\omega,k\rightarrow0$) limit. More interestingly, however, 
is that for large frequency $\omega\gg1$ Eq.~(\ref{moddisp}) does not describe
diffusive behavior, but instead propagating waves with a propagation speed 
that is finite in the limit of $k\gg 1$, 
\beq
v_T^{\rm max}\equiv \lim_{k\rightarrow \infty} \frac{d |\omega|}{d\ k}=
\sqrt{\frac{\eta_0}{(\epsilon_0+p_0)\tau_\pi}}\ ,
\label{vT}
\eeq
unless $\tau_\pi\rightarrow 0$. Interestingly, for all known fluids the 
limiting value $\sqrt{\frac{\eta_0}{(\epsilon_0+p_0)\tau_\pi}}$
has been found to be smaller than one, so that the Maxwell-Cattaneo law
seems to be an extension of the Navier-Stokes equation that preserve causality\footnote{
See the appendix for a proof of causality.}.

For heat flow, the corresponding Maxwell-Cattaneo law implies a dispersion relation equivalent 
to Eq.~(\ref{moddisp}), and there the propagating waves 
can be associated with the phenomenon of second sound \cite{Jou88}\S4, \cite{joseph89}, observed
experimentally in solid helium \cite{Ackerman66}. It is not known to me whether
propagating high frequency shear waves, as suggested by Eq.~(\ref{moddisp}), have been 
found in experiments.

While the Maxwell-Cattaneo law seems to be a successful phenomenological extension
of the Navier-Stokes equation, it is unsatisfactory that Eq.~(\ref{MC}) does 
not follow from a first-principles framework, but is rather introduced ``by hand''.
It will turn out, however, that the Maxwell-Cattaneo law -- while 
not derivable -- does seem to correctly capture some important aspects 
of relativistic viscous hydrodynamic theory, for instance that terms of higher
order in $k$ (higher order gradients) are needed to restore causality.

\subsection{M\"uller-Israel-Stewart theory, entropy-wise}
\label{sec:MISentropy}

In section \ref{sec:relns}, the Navier-Stokes equation was derived 
from the second law of thermodynamics $\partial_\mu s^\mu\ge 0$ by using
the form of the entropy current in equilibrium, $s^\mu=s u^\mu$. 
However, it is not guaranteed that the entropy current equals its equilibrium
expression for a dissipative fluid that can be out of equilibrium.
Specifically, it was suggested \cite{Mueller1,IS0a} that out of equilibrium the entropy
current can have contributions from the viscous stress tensor, which is 
sometimes referred to as ``extended irreversible thermodynamics'' \cite{Jou88,Jou99}.
Assuming that the entropy current has to be algebraic in the hydrodynamic degrees of 
freedom and that deviations from equilibrium are not too large so that high
order corrections can be neglected, the entropy current has to be of the
form \cite{Mueller1,IS0a,Muronga:2003ta}
\beq
s^\mu=s u^\mu-\frac{\beta_0}{2 T} u^\mu \Pi^2 -\frac{\beta_2}{2 T} u^\mu \pi_{\alpha \beta} \pi^{\alpha \beta} 
+ {\cal O}(\Pi^3)\ ,\label{2ndorders}
\eeq
where $\beta_0,\beta_2$ are coefficients that quantify the effect of these
second-order modifications of the entropy current.
Using again Eq.~(\ref{visceq}) to rewrite $\partial_\mu s^\mu$ as in section \ref{sec:relns}
one finds
\bqa
\partial_\mu s^\mu &=&\frac{\pi^{\alpha \beta}}{2 T}\left(\nabla_{<\alpha} u_{\beta>}
-\pi_{\alpha \beta} T D\left(\frac{\beta_2}{T}\right)-2 \beta_2 D \pi_{\alpha \beta}
-\beta_2 \pi_{\alpha \beta} \partial_\mu u^\mu\right)
\nonumber\\
&&+\frac{\Pi}{T}\left(\nabla_\alpha u^\alpha-\frac{1}{2}\Pi\ T D \left(\frac{\beta_0}{T}\right)-\beta_0 D \Pi
-\frac{1}{2}\beta_0 \Pi \partial_\mu u^\mu\right)\ge 0\ .
\label{MISineq}
\eqa
The inequality is guaranteed to be fulfilled if 
\bqa
\pi_{\alpha \beta}&=&\eta \left(\nabla_{<\alpha} u_{\beta>}
-\pi_{\alpha \beta} T D\left(\frac{\beta_2}{T}\right)-2 \beta_2 D \pi_{\alpha \beta}
-\beta_2\pi_{\alpha \beta} \partial_\mu u^\mu\right)\ ,
\nonumber\\
\Pi&=&\zeta\left(\nabla_\alpha u^\alpha-\frac{1}{2}\Pi\ T D \left(\frac{\beta_0}{T}\right)-\beta_0 D \Pi
-\frac{1}{2}\beta_0 \Pi \partial_\mu u^\mu\right)\ ,
\label{MISentropy}
\eqa
with $\eta,\zeta$ the usual bulk and shear viscosity coefficients.
Note that 
Eq.~(\ref{MISentropy}) coincides with the Navier-Stokes equation in the limit of $\beta_0,\beta_2\rightarrow 0$.
For non-vanishing $\beta_0,\beta_2$, Eq.~(\ref{MISentropy}) contains time derivatives
of $\pi_{\alpha \beta},\Pi$, which are similar (but not identical) 
to the Maxwell-Cattaneo law Eq.~(\ref{MC}) if one identifies
$\beta_2=\frac{\tau_\pi}{2 \eta}$ (and similarly, $\beta_0=\frac{\tau_\Pi}{\zeta}$). 
The set of equations (\ref{visceq}),(\ref{MISentropy}) (and some variations thereof)
are commonly referred to as ``M\"uller-Israel-Stewart'' theory 
and will be discussed more in section \ref{sec:kt}.

Similar to section \ref{sec:acprob}, one can study the causality properties
of the M\"uller-Israel-Stewart theory by considering small perturbations
around equilibrium, Eq.~(\ref{smallpert}). Keeping only perturbations
to first order, Eq.~(\ref{visceq}) and Eq.~(\ref{MISentropy}) become
\bqa
&\partial_t \delta \epsilon+(\epsilon_0+p_0) \partial_x \delta u^x =0,\quad
(\epsilon_0+p_0) \partial_t \delta u^x +\partial_x p+ \partial_\mu \delta \Pi^{\mu x}=0,&
\nonumber\\
&(\epsilon_0+p_0) \partial_t \delta u^y+\partial_\mu \delta \Pi^{\mu y}=0, \quad
\delta \Pi^{\mu \nu}=\delta \pi^{\mu \nu}+g^{\mu \nu} \delta \Pi &\nonumber\\
&\delta \pi^{xx}+\tau_\pi \partial_t \delta \pi^{xx}=-\frac{4}{3} \eta_0 \partial_x \delta u^x, \quad
\delta \pi^{xy}+\tau_\pi \partial_t \delta \pi^{xy}=-\eta_0 \partial_x \delta u^y, &\nonumber\\
&\delta \Pi+\tau_{\Pi} \partial_t \delta \Pi=\zeta_0 \partial_x \delta u^x\, .&
\label{collmodes}
\eqa
The equation of state $\epsilon=\epsilon(p)$ relates the pressure and energy density gradients,
$\partial_x p = \frac{dp}{d\epsilon} \partial_x \epsilon$, and the 
condition $u_\mu \Pi^{\mu \nu}=0$ implies $\delta \Pi^{t \nu}={\cal O}(\delta^2)$.
Using a Fourier ansatz
$$
\delta \epsilon = e^{i \omega t- i k x} \delta \epsilon_{\omega, k},\quad
\delta u^i = e^{i \omega t- i k x} \delta u^i_{\omega, k},\quad
\delta \pi^{\mu \nu}=e^{i \omega t- i k x} \delta \pi^{\mu \nu}_{\omega,k},\quad
\delta \Pi = e^{i \omega t- i k x} \delta \Pi^{\mu \nu}_{\omega,k},
$$
in Eq.~(\ref{collmodes}) gives the system of equations
\bqa
i \omega\ \delta \epsilon_{\omega, k}-ik (\epsilon_0+p_0)\  \delta u^x_{\omega, k}&=&0,\\
i \omega (\epsilon_0+p_0)\ \delta u^x_{\omega, k}-i k \frac{dp}{d\epsilon}\ \delta \epsilon_{\omega, k}
- i k \left(\frac{4}{3}\frac{i k \eta_0}{1+i \omega \tau_\pi}
+\frac{i k \zeta_0}{1 + i \omega \tau_\Pi}\right)\delta u^x_{\omega, k}&=&0,\\
i \omega (\epsilon_0+p_0)\ \delta u^y_{\omega, k}-ik
\left(\frac{i k \eta_0}{1+i \omega \tau_\pi}\right)\delta u^y_{\omega, k}&=&0\, .
\label{lastonew}
\eqa
Eq.~(\ref{lastonew}) corresponds to result from the Maxwell-Cattaneo law for the transverse
velocity perturbation $\delta u^y$, discussed in 
section \ref{sec:acprob}. The other two equations correspond to density perturbations
and longitudinal fluid velocity displacements, commonly known as sound.
The sound dispersion relation is given by
\beq
i \omega - i \frac{k^2}{\omega} \frac{dp}{d\epsilon}+
k^2\left(\frac{4}{3}\frac{\eta_0}{\epsilon_0+p_0} \frac{1}{1+i \omega \tau_\pi}
+\frac{\zeta_0}{\epsilon_0+p_0}\frac{1}{1 + i \omega \tau_\Pi}\right)=0\, ,
\label{disprelMISsound}
\eeq
and in the hydrodynamic limit ($\omega,k\ll1$) becomes
\bqa
\omega&=&\pm k c_s +i k^2 \left(\frac{2}{3} \frac{\eta_0}{\epsilon_0+p_0}
+\frac{1}{2}\frac{\zeta_0}{\epsilon_0+p_0}\right)\nonumber\\
&&\mp\frac{k^3}{2 c_s} \left[
\left(\frac{2}{3} \frac{\eta_0}{\epsilon_0+p_0}+\frac{1}{2}\frac{\zeta_0}{\epsilon_0+p_0}\right)^2
-2 c_s^2 \left( \frac{2}{3} \frac{\eta_0}{\epsilon_0+p_0} \tau_\pi+
\frac{1}{2}\frac{\zeta_0}{\epsilon_0+p_0} \tau_\Pi\right)\right]
+{\cal O}(k^4)\, .
\label{disprelsound}
\eqa
The quantity
\beq
c_s\equiv \sqrt{\frac{d\ p}{d\ \epsilon}}
\eeq
can be recognized to be the speed of sound when calculating the 
group velocity $\lim_{k\rightarrow 0}\frac{d \omega}{d k}$.
For large wavenumbers and frequencies, Eq.~(\ref{disprelMISsound})
gives a limiting sound mode group velocity of
\beq
v_L^{\rm max}\equiv\lim_{k\rightarrow \infty} \frac{d\omega}{dk}=\sqrt{c_s^2+
\frac{4}{3} \frac{\eta_0}{\tau_\pi(\epsilon_0+p_0)}
+\frac{\zeta_0}{\tau_\Pi(\epsilon_0+p_0)}}\, ,
\label{vL}
\eeq
which together with the result for the transverse mode Eq.~(\ref{vT})
suggests that the M\"uller-Israel-Stewart theory 
-- derived via an extended second law of thermodynamics -- 
constitutes a relativistic theory of viscous 
hydrodynamics that obeys causality if the relaxation times 
$\tau_\pi,\tau_\Pi$ are not too small.
Note that the requirement $v_L^{\rm max}\le1$ from Eq.~(\ref{vL}) 
is more restrictive than Eq.~(\ref{vT}) concerning the allowed values
of $c_s^2,\eta,\zeta,\tau_\pi,\tau_\Pi$.

However, many questions remain unanswerable within this formalism, e.g. how to obtain 
the value of $\tau_\pi,\tau_\Pi$, or whether the assumption that the entropy current
should be algebraic in the hydrodynamic degrees of freedom is valid
(Refs.~\cite{Loganayagam:2008is,Bhattacharyya:2008xc,Romatschke:2009kr} 
seem to indicate the contrary). Therefore,
it is necessary to have a different derivation of viscous hydrodynamics.

\section{Fluid Dynamics from Kinetic Theory}
\label{sec:kt}

\subsection{A very short introduction to kinetic theory}

Kinetic theory treats the evolution of the one-particle distribution
function $f(\vec{p},t,\vec{x})$, which can be associated with the 
number of on-shell particles per unit phase space,
\beq
f(\vec{p},t,\vec{x})\propto \frac{dN}{d^3p\, d^3x}\, .
\label{fdefprop}
\eeq
If collisions between particles can be neglected, the 
evolution of $f$ follows from Liouville's theorem,
\beq
\frac{d f}{d{\cal T}}=0=\frac{d t}{d{\cal T}}\frac{\partial f}{\partial t} +
\frac{d\vec{x}}{d{\cal T}}\cdot \frac{\partial f}{\partial \vec{x}}\label{Liouville}
\eeq
Multiplying (\ref{Liouville}) by the mass $m$ of a particle and 
recognizing $m \frac{d t}{d{\cal T}}=m\gamma(\vec{v})=p^0$, 
$m \frac{d {\vec x}}{d{\cal T}}=m{\vec v}\gamma(\vec{v})=\vec{p}$
as the particle's energy and momentum, respectively, Eq.~(\ref{Liouville})
becomes
\beq
p^\mu \partial_\mu f = 0 \, ,
\label{freestream}
\eeq
where $p^\mu$ has to fulfill the on-shell condition $p^\mu p_\mu = m^2$.

Eventually, collisions between particles cannot be neglected, and
hence Eq.~(\ref{Liouville}) is no longer valid. Taking the effect of 
collisions into account changes the evolution equation \cite{LL10}\S3 to
\beq
p^\mu \partial_\mu f = - {\cal C}[f] \label{Boltzmann},
\eeq
where ${\cal C}[f]$ is the collision term that is a functional of $f$ and 
the precise form of which depends on the particle interactions. Eq.~(\ref{Boltzmann}) is known as the
``Boltzmann-equation'' \cite{Boltzmann}. For a system in global
equilibrium $f$ is stationary, $f(\vec{p},t,\vec{x})=f_{(0)}(\vec{p})$ so that
the Boltzmann equation gives
$$
p^\mu \partial_\mu f_{(0)} = 0 = - {\cal C}[f_{(0)}],
$$
which implies that the collision term vanishes in equilibrium. Note that 
this means that Eq.~(\ref{freestream}) holds for two very different regimes,
namely firstly when one can ignore collisions (and the system is typically far from equilibrium)
and secondly when the collisions are strong enough to keep the system in equilibrium.
The first case is typically applicable if the timescale of the description is short enough so that 
the effect of particle collisions can be neglected.
Ultimately, however, particle collisions will become important and 
drive the system towards equilibrium. It is this second case, or more generally
the long time (small frequency, long wavelength) limit that corresponds to hydrodynamics
(see also the discussion in section \ref{sec:acprob}).

Given the interpretation of $f$ in Eq.~(\ref{fdefprop}), the particle number
density should be proportional to $\int d^3 p f$, or the sum of $f$ over all
momenta with weight unity. Summing instead with a weight of particle energy
$\int d^3 p\ p^0 f$, one expects a result proportional to the product 
of number density and energy, or energy density, which is a part of 
the energy-momentum tensor. More rigorously, one can define the relation
between the particle distribution function and the energy-momentum tensor 
\cite{RKT} as
\beq
\int\frac{d^4p}{(2\pi)^3} p^\mu p^\nu \delta(p^\mu p_\mu -m^2) 2 \theta(p^0) f(p,x) \equiv T^{\mu \nu}\, ,
\label{ktEMT}
\eeq
where the l.h.s.\ again can be understood as a sum over momenta, with the 
$\delta$-function imposing the condition of only counting on-shell particles and the 
step-function to restrict the sum to positive energy states.

\subsection{Ideal fluid dynamics from kinetic theory}
\label{sec:idealfluidkt}

In the following, I will limit myself to considering the ultrarelativistic
limit where all particle masses can be neglected, $m\rightarrow 0$. From Eq.~(\ref{ktEMT}),
this leads to $T^\mu_\mu=0$, or vanishing conformal anomaly. Interpreting (\ref{ktEMT})
as the fluid's energy-momentum tensor, this amounts to setting the bulk viscosity coefficient 
to zero, $\zeta=0$ (cf. Eq.~(\ref{EMTdecomp},\ref{NSrelations}) and the discussion
in section \ref{sec:nchyrdo}).

Introducing for convenience the shorthand notation 
\beq
\int d\chi \equiv \frac{d^4p}{(2\pi)^3} \delta(p^\mu p_\mu) 2 \theta(p^0),
\label{chidef}
\eeq
and taking the first moment of the Boltzmann equation, one finds
\beq
\int d\chi p^\nu p^\mu \partial_\mu f(p^\mu,x^\mu)=-\int d\chi p^\nu {\cal C}[f]=\partial_\mu \int d\chi p^\nu p^\mu f(p,x)
=\partial_\mu T^{\mu \nu}\ .
\label{BEfirstmom}
\eeq
For particle interactions that conserve energy and momentum, the integral over the collision term 
vanishes, $\int d\chi p^\nu {\cal C}[f]=0$. If $T^{\mu \nu}$ can be interpreted as a fluid's
energy-momentum tensor, then this implies that the first moment of the Boltzmann equation
corresponds to the fundamental equations of fluid dynamics (\ref{visceq}), since
these follow from $\partial_\mu T^{\mu \nu}=0$. 

The interpretation of the kinetic theory energy-momentum tensor in the fluid picture is 
most transparent in equilibrium, where $f(\vec{p},t,\vec{x})=f_{(0)}(\vec{p})$. Similar 
to the discussion in the introduction, $f_{(0)}(\vec{p})$ is not an optimal description
for a relativistic system, since it is not manifestly invariant under Lorentz transformations.
It is better to trade $f_{(0)}$ with a more convenient function,
$$
f_{(0)}(\vec{p})\rightarrow f_{\rm eq}\left(\frac{ p^\mu u_\mu}{T}\right),
$$
where $u^\mu$ is a four vector that reduces to $u^\mu\rightarrow (1,\vec{0})$ in the 
restframe of the heat bath with temperature $T$. 
Eq.~(\ref{ktEMT}) can then be written as
\beq
T^{\mu \nu}_{(0)}=\int d\chi p^\mu p^\nu f_{\rm eq}\left(\frac{ p^\mu u_\mu}{T}\right)
=a_{20} u^\mu u^\nu+a_{21} \Delta^{\mu \nu},
\label{ktemt0}
\eeq
where in hindsight it is more convenient to choose $u^\mu u^\nu, \Delta^{\mu\nu}$ as a tensor basis
then $u^\mu u^\nu, g^{\mu\nu}$. The coefficients $a_{20},a_{21}$ are functions of the 
temperature only and are determined by contracting (\ref{ktemt0}) with $u^\mu u^\nu$ and
$\Delta^{\mu \nu}$, respectively,
\beq
a_{20}=\int d\chi (p^\mu u_\mu)^2 f_{\rm eq}\left(\frac{ p^\mu u_\mu}{T}\right), \quad
a_{21}=\frac{1}{3}\int d\chi \left(p^\mu p_\mu - (p^\mu u_\mu)^2\right)f_{\rm eq}\left(\frac{ p^\mu u_\mu}{T}\right) \, .
\eeq

Identifying $u^\mu$ with the fluid four velocity, Eq.~(\ref{ktemt0}) corresponds
to the ideal fluid energy-momentum tensor Eq.~(\ref{idTmunu}) with $\epsilon=a_{20},p=-a_{21}$,
and the equation of state $\epsilon=3\ p$ (or speed of sound squared $c_s^2=\frac{1}{3}$)
following from on-shell condition in
the massless limit, $p^\mu p_\mu=0$.

To calculate $a_{20},a_{21}$, one has to specify the equilibrium distribution
function $f_{\rm eq}$. A concrete example where the evaluation is straightforward
is for a single species of particles that obey Boltzmann statistics, so that
$f_{\rm eq}\left(\frac{ p^\mu u_\mu}{T}\right)=\exp{\left[-\left(\frac{ p^\mu u_\mu}{T}\right)\right]}$.
In this case, $a_{20}$ is easily calculated by choosing the convenient frame $u^\mu=(1,\vec{0})$,
so that 
$$
a_{20}=\int \frac{d^4 p}{(2\pi)^3} (p^0)^2 \delta\left((p^0)^2-\vec{p}^{\, 2}\right)2\theta(p^0) e^{-p^0/T}
=\frac{1}{2 \pi^2} \int_0^\infty dp\ p^3 e^{-p/T}= \frac{3 T^4}{\pi^2},
$$
and $a_{21}=-\frac{1}{3}a_{20}=-\frac{T^4}{\pi^2}$. 
For a single species of particles obeying Bose-Einstein statistics, $f_{\rm eq}(x)=\left[e^x-1\right]^{-1}$, 
the result would be $a_{20}=\frac{3 T^4 \pi^2}{90}$.
The relation between $a_{20}$ and $\epsilon$
can be used to re-express the temperature in terms of the energy density.

\subsection{Out of equilibrium}

From Eq.~(\ref{ktemt0}) it is evident that when the argument of the distribution function $f$ 
depends only on scalars and one Lorentz vector $u^\mu$, the form of the 
energy-momentum tensor for kinetic theory is the same as for ideal fluid dynamics.
If the system is \emph{locally} in equilibrium, $f_{\rm eq}$ is completely characterized
by a vector-valued function that specifies the local rest frame of the heat bath, $u^\mu(x)$,
and the local temperature (or energy density). Therefore, a system that is in perfect
local equilibrium is described by ideal fluid dynamics. Departures from equilibrium
result in departures from the ideal fluid dynamics picture, and hence can only be
captured with dissipative (viscous) fluid dynamics. One can derive the correspondence 
between kinetic theory out of equilibrium and viscous hydrodynamics by
considering small departures from equilibrium where 
\beq
f(p^\mu,x^\mu)=f_{\rm eq}\left(\frac{ p^\mu u_\mu}{T}\right)\left[1+\delta f(p^\mu,x^\mu)\right]\, ,
\label{foutofeq}
\eeq
and $\delta f(p^\mu,x^\mu)\ll 1$. Using Eq.~(\ref{foutofeq}) in the definition of the energy momentum
tensor Eq.~(\ref{ktEMT}) and demanding that it should correspond to Eq.~(\ref{EMTdecomp})
from viscous hydrodynamics, one finds
\bqa
T^{\mu \nu}&=&T^{\mu \nu}_{(0)}+\int d\chi p^\mu p^\nu f_{\rm eq} \delta f=T^{\mu \nu}_{(0)}+\pi^{\mu \nu}\, ,\nonumber\\
\pi^{\mu \nu}&=&\int d\chi p^\mu p^\nu f_{\rm eq} \delta f\, , \label{consistrel}
\eqa
where again the contribution $\Pi$ proportional to bulk viscosity was dropped 
because of the ultrarelativistic limit in Eq.~(\ref{chidef}). The out-of-equilibrium
correction to the distribution function $\delta f$ may depend on scalars, 
the heat bath vector $u^\mu$, the metric $g_{\mu \nu}$, and gradients thereof. 
To make progress, it is convenient to make the dependence of $\delta f$ on the momentum $p^\mu$
explicit, e.g. by using a truncated expansion in a Taylor-like series \cite{IS0b}
\beq
\delta f(p^\mu,x^\mu)=c+p^\alpha c_{\alpha} + p^\alpha p^\beta c_{\alpha \beta}+{\cal O}(p^3)\,,
\eeq
or using a different basis \cite{RKT}. Using this expression in Eq.~(\ref{consistrel}) and
integrating over momenta, one can proceed to obtain the coefficients
$c,c_\alpha,c_{\alpha \beta}$ in a (somewhat tedious) calculation \cite{IS0b}. A more 
direct way (that gives the same result) is to assume -- similar to section \ref{sec:MISentropy} -- that $\delta f$ must  
be an algebraic function of the hydrodynamic degrees of freedom, $\epsilon,p,u^\mu,g^{\mu\nu},\pi^{\mu\nu}$.
Then the requirement that $\delta f$ vanishes in equilibrium implies that $c=0,c_\alpha=0$,
and $c_{\alpha \beta}=c_2 \pi_{\alpha \beta}$ with $c_2$ a function of the thermodynamic
variables $\epsilon,p$. The relation Eq.~(\ref{consistrel}) then leads to 
\beq
\pi^{\mu \nu}=\pi_{\alpha \beta} c_2 I^{\mu \nu \alpha \beta}\, ,
\label{pirel}
\eeq
where $I^{\mu \nu \alpha \beta}$ corresponds to the $n=4$ case of the integral definition
\beq
I^{\mu_1 \mu_2 \ldots \mu_n}=\int d\chi p^{\mu_1} p^{\mu_2}\ldots p^{\mu_n} f_{\rm eq}\, .
\label{Imunudef}
\eeq
Note that for the special case of two indices $I^{\mu \nu}=T^{\mu \nu}_{\rm (0)}$, and 
a decomposition into Lorentz tensors similar to Eq.~(\ref{ktemt0}) can be done for 
each of the integrals (\ref{Imunudef}). In particular, one finds
\beq
I^{\mu \nu \alpha \beta}=a_{40} u^\mu u^\nu u^\alpha u^\beta+a_{41} \left(u^\mu u^\nu \Delta^{\alpha \beta}
+{\rm perm.}\right) + a_{42} \left(\Delta^{\mu \nu} \Delta^{\alpha \beta}+\Delta^{\mu \alpha} \Delta^{\nu \beta}
+\Delta^{\mu \beta} \Delta^{\nu \alpha}\right)\, ,
\label{I4}
\eeq
where ``${\rm perm.}$'' denotes all non-trivial index permutations.
Contracting the indices in Eq.~(\ref{pirel}) using the properties of the shear part
of the viscous stress tensor, $u_\mu \pi^{\mu \nu}=0,\pi^\mu_\mu=0$, one finds $c_2=\frac{1}{2 a_{42}}$
and, with Eq.~(\ref{foutofeq}), the distribution function for small departures from equilibrium
takes the form
\beq
f(p^\mu,x^\mu)=f_{\rm eq}\left(\frac{ p^\mu u_\mu}{T}\right)
\left[1+\frac{1}{2 a_{42}} p^\alpha p^\beta \pi_{\alpha \beta}\right]\, .
\label{dfrel}
\eeq
The coefficients $a_{40},a_{41},a_{42}$ can be calculated the same way as $a_{20},a_{21}$ 
in section \ref{sec:idealfluidkt} once $f_{\rm eq}$ has been specified.
For the special case of a Boltzmann gas where $f_{\rm eq}(x)=e^{-x}$, 
a straightforward calculation gives the relation
$$
a_{42}=(\epsilon+p) T^2\, ,
$$
which holds also when allowing for nonzero particle masses.

The equation (\ref{dfrel}) establishes the relation of the 
particle distribution function out of (but close to) equilibrium and viscous hydrodynamics.
Still missing is for a relation of the Boltzmann equation (\ref{Boltzmann}) 
and viscous hydrodynamics is an expression for the collision term. 
Depending on the nature of the particle interactions, ${\cal C}[f]$ will have 
a particular, and sometimes complicated, form that can be simplified by 
assuming small departures from equilibrium, cf. Eq.~(\ref{foutofeq}).

If one identifies the magnitude of $\delta f$ with the size of 
gradients of hydrodynamic degrees of freedom,  
a shortcut to obtain ${\cal C}[f]$ to lowest order in a gradient
expansion is to insert (\ref{foutofeq}) into the Boltzmann equation:
\beq
{\cal C}[f]=-p^\mu\partial_\mu \left[f_{\rm eq}\left(1+\delta f\right)\right]=
-p^\mu\partial_\mu f_{\rm eq} + {\cal O}(\delta^2)\, .
\label{CfromB}
\eeq
This approach is similar to the Chapman-Enskog approach to fluid dynamics
\cite{CE}. 

For the special case of particles obeying Boltzmann statistics, $f_{\rm eq}=e^{-x}$,
the calculation of ${\cal C}[f]$ from Eq.~(\ref{CfromB}) to first 
order in gradients is simple and will be given here as an
illustrative example. Using the fundamental equations of viscous fluid 
dynamics (\ref{visceq}), one can rewrite
\bqa
p^\mu \partial_\mu f_{\rm eq}\left(\frac{ p^\mu u_\mu}{T}\right)&=&
-p^\mu p^\nu f_{\rm eq}\left(\nabla_\mu +u_\mu D\right)\frac{u_\nu}{T}\nonumber\\
&=&-\frac{p^\mu p^\nu}{T} f_{\rm eq}\left(\nabla_{\mu} u_{\nu}+u_{[\mu} \nabla_{\nu]} \ln T
+\frac{1}{3} u_{\mu} u_\nu \nabla_\alpha u^\alpha\right) + {\cal O}(\delta^2),
\label{intermed}
\eqa
where here and in the following $[\dots]$ denote antisymmetrization, e.g.
$$
A_{[\mu} B_{\nu]} = \frac{1}{2} \left(A_\mu B_\nu-A_\nu B_\mu\right)\ .
$$
Since the structure $p^\mu p^\nu$ in Eq.~(\ref{intermed}) is symmetric in the indices
and vanishes when contracted with $g_{\mu \nu}$ because of the on-shell condition,
Eq.~(\ref{intermed}) reduces to
\beq
p^\mu \partial_\mu f_{\rm eq}=-\frac{p^\mu p^\nu}{2 T} \nabla_{<\mu} u_{\nu>} f_{\rm eq}+{\cal O}(\delta^2)\, ,
\eeq
where $\nabla_{<\mu} u_{\nu>}$ was defined in Eq.~(\ref{pointbrackdef}). To first 
order in gradients, the Navier-Stokes equation (\ref{NSrelations}) is valid and 
as a consequence one finds
\beq
{\cal C}[f]=\frac{p^\mu p^\nu}{2 T \eta} \pi_{\mu \nu} f_{\rm eq}+{\cal O}(\delta^2)\ ,
\label{colltermexp}
\eeq
which establishes the relation between the collision term and viscous hydrodynamics
to first order in gradients.

\subsection{M\"uller-Israel-Stewart theory, kinetic theory-wise}
\label{sec:MISkt}

In a theory with conserved charges where $\int d\chi {\cal C}=0$, the integral over momenta (or zeroth moment)
of the Boltzmann equation gives
\beq
\int d\chi p^\mu \partial_\mu f= \partial_\mu \int d\chi p^\mu f= \partial_\mu n^\mu = 0,
\eeq
or conservation of charge current $n^\mu$. The first moment of the Boltzmann
equation (shown in Eq.~(\ref{BEfirstmom})) 
gives the conservation of the energy-momentum tensor, since $\int d\chi p^\alpha {\cal C}=0$.
However, the integral $\int d\chi p^\alpha p^\beta {\cal C}$ does not trivially vanish,
unless the system is in equilibrium. Therefore, the second moment of the Boltzmann
equation 
\beq
\int d\chi p^\alpha p^\beta p^\mu \partial_\mu f = - \int d\chi p^\alpha p^\beta {\cal C}[f]\,,
\label{BEsecmom}
\eeq
will carry some information about the non-equilibrium (or viscous) dynamics
of the system \cite{Grad}. Considering again small departures from equilibrium, 
Eq.~(\ref{dfrel}) implies
\beq
\int d\chi p^\alpha p^\beta p^\mu \partial_\mu f = 
\partial_\mu \left(I^{\alpha \beta \mu}+\frac{\pi_{\gamma \delta}}{2 a_{42}} I^{\alpha \beta \mu \gamma \delta}\right)\, ,
\label{intermed2}
\eeq
where the integrals $I^{\mu_1\mu_2\ldots\mu_n}$ were defined in Eq.~(\ref{Imunudef}). Similar to
Eq.~(\ref{I4}) one can do a tensor decomposition of $I^{\alpha \beta \mu}$, $I^{\alpha \beta \mu \gamma \delta}$,
with coefficients $a_{30},a_{31}$ and $a_{50},a_{51},a_{52}$, respectively.
To extract the relevant terms from Eq.~(\ref{intermed2}), it is useful to use a tensor projector
on the part that is symmetric and traceless,
\beq
P^{\mu \nu}_{\alpha \beta}=\Delta^{\mu}_\alpha \Delta^\nu_\beta+\Delta^{\mu}_\beta \Delta^\nu_\alpha
-\frac{2}{3} \Delta^{\mu \nu} \Delta_{\alpha \beta}\, ,
\label{tensorproj}
\eeq
with properties $u^\alpha P^{\mu \nu}_{\alpha \beta}=u^\beta P^{\mu \nu}_{\alpha \beta}=0$,
$\Delta^{\alpha \beta}P^{\mu \nu}_{\alpha \beta}=0$ and $\pi^{\alpha \beta} P^{\mu \nu}_{\alpha \beta}=2 \pi^{\mu \nu}$.
Using this projector on Eq.~(\ref{intermed2}), one finds after some algebra
\bqa
P^{\mu \nu}_{\alpha \beta} \partial_\phi I^{\alpha \beta \phi} &=&
P^{\mu \nu}_{\alpha \beta} a_{31} \left[D \Delta^{\alpha \beta}+2 \nabla^{(\alpha} u^{\beta)}\right]=
2 a_{31} \nabla^{<\mu} u^{\nu>}\, ,\nonumber\\
P^{\mu \nu}_{\alpha \beta} \partial_\phi \left[\frac{\pi_{\gamma \delta}}{2 a_{42}} I^{\alpha \beta \phi \gamma \delta}\right]
&=&P^{\mu \nu}_{\alpha \beta} \partial_\phi \left[\frac{a_{52}}{a_{42}} 3 \pi^{(\alpha \beta} u^{\phi)}\right]
\\
&=&2 \pi^{\mu \nu}D \left(\frac{a_{52}}{a_{42}}\right)
+2 \frac{a_{52}}{a_{42}} \left(\Delta^\mu_\alpha \Delta^\nu_\beta D \pi^{\alpha \beta}+
P^{\mu \nu}_{\alpha \beta} \pi^{\phi \beta} \nabla_\phi u^\alpha+\pi^{\mu\nu}\partial_\alpha u^\alpha\right)\, .
\nonumber
\label{intermed3}
\eqa

To calculate the r.h.s.\ of Eq.~(\ref{BEsecmom}) one would have to specify
the precise form of the collision integral. If one is only interested
in the form of the equation, not the coefficients of the individual terms,
it is convenient to again assume Boltzmann statistics, $f_{\rm eq}(x)=e^{-x}$,
because then the form of the collision term is given by Eq.~(\ref{colltermexp})
and one finds
\beq
P^{\mu \nu}_{\alpha \beta}\int d\chi p^\alpha p^\beta {\cal C}[f]=P^{\mu \nu}_{\alpha \beta}\frac{\pi_{\gamma \delta}}{2 T \eta} I^{\alpha \beta \gamma \delta}
=P^{\mu \nu}_{\alpha \beta}\frac{a_{42} \pi^{\alpha \beta}}{T \eta}=\frac{2 a_{42} \pi^{\mu \nu}}{T \eta}\, .
\label{intermed4}
\eeq
The coefficients $a_{31},a_{42},a_{52}$ are readily evaluated for a massless Boltzmann gas,
$$
a_{31}=-\frac{4 T^5}{\pi^2},\quad a_{42}=\frac{4T^6}{\pi^2},\quad a_{52}=\frac{24 T^7}{\pi^2}\, ,
$$
and after collecting the terms from Eq.~(\ref{intermed3}) and Eq.~(\ref{intermed4}) one finds
for the second moment of the Boltzmann equation (\ref{BEsecmom}) the result
\beq
\pi^{\mu \nu}+\frac{a_{52} T \eta}{a_{42}^2} \left[\Delta^\mu_\alpha \Delta^\nu_\beta D \pi^{\alpha \beta}+
P^{\mu \nu}_{\alpha \beta} \pi^{\phi \beta} \nabla_\phi u^\alpha+\pi^{\mu\nu}\partial_\alpha u^\alpha
+\pi^{\mu \nu} D\ln{T}\right]=\eta \nabla^{<\mu} u^{\nu>}\, .
\label{intermed5}
\eeq
It is useful to rewrite the expression $P^{\mu \nu}_{\alpha \beta} \pi^{\phi \beta} \nabla_\phi u^\alpha$
in this equation by introducing the fluid vorticity
\beq
\Omega_{\alpha \beta}=\nabla_{[\alpha}u_{\beta]}\, ,
\eeq
which is antisymmetric, $\Omega_{\alpha \beta}=-\Omega_{\beta \alpha}$.
After some algebra one finds the relation
\bqa
P^{\mu \nu}_{\alpha \beta} \pi^{\phi \beta} \nabla_\phi u^\alpha&=&
P^{\mu \nu}_{\alpha \beta} \Delta^{\alpha \gamma} \pi^{\phi \beta} \left[
\Omega_{\phi \gamma}+\frac{1}{2}\nabla_{<\phi} u_{\gamma>}+\frac{1}{3} \Delta_{\phi \gamma} \nabla_\delta u^\delta\right]
\nonumber\\
&=&-2 \pi^{\phi (\mu}\Omega^{\nu)}_{\ \phi}+\frac{\pi^{\phi <\mu} \pi^{\nu>}_\phi}{2 \eta} + \frac{2}{3} \pi^{\mu \nu}
\nabla_\delta u^\delta\, +{\cal O}(\delta^3),
\eqa
where (\ref{NSrelations}) was used to rewrite $\nabla_{<\phi} u_{\gamma>}$
to first order in gradients.
For the massless Boltzmann gas, one furthermore
has $D\ln{T}=D \ln \epsilon^{1/4}=-\frac{1}{3}\nabla_\alpha u^\alpha+{\cal O}(\delta^2)$, so that 
Eq.~(\ref{intermed5}) becomes
\beq
\pi^{\mu \nu}+\tau_\pi \left[\Delta^\mu_\alpha \Delta^\nu_\beta D \pi^{\alpha \beta}+
\frac{4}{3}\pi^{\mu\nu}\nabla_\alpha u^\alpha-2 \pi^{\phi (\mu}\Omega^{\nu)}_{\ \phi}
+\frac{\pi^{\phi <\mu} \pi^{\nu>}_\phi}{2 \eta}\right]
=\eta \nabla^{<\mu} u^{\nu>}+{\cal O}(\delta^2)\, ,
\label{MISktwise}
\eeq
where the expression $\frac{a_{52} T \eta}{a_{42}^2}$ was labeled $\tau_\pi$
to make the connection to the Maxwell-Cattaneo law Eq.~(\ref{MC}) explicit.
Eq.~(\ref{MISktwise}) constitutes a different variant of the M\"uller-Israel-Stewart
theory, and the connection between this equation and Eq.~(\ref{MISentropy}), which
was derived earlier in section \ref{sec:MISentropy} via the second law
of thermodynamics, will be discussed below.

Since $\tau_\pi$ multiplies all the terms in Eq.~(\ref{MISktwise}) which are at least of second
order in gradients, it is a generalization of the concept of hydrodynamic transport
coefficients (such as shear viscosity $\eta$), and is accordingly referred to
as a second order transport coefficient. For a Boltzmann gas, the known values of
$a_{42},a_{52}$ imply the relation 
\beq
\frac{\eta}{\tau_{\pi}}=\frac{2 T^4}{3 \pi^2}=\frac{2}{3}p\, ,
\label{simpletprel}
\eeq
which together with $c_s^2=\frac{1}{3}$ give the 
definite values $v_T^{\rm max}=\sqrt{\frac{1}{6}}$ and $v_L^{\rm max}=\sqrt{\frac{5}{9}}$ for the propagation speeds 
Eq.~(\ref{vT},\ref{vL}). This indicates that the theory by M\"uller, Israel and Stewart
does indeed preserve causality since signal propagation is subluminal.

For a massive Boltzmann gas, one can recalculate the coefficients $a_{52},a_{42}$ to
show that the more general relation
$$
\frac{\eta}{\tau_{\pi}}=\frac{\epsilon+p}{3+\frac{T}{s}\frac{ds}{dT}}
$$
holds. Also, for Bose-Einstein statistics, the proportionality factor $\frac{2}{3}$
in Eq.~(\ref{simpletprel}) is only changed by a few percent \cite{Baier:2006um}.
Thus it seems that the property of causality of the viscous fluid dynamic
equations (\ref{visceq}),(\ref{MISktwise}) is fairly robust 
whenever kinetic theory is applicable.

\subsection{Discussion of M\"uller-Israel-Stewart theory}
\label{sec:discMIS}

Eq.~(\ref{MISktwise}) contains the Navier-Stokes equation (\ref{NSrelations})
in the limit of small departures from equilibrium where second
order gradients (all the terms multiplied by $\tau_\pi$ in Eq.~(\ref{MISktwise}))
can be neglected. However, the form of the terms to second order 
in gradients is such that Eq.~(\ref{MISktwise}) reproduces the phenomenologically
attractive feature of the Maxwell-Cattaneo law, namely finite
signal propagation speed. In addition, the kinetic theory derivation of 
Eq.~(\ref{MISktwise}) also gives a definite relation between shear viscosity
and relaxation time, the first and second order transport coefficients, 
respectively, which implies not only finite, but subluminal signal
propagation.

However, the evolution equation for the shear stress $\pi^{\mu \nu}$ differs
between the derivation from kinetic theory Eq.~(\ref{MISktwise}) and 
the second law of thermodynamics, Eq.~(\ref{MISentropy}), respectively.
To make this more apparent, it is useful to rewrite Eq.~(\ref{MISentropy})
for the case of a Boltzmann gas with $\beta_2=\frac{\tau_\pi}{2\eta}=\frac{3}{4p}$,
\beq
\pi^{\mu \nu}+\tau_\pi \left[D \pi^{\mu \nu}+\frac{4}{3}\pi^{\mu\nu}\nabla_\alpha u^\alpha
\right]
=\eta \nabla^{<\mu} u^{\nu>}+{\cal O}(\delta^2)\, ,
\label{MISentropyagain}
\eeq
where Eq.~(\ref{visceq}) was used to 
rewrite $D \ln{T}=-\frac{1}{3}\nabla_\alpha u^\alpha+{\cal O}(\delta^2)$.
One first notes that the terms involving the time derivative $D\pi^{\alpha \beta}$ 
differ between Eq.~(\ref{MISentropyagain}) and Eq.~(\ref{MISktwise}),
$$
\Delta^\mu_\alpha \Delta^\nu_\beta D \pi^{\alpha \beta}-D \pi^{\mu \nu}=
-u^\mu u_\alpha D\pi^{\alpha \nu}-u^\nu u_\beta D\pi^{\mu \beta}+u^\mu u^\nu u_\alpha u_\beta D \pi^{\alpha \beta}\, .
$$
This difference is easily explained by noting that for the derivation 
of Eq.~(\ref{MISentropyagain}), only the projection of $\pi_{\mu \nu}$ on 
Eq.~(\ref{MISentropyagain}) was required to have well defined sign (\ref{MISineq}).
But the difference between Eq.~(\ref{MISentropyagain}) and Eq.~(\ref{MISktwise}) vanishes
when contracted with $\pi_{\mu \nu}$, so these terms do not actually contribute to entropy production
and therefore are not naturally captured by the derivation in section \ref{sec:MISentropy}. Nevertheless, 
one can convince oneself that these terms are necessary and important by contracting 
Eq.~(\ref{MISktwise}) and (\ref{MISentropyagain}) with $u_\mu$: unlike Eq.~(\ref{MISktwise}),
the contraction does not vanish for (\ref{MISentropyagain}), but instead gives
$u_\mu D \pi^{\mu \nu}=0$ which amounts to an extra (unphysical) constraint on the evolution
of the shear stress tensor \cite{Baier:2006sr}. Therefore, the variant Eq.~(\ref{MISktwise})
of M\"uller-Israel-Stewart theory derived from kinetic theory is superior to
Eq.~(\ref{MISentropyagain}) in this respect. 

However, when inserting the kinetic theory result Eq.~(\ref{MISktwise}) into
the conservation equation for the entropy current (\ref{MISineq}), one 
finds for the shear viscosity contribution the requirement
\beq
\frac{\pi_{\mu \nu}}{2 T} \left[\frac{\pi^{\mu \nu}}{\eta}
+\frac{\tau_\pi}{2 \eta^2} \pi^{\phi <\mu} \pi^{\nu>}_\phi\right]\ge0\, ,
\label{entropyviol}
\eeq
where the identity $\pi_{\mu \nu} \pi^{\phi (\mu}\Omega^{\nu)}_{\ \phi} =0$ has been used.
On the one hand, there is no obvious reason why Eq.~(\ref{entropyviol}) should be fulfilled for all
values of $\pi^{\mu \nu}$, but on the other hand the second law of thermodynamics $\partial_\mu s^\mu\ge 0$
should not be violated. 

One solution to the problem is that Eq.~(\ref{entropyviol}) may still be fulfilled if
departures from equilibrium are small enough, so that the first 
term in Eq.~(\ref{entropyviol}) ---being second order in gradients and manifestly 
positive \cite{Baier:2007ix}--- is larger than the other term, which is third order in gradients,
$\pi^{\mu \nu}\pi_{\mu \nu}\sim {\cal O}(\delta^2)\gg {\cal O}(\delta^3)$.
In other words, the region of applicability of viscous hydrodynamics would coincide with 
the region of applicability of the gradient expansion used to derive it.

However, most likely one also has to give up the assumption made in Eq.~(\ref{2ndorders})
about the particular form of the generalized entropy
current. Indeed, a different form for $s^\mu$ allowing for gradients 
\cite{Loganayagam:2008is,Bhattacharyya:2008xc,Romatschke:2009kr} does
seem to imply $\partial_\mu s^\mu\ge0$ for evolution equations of $\pi^{\mu\nu}$
that are more general than Eq.~(\ref{MISentropyagain}). 

While this implies that the correct theory is more complicated,
Eq.~(\ref{MISktwise}) is a good candidate for a
theory of relativistic viscous hydrodynamics that fulfills the necessary
minimal 
requirements, namely reduction to Navier-Stokes equation in the limit
of long wavelengths, and causal signal propagation. The shortcoming
of Eq.~(\ref{MISktwise}) is that the equation has unknown corrections
to second order in gradients ${\cal O}(\delta^2)$, stemming from the unknown form
of the collision term. Since the second-order gradients on the l.h.s.\ of
Eq.~(\ref{MISktwise}) are needed to guarantee finite signal propagation, 
it does not seem to be consistent to ignore terms of second order on the r.h.s.\
Rather, one would want to have a more general theory that includes all terms to second order 
in gradients.

\section{A new theory of relativistic viscous hydrodynamics}

\subsection{Hydrodynamics as a gradient expansion}

All the hydrodynamic results discussed so far can be classified in terms of a
gradient expansion of the fluid's energy momentum tensor\footnote{Again for simplicity
only the case of shear viscosity is discussed, where $\Pi^{\mu \nu}=\pi^{\mu \nu}$.}
$T^{\mu \nu}=T^{\mu \nu}_{(0)}+\pi^{\mu \nu}$,
namely
\begin{itemize}
\item
Ideal Hydrodynamics: contains no gradients (zeroth order), 
$$\pi^{\mu \nu}=0$$
\item
Navier-Stokes Equation: contains first order gradients, $$\pi^{\mu \nu}=\eta \nabla^{<\mu }u^{\nu>}$$
\item
M\"uller-Israel-Stewart theory: contains second order gradients, 
$$\pi^{\mu \nu}=\eta \nabla^{<\mu }u^{\nu>}+\tau_\pi\left[\Delta^\mu_\alpha \Delta^\nu_\beta D \pi^{\alpha \beta}
\ldots\right]+{\cal O}(\delta^2)\, .$$
\end{itemize}
As discussed in the introduction, the ideal hydrodynamic energy-momentum tensor 
is the most general structure allowed by symmetry, and therefore the zeroth order gradient
expansion is complete. On the other hand, section \ref{sec:discMIS} indicates that
the M\"uller-Israel-Stewart theory potentially misses terms of second order in gradients,
and hence the gradient expansion may not be complete to this order.  
To obtain the most general structure of viscous hydrodynamics to second order,
one has to completely classify all possible terms in $\pi^{\mu \nu}$ to first and second order 
gradients of the hydrodynamic degrees of freedom \cite{Baier:2007ix}.

To first order, since the equation of state links the pressure to the energy density,
the only independent gradients are $\partial_\mu u^\alpha, \partial_\mu \epsilon$.
Decomposing $\partial_\mu=\nabla_\mu + u_\mu D$, the fundamental equations (\ref{visceq})
can be used to express all time-like derivatives $D u^\alpha, D\epsilon$ in terms of
space-like gradients $\nabla_\mu$, so only the latter are independent. This implies
that the shear-stress tensor should have the structure
\beq
\pi^{\mu \nu}=c_4 \nabla^{(\mu } u^{\nu)}+c_5 \Delta^{\mu \nu} \nabla_\alpha u^\alpha+c_6\, u^{(\mu} \nabla^{\nu)} \epsilon\, ,
\eeq
where $c_4,c_5,c_6$ are functions of $\epsilon$ only. 
The Landau-Lifshitz frame condition $u_\mu \pi^{\mu \nu}=0$ implies that $c_6=0$,
or the absence heat flow (see section \ref{sec:relns}). Furthermore, since 
effects from bulk viscosity have been ignored, the stress tensor is traceless, which
gives $c_5=-\frac{1}{3}c_4$. Choosing the proportionality constant $c_4=2 \eta$,
one finds $\pi^{\mu \nu}=\eta \nabla^{<\mu } u^{\nu>}$, which shows that 
the Navier-Stokes equation corresponds to a complete gradient expansion to first order.

To second order in gradients, the analysis proceeds similar to the one above,
but there are more terms to consider. It turns out that for the case
of only shear viscosity, there is an additional restriction for $\pi^{\mu \nu}$ besides
$u_\mu \pi^{\mu\nu}=0$ and $\pi^\mu_\mu=0$, namely conformal symmetry, 
that can be used to reduce the number of possible structures.

\subsection{Conformal viscous hydrodynamics}
\label{sec:cvh}

A theory is said to be conformally symmetric if its
action is invariant under Weyl transformations of the metric,
\beq
g_{\mu \nu}\rightarrow \bar{g}_{\mu \nu}=e^{-2 w(x)} g_{\mu \nu}\, ,
\label{Weylresc}
\eeq
where $w(x)$ can be an arbitrary function of the spacetime coordinates,
and hence $g_{\mu\nu}$ is the metric of curved rather
than flat spacetime. While on the classical level many theories
obey this invariance, quantum correction typically spoil
the symmetry, giving rise to a non-vanishing 
trace of the energy momentum tensor. One distinguishes between 
theories where in flat space
quantum corrections generate $T^\mu_\mu\neq 0$---such 
as SU(N) gauge theories (``non-conformal'')---and 
those where conformal symmetry is unbroken,
such as ${\cal N}=4$ Super Yang-Mills (``conformal'').
Note that even for ``conformal'' theories quantum
corrections may couple to gravity, such 
that the trace of the energy-momentum tensor is
non-vanishing in curved space (``Weyl anomaly'') \cite{Duff:1993wm},
\beq
g_{\mu \nu} T^{\mu \nu}=T^{\mu}_\mu=W[g_{\mu \nu}]\, .
\eeq
The Weyl anomly $W[g_{\mu \nu}]$ in four dimensions is a function of the
product of either two Riemann tensors $R_{\mu\nu\lambda\rho}$,
two Ricci tensors $R_{\mu \nu}$ or two Ricci scalars $R$,
and hence is of fourth order in derivatives of $g_{\mu \nu}$,
since $R_{\mu\nu\lambda\rho}$, $R_{\mu \nu}$ and $R$ are all second
order in derivatives \cite{Aharony:1999ti}. Being interested in a gradient expansion
to second order, one may therefore effectively ignore the presence of the Weyl
anomaly. To second order in gradients, conformally invariant theories
thus have a traceless energy-momentum tensor, which in addition transforms
as 
\beq
T^{\mu\nu}\rightarrow \bar{T}^{\mu \nu}=e^{6 w(x)} T^{\mu\nu}
\eeq
under a Weyl rescaling in four dimensions \cite{Baier:2007ix} (see also
the discussion in section \ref{sec:nchyrdo}). 
It is this additional symmetry of conformal theories that helps
to restrict the possible second order gradient terms in a 
theory of hydrodynamics in the presence of shear viscosity.
For curved space, there are 8 possible contributions
of second order in gradients to $\pi^{\mu \nu}$ that
obey $\pi^\mu_\mu=0,\ u_\mu \pi^{\mu \nu}=0$,
\bqa
&D^{<\mu} \ln \epsilon\, D^{\nu>} \ln \epsilon, \quad D^{<\mu} D^{\nu>} \ln \epsilon, \quad 
\nabla^{<\mu} u^{\nu>} \left(\nabla_\alpha u^\alpha\right), \quad
P^{\mu \nu}_{\alpha \beta}\ \nabla^{<\alpha} u^{\gamma>} g_{\gamma \delta}
\nabla^{<\delta} u^{\beta>}&\nonumber\\
&P^{\mu \nu}_{\alpha \beta}\ \nabla^{<\alpha} u^{\gamma>} g_{\gamma \delta}
\Omega^{\beta \delta},\quad 
P^{\mu \nu}_{\alpha \beta}\ \Omega^{\alpha \gamma} g_{\gamma \delta}
\Omega^{\beta \delta},\quad u_\gamma R^{\gamma <\mu \nu> \delta} u_\delta,\quad
R^{<\mu \nu>}\, ,&
\label{8terms}
\eqa
but only five combinations of those transform homogeneously under Weyl rescalings,
$\pi^{\mu \nu}\rightarrow e^{6 w(x)} \pi^{\mu \nu}$ (here and in the following
$D_\alpha$ denotes the (geometric) covariant derivative in curved space).
The calculation is straightforward but somewhat lengthy, so I only demonstrate
the ingredients by studying again the first order result, $\pi^{\mu \nu}=\eta \nabla^{<\mu } u^{\nu>}$.
Under conformal transformations, dimensionless scalars are 
invariant, $u^\mu g_{\mu \nu} u^\nu = 1 = \bar{u}^\mu \bar{g}_{\mu \nu} \bar{u}^\nu\, $
which implies 
$$u^\mu \rightarrow \bar{u}^\mu = e^{w(x)} u^\mu$$ 
under Weyl
rescalings. Furthermore, the transformation of the ideal fluid's energy momentum
tensor, 
$T^{\mu \nu}_{(0)}=\epsilon\ u^\mu u^\nu - p\, \Delta^{\mu \nu}\rightarrow \bar{T}^{\mu \nu}_{(0)} =
e^{6 w(x)} T^{\mu \nu}_{(0)}$ then requires
\beq
\epsilon \rightarrow \bar{\epsilon}=e^{4 w(x)} \epsilon\, .
\eeq
For conformal fluids, the shear viscosity coefficient is related to 
the energy density by \hbox{$\eta \propto \epsilon^{3/4}$}, so that one has
$\eta\rightarrow \bar{\eta}=e^{3 w(x)}\eta$.
Since $\pi^{\mu \nu}=\eta \nabla^{<\mu } u^{\nu>}$, one then has to verify
that the first order derivative transforms homogeneously as 
$\nabla^{<\mu } u^{\nu>}\stackrel{?}{\rightarrow} e^{3 w(x)} \nabla^{<\mu } u^{\nu>}$.
From the expansion
\beq
\nabla^{<\mu } u^{\nu>}=\nabla^\mu u^\nu+\nabla^\nu u^\mu-\frac{2}{3} \Delta^{\mu \nu}
\nabla_\alpha u^\alpha=\Delta^{\mu \alpha} D_\alpha u^\nu
+\Delta^{\nu \alpha} D_\alpha u^\mu-\frac{2}{3} \Delta^{\mu \nu} D_\alpha u^\alpha
\label{intermed10}
\eeq
it becomes clear that one has to study the transformation property of
the covariant derivative of the fluid velocity, 
$$
D_\alpha u^\nu = \partial_\alpha u^\nu + \Gamma^\nu_{\alpha \beta} u^\beta\, ,
$$
where $\Gamma^\nu_{\alpha \beta}$ are the Christoffel symbols given by
$$
\Gamma^\nu_{\alpha \beta}=\frac{1}{2} g^{\nu \rho} \left(
\partial_\alpha g_{\rho \beta}+\partial_\beta g_{\rho \alpha}-\partial_\rho g_{\alpha \beta}\right)\, .
\label{Christoffel}
$$
The transformation of the Christoffel is readily calculated 
from the transformation of the metric (\ref{Weylresc}),
$$
\Gamma^\nu_{\alpha \beta}\rightarrow \bar{\Gamma}^\nu_{\alpha \beta}=
\Gamma^\nu_{\alpha \beta}- \left(g^\nu_\beta \partial_\alpha w
+g^\nu_\alpha \partial_\beta w- g_{\alpha \beta} \partial^\nu w\right)\, ,
$$
so that together with the transformation property of the fluid velocity
one finds
$$
D_\alpha u^\nu \rightarrow e^{w} \left(D_\alpha u^\nu -
g^\nu_\alpha u^\beta \partial_\beta w+u_\alpha \partial^\nu w\right)\, .
$$
Using this result in Eq.~(\ref{intermed10}) one finds that all the terms
involving derivatives of the scale factor $w(x)$ cancel, 
\beq
\nabla^{<\mu } u^{\nu>}\rightarrow e^{3 w(x)} \nabla^{<\mu } u^{\nu>}\, ,
\eeq
so that indeed the first order expression for $\pi^{\mu \nu}$ transforms
homogeneously under Weyl transformations.

To second order, one repeats the above analysis for all of the eight
terms in Eq.~(\ref{8terms}), combining them in such a way that
all the derivatives of $w(x)$ cancel. One finds the result
\begin{center}
\fbox{\parbox{14cm}
{\bqa
\pi^{\mu\nu} &=& \eta \nabla^{\langle \mu} u^{\nu\rangle}
- \tau_\pi \left[ \Delta^\mu_\alpha \Delta^\nu_\beta D\pi^{\alpha\beta} 
 + \frac 4{3} \pi^{\mu\nu}
    (\nabla_\alpha u^\alpha) \right] \nonumber\\
  &&\quad 
  + \frac{\kappa}{2}\left[R^{<\mu\nu>}+2 u_\alpha R^{\alpha<\mu\nu>\beta} 
      u_\beta\right]\nonumber\\
  && -\frac{\lambda_1}{2\eta^2} {\pi^{<\mu}}_\lambda \pi^{\nu>\lambda}
  -\frac{\lambda_2}{2\eta} {\pi^{<\mu}}_\lambda \Omega^{\nu>\lambda}
  - \frac{\lambda_3}{2} {\Omega^{<\mu}}_\lambda \Omega^{\nu>\lambda}\, ,
\label{maineq}
\eqa}}
\end{center}
where $\tau_{\pi},\kappa,\lambda_1,\lambda_2,\lambda_3$ are five
independent second order transport coefficients,
and Eq.~(\ref{NSrelations}) has been used to rewrite 
some expressions, disregarding correction terms of third order in gradients.
Eq.~(\ref{maineq}) is the most general expression for $\pi^{\mu \nu}$ 
to second order in a gradient expansion in curved space for a conformal
theory.

\subsection{Hydrodynamics of strongly coupled systems}
\label{sec:hystrong}

Particularly interesting examples of conformal quantum-field theories are those that
have known supergravity duals in the limit of infinitely strong coupling \cite{Maldacena:1997re}.
Since fluid dynamics is a gradient expansion around the equilibrium of the system,
Eq.~(\ref{visceq}),(\ref{maineq}) should be general enough to also capture the 
dynamics of these strongly coupled quantum systems in the hydrodynamic limit.
These systems will in general not allow for a quasiparticle interpretation, since
the notion of a (quasi-)particle hinges on the presence of a well-defined
peak in the spectral density, which may not exist at strong coupling.
Therefore, infinitely strongly coupled system are very different than
systems described by kinetic theory (which relies on the presence of
quasiparticles), making their hydrodynamic limit interesting to study.

If a known supergravity dual to a strongly coupled field theory is 
known, one can calculate Green's functions in these theories
(for a review, see for instance \cite{Son:2007vk}). A particular 
example is the Green's function for the sound mode in strongly 
coupled ${\cal N}=4$ SYM theory, with gravity dual on a $AdS_5\times S_5$ background, which gives rise to
sound dispersion relation \cite{Baier:2007ix}
\beq
\omega=\pm\frac{k}{\sqrt{3}}+\frac{i k^2}{6 \pi T}\pm \frac{3-2 \ln 2}{6 \sqrt{3}
(2 \pi T)^2} k^3+{\cal O}(k^4)\, .
\eeq
By comparing to the hydrodynamic sound dispersion relation Eq.~(\ref{disprelsound}),
one finds the values for the speed of sound, shear viscosity and relaxation time
for strongly coupled ${\cal N}=4$ SYM,
\beq
c_s=\sqrt{\frac{1}{3}},\quad \frac{\eta}{s}=\frac{\eta T}{(\epsilon+p)}=\frac{1}{4 \pi},\quad
\tau_\pi = \frac{2 - \ln 2}{2 \pi T}\, .
\label{N4vals}
\eeq
Calculating other quantities both in ${\cal N}=4$ SYM and hydrodynamics \cite{Baier:2007ix}
and rederiving the fluid dynamic equations from the supergravity dual of ${\cal N}=4$ SYM 
\cite{Bhattacharyya:2008jc},
one additionally finds
\beq
\kappa=\frac{\eta}{\pi T},\quad \lambda_1=\frac{\eta}{2 \pi T},\quad \lambda_2=-\ln{2} \frac{\eta}{\pi T},\quad \lambda_3=0\, .
\eeq
As a side remark, note that the dispersion relation for transverse perturbations 
(the shear mode) discussed in section \ref{sec:acprob}, is ill-suited to determine
the second order transport coefficients such as $\tau_\Pi$, because information
about $\tau_\Pi$ enters only at fourth order in gradients (\ref{moddisp}), and therefore
receives corrections from terms not captured by second-order conformal 
hydrodynamics \cite{Baier:2007ix,Natsuume:2007ty}.

As expected, the hydrodynamic limit of strongly coupled ${\cal N}=4$ SYM 
reproduces the structure of Eq.~(\ref{visceq}),(\ref{maineq}), which 
had to be true if these equations are truly universal.
Furthermore, plugging the values (\ref{N4vals}) into the sound mode
group velocity for large wavenumbers (\ref{vL}), one finds 
$v_L^{\rm max}\sim 0.92$; this suggests that the hydrodynamic theory 
Eq.~(\ref{visceq}),(\ref{maineq}) obeys causality for 
strongly coupled ${\cal N}=4$ SYM. Interestingly, this seems
to be also the case for other known gravity duals,
for instance $AdS_{D+1}$, for $D>2$, corresponding to 
strongly coupled conformal field theories in $D$ spacetime dimensions.
There has been an extensive amount of work on calculating
the second-order transport coefficients in these theories 
\cite{Baier:2007ix,Bhattacharyya:2008jc,Natsuume:2007ty,Natsuume:2008iy,VanRaamsdonk:2008fp,Haack:2008cp}, 
which are now known analytically for all $D>2$ \cite{Bhattacharyya:2008mz} 
\beq
\tau_\pi=\frac{D+{\cal H}[2/D-1]}{4 \pi T},\quad
\lambda_1=\frac{\eta D}{8 \pi T},\quad
\lambda_2=\frac{\eta {\cal H}[2/D-1]}{2 \pi T} ,\quad
\lambda_3=0,\quad
\kappa=\frac{\eta D}{2 \pi T (D-2)}\, ,
\label{allknown}
\eeq
with ${\cal H}[x]$ the harmonic number function \cite{Natsuume:2008gy,Bhattacharyya:2008mz}
$$
{\cal H}[x]=\int_0^{1}dz \frac{1-z^x}{1-z}=\gamma_E+\left.\frac{d\ln{\Gamma(z)}}{dz}\right|_{z=x+1}\, .
$$
Note that the special case $D=4$ corresponds to the results (\ref{N4vals})
for strongly coupled ${\cal N}=4$ SYM, and that 
the ratio $\frac{\eta}{s}=\frac{1}{4\pi}$ is universal for all of these,
in line with the observation of Ref.~\cite{Kovtun:2004de}.
Also, there seems to be some universality for the second order transport coefficients:
for instance, it has been found that $4\lambda_1+\lambda_2=2 \eta \tau_\pi$ 
for a class of strongly coupled field theories \cite{Erdmenger:2008rm,Haack:2008xx}.

\begin{figure}[t]
\center
\includegraphics[width=.6\linewidth]{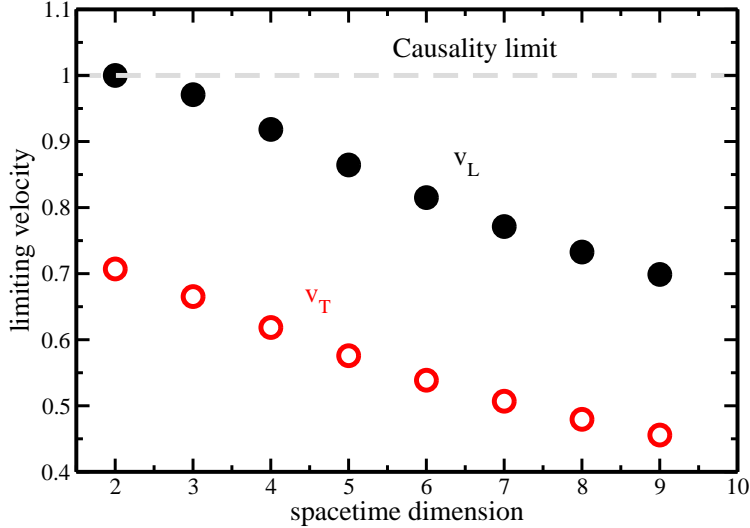}
\caption{The limiting velocities for longitudinal (\ref{vLd}) and transverse (\ref{vT})
perturbations as a function of spacetime dimension  in conformal second order hydrodynamics.}
\label{fig:vTvL}
\end{figure}

Generalizing Eq.~(\ref{vL}) to arbitrary spacetime dimension
gives
\beq
v_L^{\rm max}=\lim_{k\rightarrow \infty} \frac{d\omega}{dk}=\sqrt{c_s^2+
\frac{2 (D-2)}{(D-1)} \frac{\eta}{\tau_\pi(\epsilon+p)}
+\frac{\zeta}{\tau_\Pi(\epsilon+p)}}\, .
\label{vLd}
\eeq
For conformal theories, $\zeta=0$ and $c_s^2=\frac{1}{D-1}$, and using the values (\ref{allknown}),
one finds that that $v_L^{\rm max}$ is
decreasing monotonously with $D>2$ from its maximum at $D=2$,
where $v_L^{\rm max}$ would reach unity\footnote{For $D=2$, the conformal
field theory does not have any (first or second order) transport coefficients, but is completely
characterized by ideal fluid dynamics \cite{Haack:2008cp}.}. 
The values for $v_L^{\rm max}$ and $v_T^{\rm max}$ 
(Eqns.~(\ref{vLd},\ref{vT}), respectively) for spacetime dimensions 
$D<10$ are shown in Figure \ref{fig:vTvL}).

For $D=4$, corresponding to strongly coupled ${\cal N}=4$ SYM,
also the corrections at finite (but large) coupling strength $\lambda$
to the transport coefficients have been calculated \cite{Buchel:2008sh,Buchel:2008bz,Buchel:2008kd},
\bqa
&\frac{\eta}{s}=\frac{1}{4\pi}\left(1+\frac{120}{8} \zeta(3) \lambda^{-3/2}+\ldots\right),\quad
\tau_{\pi} T=\frac{2-\ln 2}{2 \pi}+\frac{375}{32 \pi} \zeta(3) \lambda^{-3/2}+\ldots,&\nonumber\\
&\kappa=\frac{\eta}{\pi T}\left(1-\frac{145}{8} \zeta(3) \lambda^{-3/2}+\ldots\right),\quad
\lambda_1=\frac{\eta}{2 \pi T}\left(1+\frac{215}{8} \zeta(3) \lambda^{-3/2}+\ldots\right)\, ,
\eqa
which lead to $v_L^{\rm max}\simeq 0.92-0.9796 \lambda^{-3/2}$.

\subsection{Hydrodynamics of weakly coupled systems and discussion}

Weakly coupled theories in general have a well defined quasiparticle
structure and hence it is expected that the hydrodynamic properties
of these theories are captured by kinetic theory. In particular,
it is known that kinetic theory correctly reproduces
the results from finite temperature quantum field theories,
in the hard-thermal-loop (resummed one-loop) approximation \cite{Blaizot:2001nr}.
As a consequence, one would expect that the dynamics of weakly
coupled quantum field theories in the hydrodynamic limit are well captured by 
the M\"uller-Israel-Stewart theory derived via kinetic theory in section \ref{sec:kt}.
Comparing Eq.~(\ref{maineq}) to Eq.~(\ref{MISktwise}) --- and recalling the approximation
used to derive (\ref{MISktwise}) --- one finds
\beq
\tau_\pi=\frac{6}{T}\frac{\eta}{s},\quad\lambda_1=\eta \tau_\pi+T^2{\cal O}(1),\quad
\lambda_2=-2 \eta \tau_\pi,\quad \lambda_3=0,\quad \kappa=0\, ,
\eeq
where ${\cal O}(1)$ reflects the unknown contribution to $\lambda_1$ from the
collision term and $\kappa=0$ stems from 
rederiving Eq.~(\ref{MISktwise}) in curved space \cite{Baier:2007ix}.
First order transport coefficients have been calculated in the weak-coupling
limit for high temperature gauge theories \cite{Arnold:2003zc}, 
in particular ${\cal N}=4$ SYM \cite{Huot:2006ys}
\beq
\frac{\eta}{s}=\frac{6.174}{\lambda^2 \ln{\left(2.36 \lambda^{-1/2}\right)}}\, .
\eeq
More recently, all second-order transport coefficients were evaluated
consistently in QCD and scalar field theories at weak coupling \cite{York:2008rr,Romatschke:2009ng},
\beq
\tau_\pi=\frac{5.0\ldots 5.9}{T}\frac{\eta}{s},\quad\lambda_1=\frac{4.1\ldots5.2}{T}
\frac{\eta^2}{s},\quad
\lambda_2=-2 \eta \tau_\pi,\quad \lambda_3=0,\quad
\kappa=\frac{5\ s}{8 \pi^2 T}\,,
\label{weakcouplingres}
\eeq
where the range of values indicate the dependence on the coupling constant $g$
(see Ref.~\cite{York:2008rr} for details).
Note that the results from Eq.~(\ref{MISktwise}) agree 
reasonably well with the full calculation Eq.~(\ref{weakcouplingres}). 
In particular, the value of 
the relaxation time $\tau_\pi$ is such that the limiting velocity $v_L^{\rm max}$
is smaller than for strongly coupled systems.

Comparing the results (\ref{weakcouplingres}) to those obtained 
for strongly coupled theories (\ref{allknown}), one finds that
$\lambda_3$ always vanishes. This could indicate that there is an additional,
unidentified symmetry in conformal hydrodynamics that forces this coefficient to
be zero. Moreover, a direct calculation shows that 
the value of $\kappa$ is beyond the accuracy of kinetic theory \cite{York:2008rr,Moore,Romatschke:2009ng}. 
This indicates that the kinetic theory result is not general enough to capture the dynamics
of conformal fluids in the hydrodynamic limit for arbitrary coupling,
at least when spacetime is curved (since $\kappa$ couples only to the Riemann
and Ricci tensor, it does not contribute to Eq.~(\ref{visceq}) when spacetime is flat;
however, $\kappa$ \emph{does} enter in correlators for the energy-momentum tensor in flat space 
\cite{Baier:2007ix}).
A possible reason for this could be the fact that kinetic theory
itself is only a gradient expansion to first order of the underlying quantum field
theory \cite{Blaizot:2001nr}, thereby possibly missing second-order contributions.

Furthermore, one finds that $\lambda_{1,2}$ have the same sign
for both kinetic theory and the strongly coupled systems studied, which could indicate that
the sign of these coefficients does not depend on the coupling. 
Finally, the fact that $v_L^{\rm max}$ never exceeds unity for 
infinitely strongly coupled theories,
for theories at large (but finite) coupling, and at weak coupling, suggests, but does not prove,
that causality in a second-order conformal hydrodynamics description is obeyed.
At this time, there is only a proof for theories that have
dual description in terms of Gauss-Bonnet gravity, where it has been
shown that causality in second order hydrodynamics follows from the causality
of the field theory itself \cite{Buchel:2009tt}.
As a consequence, one may hope that the system of equations (\ref{visceq}),(\ref{maineq})
constitutes a valid starting point to attempt a description of real (but nearly
conformal) laboratory fluids at relativistic speeds. This application
of the viscous hydrodynamic theory to high energy nuclear physics
will be discussed in section \ref{sec:rhic}.

\subsection{Non-conformal hydrodynamics}
\label{sec:nchyrdo}

Since most quantum field theories that successfully describe nature
are not conformal theories, one may ask how deviations from
conformality change the hydrodynamic equations. In particular,
one may ask how important non-conformal terms not included in Eq.~(\ref{maineq}) 
are once conformal symmetry is slightly violated.
To this end, consider the specific example of a SU(N) gauge theory at
high temperature which has a trace anomaly \cite{Braun:2003rp},
\beq
T^\mu_\mu=g_{\mu \nu} T^{\mu \nu}=\langle \frac{\beta(g_{\rm YM})}{2 g_{\rm YM}} g^{\mu \alpha} g^{\nu \beta} F^a_{\mu \nu} F^a_{\alpha \beta}\rangle_T+W[g_{\mu \nu}]\,, \quad F^a_{\mu \nu}=\partial_\mu A_\nu^a-\partial_\nu A_\mu^a
-g_{\rm YM}\,f_{abc}  A^b_\mu A^c_\nu\, ,
\label{traceano}
\eeq
where $f_{abc}$ are the SU(N) structure constants, $A_\mu^a$ are the gauge fields
and $\langle \rangle_T$ denotes the thermal quantum field theory average.
Similar to section \ref{sec:cvh}, the Weyl anomaly $W[g_{\mu \nu}]$ is not important 
for what follows and will be ignored.
The change of the gauge theory coupling $g_{\rm YM}$ when changing
the renormalization scale $\Lambda$ is given by the beta-function,
\beq
\Lambda \frac{\partial g_{\rm YM}}{\partial \Lambda}=\beta(g_{\rm YM})\,
\eeq
which for weakly coupled SU(N) gauge theories is given by \cite{Gross:1973id}
\beq
\beta(g_{\rm YM})=-\frac{11\, N}{3}  \frac{g_{\rm YM}^3}{16 \pi^2}+{\cal O}(g_{\rm YM}^5)\, .
\eeq
In fact, Eq.~(\ref{traceano}) can be derived from the gauge theory action
when performing a Weyl transformation (\ref{Weylresc}) of the partition function and 
noting that the renormalization scale changes according to 
$\Lambda\rightarrow e^{w(x)} \Lambda $,
\beq
\frac{\delta \ln Z}{\delta g_{\mu \nu}}\propto \sqrt{-g} T^{\mu \nu}\, ,
\label{gravtdef}
\eeq
where $g$ is the determinant of the metric $g_{\mu \nu}$ (not to be confused with $g_{\rm YM}$).
Taking another functional derivative of the trace anomaly \cite{Baier:2007ix} leads to 
\bqa
\frac{\delta}{\delta g_{\alpha \beta}(y)} \left(\sqrt{-g} g_{\mu \nu}(x) T^{\mu \nu}\right)&=&
\sqrt{-g} T^{\alpha \beta} \delta(x-y)+g_{\mu \nu}(x) \frac{\delta}{\delta g_{\alpha \beta}(y)}
\left(\sqrt{-g}  T^{\mu \nu}\right)\nonumber\\
&=&\sqrt{-g} T^{\alpha \beta} \delta(x-y)+g_{\mu \nu}(x) \frac{\delta}{\delta g_{\mu \nu}(y)}
\left(\sqrt{-g}  T^{\alpha \beta}\right)\nonumber\\
&=&\sqrt{-g} \left(3 T^{\alpha \beta} \delta(x-y)+g_{\mu \nu}\frac{\delta T^{\alpha \beta}(x)}{\delta g^{\mu \nu}(y)}\right)\,,
\eqa
where the symmetry of the second derivative of the partition function with
respect to the metric (cf. Eq.~(\ref{gravtdef})) was used. On the other hand,
using Eq.(\ref{traceano}) one finds
\bqa
\frac{\delta}{\delta g_{\alpha \beta}(y)} \left(\sqrt{-g} g_{\mu \nu}(x) T^{\mu \nu}\right)
&=&\sqrt{-g} \langle
\frac{\beta(g_{\rm YM})}{g_{\rm YM}}\left(\frac{g^{\alpha \beta}}{4} F^{\mu \nu}_a F_{\mu \nu}^a
-F_a^{\alpha \lambda} F^{\beta}_{\ \lambda, a}\right)\rangle_T \delta(x-y)+{\cal O}(g_{\rm YM}^6)\,,
\nonumber\\
&=&{\cal O}(g_{\rm YM}^2)\, ,
\eqa
so that for Weyl transformations (\ref{Weylresc}) this implies
\beq
-2 g_{\mu \nu}\frac{\delta T^{\alpha \beta}(x)}{\delta g^{\mu \nu}(y)}=
\frac{\delta T^{\alpha \beta}}{\delta w(y)} = 6 T^{\alpha \beta} \delta(x-y)+{\cal O}(g_{\rm YM}^2)\,,\quad 
T^{\alpha}_\alpha=\epsilon-3 p={\cal O}(g_{\rm YM}^2)\, .
\label{confano}
\eeq
Note that an exact calculation gives
$T^{\alpha}_\alpha={\cal O}(g_{\rm YM}^4)$
for weakly coupled SU(N) gauge theories \cite{Moore:2008ws}.
Recalling that all terms in Eq.~(\ref{maineq}) transform as 
$\delta T^{\alpha \beta}/\delta w(y)\rightarrow 6 T^{\alpha \beta} \delta(x-y)$,
it becomes clear that terms not included in Eq.~(\ref{maineq}) must 
be ${\cal O}(g_{\rm YM}^2)$, or in other words are small 
in the weak-coupling limit where SU(N) gauge theory is almost conformal.
For instance, when neglecting quark masses,
bulk viscosity in QCD turns out to be smaller 
than shear viscosity by a factor of $g_{\rm YM}^8$ \cite{Arnold:2006fz}.

For weakly coupled systems, the form of the non-conformal hydrodynamic
equations has been investigated in \cite{Betz:2008me} from kinetic theory,
but the second-order transport coefficients are not known to date.

For strongly coupled systems, Ref.~\cite{Kanitscheider:2009as} offers
a beautiful example of non-conformal theories obtained by dimensional
reduction of conformal theories. Starting with a conformal theory
in $2 \sigma>D$, and reducing to $D$ spacetime dimensions, gives
an explicit realization of a relativistic hydrodynamic theory where 
the conformal invariance is (strongly) broken.
In particular, for this theory the bulk viscosity coefficient $\zeta$ is related to 
shear shear viscosity as
\beq
\frac{\zeta}{s}=2 \frac{\eta}{s} \left(\frac{1}{D-1}-c_s^2\right),
\eeq
and the speed of sound depends on the dimension of the original theory,
$c_s=\sqrt{\frac{1}{2 \sigma-1}}$. The relaxation time in
the bulk sector $\tau_\Pi$ equals that for the shear sector,
\beq
\tau_\Pi=\tau_\pi,
\eeq
so that one obtains for the limiting velocity Eq.~(\ref{vLd})
\beq
v_L^{\rm max}=\sqrt{c_s^2\left(1-\frac{2 \eta}{\tau_\pi (\epsilon+p)}\right) +\frac{2 \eta}{\tau_\pi (\epsilon+p)}}\, .
\eeq
Using the results found in section \ref{sec:hystrong}, $\frac{2 \eta}{\tau_\pi (\epsilon+p)}$ is
maximal in the limit of $D\rightarrow 2$, where $\frac{2 \eta}{\tau_\pi (\epsilon+p)}\rightarrow 1$.
In this limit, $v_L^{\rm max}\rightarrow 1$, regardless of the value of $c_s^2$. For $D>2$, $\frac{2 \eta}{\tau_\pi (\epsilon+p)}<1$
and hence $v_L^{\rm max}$ is maximal for the largest possible value of $c_s^2$, which is $c_s^2=\frac{1}{D-1}$.
As a consequence, one finds that for this class of strongly coupled theories where conformal symmetry 
is broken ($\zeta>0$) the limiting velocity  --- despite the appearance of Eq.~(\ref{vLd}) --- 
is actually smaller than for a conformal theory in the same number of spacetime dimensions,
and in particular always smaller than the speed of light. Again, while this does not
proof that causality is always obeyed in second-order hydrodynamics, it adds to the list
of known theories where ``by coincidence'' this turns out to be the case.

See Ref.~\cite{Romatschke:2009kr} for a complete classification of all second-order structures
in the energy-momentum tensor for non-conformal fluids.

\section{Applying hydrodynamics to high energy nuclear collisions}
\label{sec:rhic}

\subsection{Heavy-Ion Collision Primer}

\begin{figure}
\begin{minipage}[t]{.4\linewidth}
\includegraphics[width=.8\linewidth]{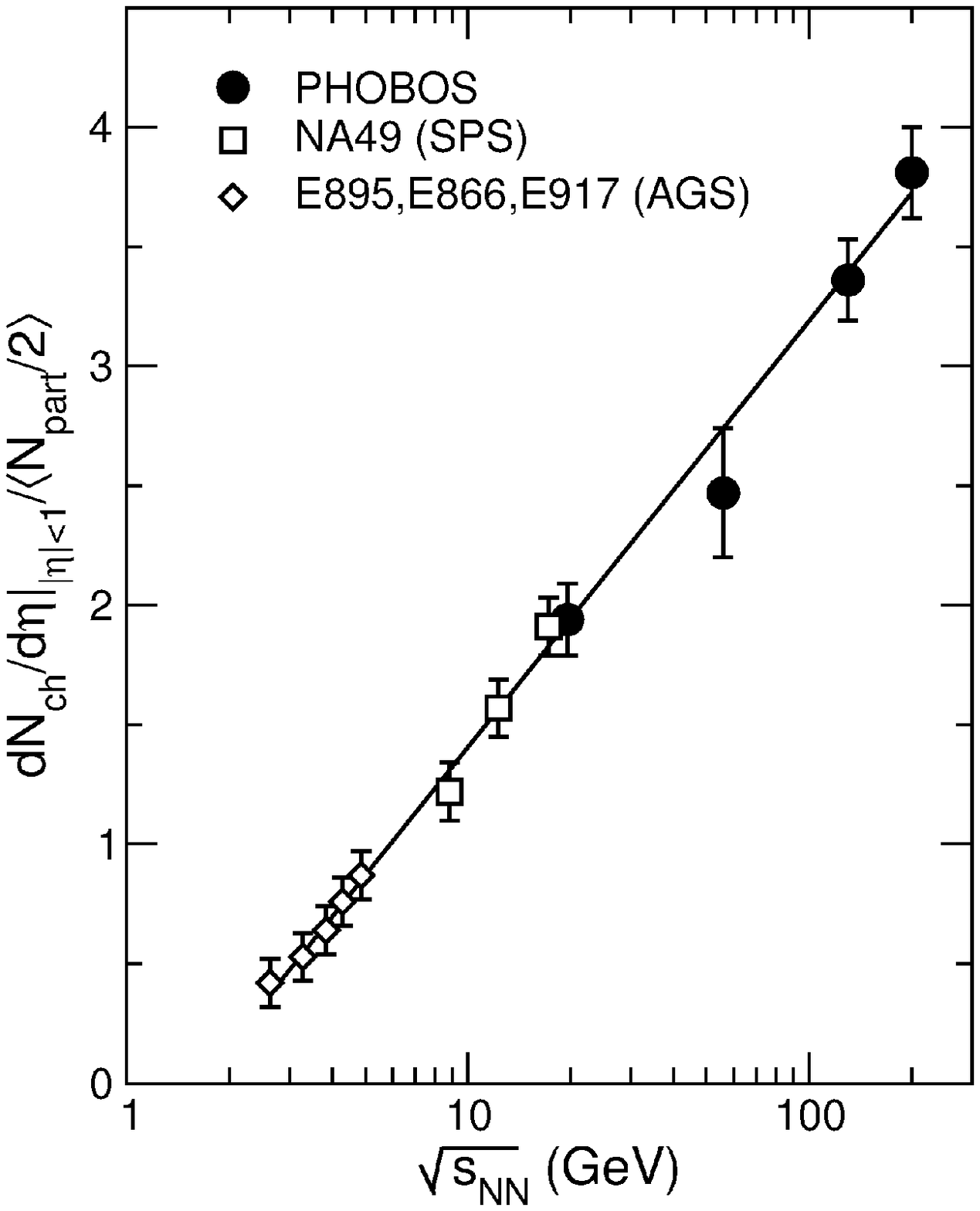}
\end{minipage}
\hfill
\begin{minipage}[t]{.55\linewidth}
\includegraphics[width=\linewidth]{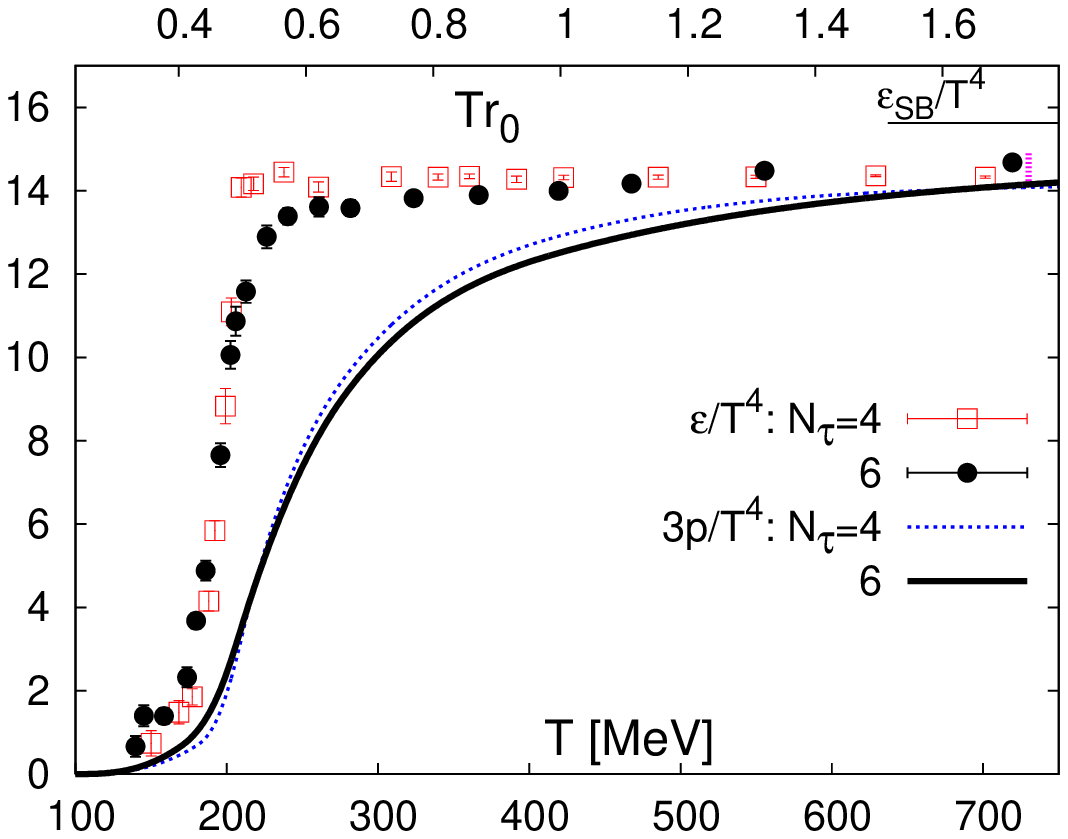}
\end{minipage}
\caption{Left: Particle density (number of particles per unit rapidity, normalized by system size) as a function
of collision energy (figure from Ref.~\cite{Back:2004je}). Right: The energy density
of QCD, calculated using lattice gauge theory, shows a strong rise close to the QCD deconfinement temperature (figure
from \cite{Cheng:2007jq}).}
\label{fig:phobosplot}
\end{figure}

Relativistic collisions of heavy ions (nuclei with an atomic weight heavier than carbon)
offer one of the few possibilities to study nuclear matter under extreme
conditions in a laboratory. The defining parameters for heavy-ion collisions
are the center-of-mass collision energy per nucleon pair $\sqrt{s}$ and the
geometry of the colliding nuclei (gold nuclei are typically larger than
copper, and uranium nuclei are not spherically symmetric).
The collisions are said to be relativistic once the 
center-of-mass energy is larger than the rest mass of the nuclei,
or equivalently if $\sqrt{s}/2$ is larger than the nucleon mass. 
For the Lorentz $\gamma$ factor of the collision, this implies
\beq
\gamma=\frac{m \gamma c^2}{m c^2}=\frac{E^{\rm total}}{m}\simeq\frac{\sqrt{s}}{2 {\rm GeV}}\, ,
\label{gammafactor}
\eeq
so typically $\gamma>1$. Experiments at Brookhaven National Laboratory (AGS, RHIC) and
CERN (SPS) have provided a wealth of data for Au+Au collisions (AGS, RHIC) and Pb+Pb collisions (SPS)
ranging in energy from $\sqrt{s}\sim 2.5-4.3$ GeV at the AGS over $\sqrt{s}\sim 8-17.3$ GeV at the SPS to
$\sqrt{s}\sim 130-200$ GeV at RHIC. It was found that the number density of particles produced 
in these collisions increases substantially for larger $\sqrt{s}$, indicating a similar rise
in the energy density \cite{Back:2004je}, that may allow the study of nuclear
matter above the deconfinement transition (see Figure \ref{fig:phobosplot}).

For Au+Au collisions at RHIC, the highest energy heavy-ion collisions achieved so far,
two beams of gold nuclei were accelerated in opposite directions in the RHIC ring
and brought to collide once they reached their design energies.
For an energy of $\sqrt{s}=200$ GeV, Eq.~(\ref{gammafactor}) indicates that before the 
collision the individual gold nuclei are highly Lorentz-contracted in the laboratory frame.
Thus, rather than picturing the collisions of two spheres, one
can should think of two ``pancake''-like objects approaching and ultimately colliding
with each other. As a consequence, the duration of the collision itself 
(which is on the order of the nuclear radius divided by the Lorentz gamma factor)
is much shorter than the nuclear radius divided by the speed of light.
Therefore, early after the collision the evolution in the directions
transverse to the initial beam direction (the ``transverse plane'') 
can be assumed to be static, and the dynamics is dominated by 
the longitudinal expansion of the system.

\begin{figure}[t]
\includegraphics[width=.5\linewidth]{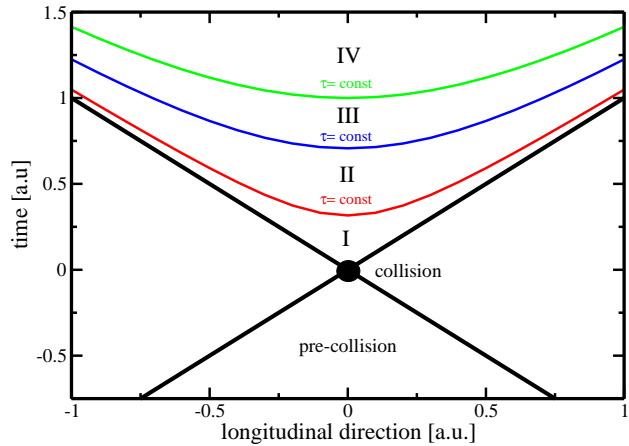}
\caption{Schematic view of a relativistic heavy-ion collision. See text for details.}
\label{fig:hyperbolas}
\end{figure}

Being interested in the bulk dynamics of the matter created in a relativistic
heavy-ion collision, one can divide the evolution into four stages
in proper time $\tau=\sqrt{t^2-z^2}$, shown schematically in figure \ref{fig:hyperbolas}.
Stage {\bf I} immediately following the collision is the pre-equilibrium stage 
characterized by strong gradients and possibly strong gauge fields \cite{Iancu:2003xm},
where a hydrodynamic description is not applicable. The duration of this stage
is unknown since the process of equilibration in QCD at realistic values of 
the coupling is not understood, but it is generally assumed to last about
$1$ fm/c.
Stage {\bf II} is the near-equilibrium regime characterized by small gradients
where hydrodynamics should be applicable if the local temperature
is well above the deconfinement transition. This stage lasts about $5-10$ fm/c, 
until the system becomes too dilute for equilibrium to be maintained
and enters stage {\bf III}, the hadron gas regime. The hadron gas 
is characterized by a comparatively large viscosity coefficient \cite{Prakash:1993bt},
making it ill suited to be described by hydrodynamics, but 
well approximated by kinetic theory \cite{Bass:2000ib}. 
This stage ends when the hadron scattering cross sections
become too low and particles stop interacting. In stage {\bf IV}, 
hadrons then fly on straight lines (free streaming) until
they reach the detector.


Assuming the system created by a relativistic collision of two heavy ions becomes
nearly equilibrated at some instant $\tau=\tau_0$ in proper time, the subsequent bulk
dynamics in stage {\bf II} should be governed by the hydrodynamic equations (\ref{visceq}),(\ref{maineq}),
amended by relevant non-conformal terms. To simplify the discussion, in the 
following these non-conformal terms will be neglected, and thus strictly speaking
I will not be dealing with real heavy-ion collisions but QCD matter 
in the conformal approximation. However, since the conformal anomaly Eq.~(\ref{confano})
is small except for a region close to the QCD phase transition \cite{Boyd:1996bx}, there is some
hope that this approximation does capture most of the important dynamics of real 
heavy-ion collisions.

To describe the fluid dynamics stage following a heavy-ion collision, 
one needs to specify the value of the hydrodynamic
degrees of freedom $\epsilon,p,u^\mu,\pi^{\mu \nu}$ at $\tau=\tau_0$,
the equation of state $p=p(\epsilon)$, the 
transport coefficients $\eta,\tau_\pi,\lambda_{1,2,3}$ governing the evolution (\ref{maineq})
as well as a decoupling procedure to the hadron gas stage at the end of the
hydrodynamic evolution. None of these are known from first principles, so
one typically has to resort to models which will be described in the following sections.

\subsection{Bjorken flow}
\label{sec:BF}

The physics of relativistic heavy-ion collisions has been strongly
influenced by Bjorken's notion of ``boost-invariance'' \cite{Bjorken:1982qr}, or the
statement that at a (longitudinal) distance $z$ away from the point
of (and time $t$ after) the collision, the matter should be moving with a velocity $v^z=z/t$.
Neglecting transverse dynamics ($v^x=v^y=0$) and introducing Milne coordinates proper time $\tau=\sqrt{t^2-z^2}$
and spacetime rapidity $\xi={\rm arctanh}(z/t)$, boost-invariance for
hydrodynamics simply translates into 
\beq
u^z=\frac{z}{\tau}\,,\quad u^\xi=-u^t \frac{\sinh{\xi}}{\tau}+ u^z \frac{\cosh{\xi}}{\tau}=0
\eeq 
and as a consequence
$\epsilon,p,u^\mu,\pi^{\mu \nu}$ are all independent of $\xi$, and 
therefore unchanged when performing a Lorentz-boost.

Even though in this highly simplified model the hydrodynamic
degrees of freedom now only depend on proper time $\tau$, the system
dynamics is not entirely trivial.
The reason for this is that in Milne coordinates, the metric is given
by $g_{\mu \nu}={\rm diag}(1,-1,-1,-\tau^2)$ and hence is no longer coordinate-invariant.
Indeed, one finds that the Christoffel symbols (\ref{Christoffel}) are non-zero,
\beq
\Gamma^\xi_{\xi \tau}=\frac{1}{\tau}\,,\qquad \Gamma^{\tau}_{\xi \xi}=\tau
\eeq
so as a consequence the covariant fluid gradients are non-vanishing
\beq
\nabla_\mu u^\mu= \partial_\mu u^\mu+ \Gamma^\mu_{\mu \nu} u^\nu=\Gamma^\xi_{\xi \tau} u^\tau =\frac{1}{\tau}\neq0\,,
\eeq
even though the fluid velocities are constant $u^{\mu}=(1,\vec{0})$ !
In essence, the Milne coordinate system describes a space-time that is
expanding one-dimensionally, so that a system at rest within these coordinates
``feels'' gradients from the ``stretching'' of spacetime, akin
to the effect of Hubble expansion in cosmology. Unlike in cosmology,
however, the spacetime described by Milne coordinate is flat,
as can be verified by showing that the Ricci scalar $R=0$.
This is important, since one does not want to describe 
heavy-ion collisions in curved spacetime, but rather use the 
Milne coordinates as a convenient way to implement the rapid longitudinal
expansion following heavy-ion collisions. Indeed, the covariant fluid
gradient in Milne coordinates is precisely the same as the
one from Bjorken's boost-invariance hypothesis,
\beq
\nabla_\mu u^\mu=\frac{1}{\tau}=\partial_t \frac{t}{\tau}+\partial_z \frac{z}{\tau}.
\eeq

This longitudinal flow (or ``Bjorken flow''), together with the assumption 
$u^x=u^y=0$, can be seen as 
a toy model for the hydrodynamic stage following the collision of
two infinitely large, homogeneous nuclei. The initial conditions 
for hydrodynamics at $\tau=\tau_0$ are then completely specified by 
two numbers: the initial energy density $\epsilon(\tau_0)$ and one component of the
viscous stress tensor, e.g. $\pi^\xi_\xi(\tau_0)$ (the other components of $\pi^\mu_\mu$
are completely determined by symmetries as well as $u_\mu \pi^{\mu\nu}=\pi^\mu_\mu=0$).
For example, in ideal hydrodynamics one finds (cf. Eq.~(\ref{visceq}))
\beq
D \epsilon+(\epsilon+p)\nabla_\mu u^\mu =\partial_\tau \epsilon+\frac{\epsilon+p}{\tau}=0\,
\label{simpeleq}
\eeq
for the evolution of the energy density (the evolution equations for $Du^\alpha$ are trivially satisfied). 
For an equation of state 
with a constant speed of sound $c_s$ this can be solved analytically
to give
\beq
\epsilon(\tau)=\epsilon (\tau_0) \left(\frac{\tau_0}{\tau}\right)^{1+c_s^2}.
\label{epsbjorkid}
\eeq
Therefore, the energy density is decreasing from its starting value
because of the longitudinal system expansion, with an exponent 
that depends on the value of the speed of sound. For an ideal gas of 
relativistic particles $c_s^2=1/3$, giving rise to the 
behavior $\epsilon \propto \tau^{-4/3}$ that is sometimes used in
heavy-ion phenomenology.

\begin{figure}[t]
\includegraphics[width=.5\linewidth]{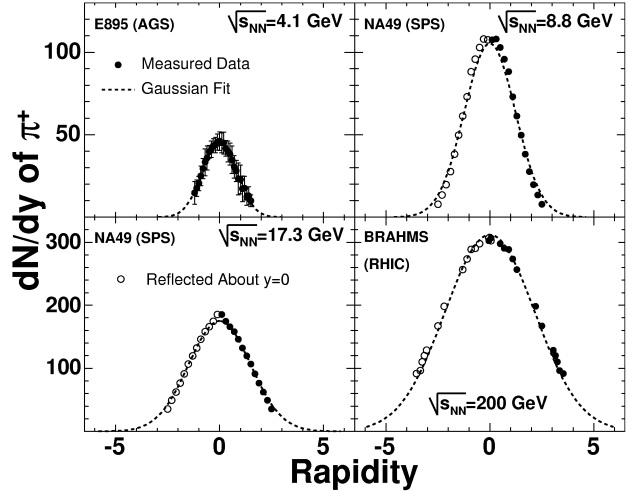}
\caption{Rapidity dependence of produced particles (pions), for different collision energies
(figure from Ref.~\cite{Back:2004je}). Even at highest energies, a plateau 
at $Y=0$ (boost-invariance) does not seem to emerge.}
\label{fig:gauss}
\end{figure}

Viscous corrections to Eq.~(\ref{epsbjorkid}) have been calculated
in first order viscous hydrodynamics \cite{Danielewicz:1984ww,Kouno:1989ps}
(the acausality problem discussed in section \ref{sec:acprob} does not appear for Bjorken flow 
due to the trivial fluid velocities), as well as second order viscous 
hydrodynamics \cite{Muronga:2001zk,Baier:2006um,Luzum:2008cw}, where 
the equations become
\bqa
\partial_\tau \epsilon&=&-\frac{\epsilon+p}{\tau}+\frac{\pi^\xi_\xi}{\tau}
\nonumber\\
\partial_\tau \pi^\xi_\xi &=& -\frac{\pi^\xi_\xi}{\tau_\pi}
+\frac{4 \eta}{3 \tau_\pi \tau}-\frac{4}{3 \tau} \pi^\xi_\xi
-\frac{\lambda_1}{2\tau_\pi\eta^2} \left(\pi^\xi_\xi\right)^2.
\label{0+1dsystem}
\eqa
Moreover, higher order
corrections are accessible for known supergravity duals to gauge theories
\cite{Heller:2008mb,Kinoshita:2008dq}. Due to its simplicity,
one can expect that Bjorken flow will continue to be used as a toy model of a heavy-ion collisions 
also in the future, and indeed also I will assume rapidity-independence for the remainder
of the discussion on the hydrodynamic descriptions for simplification. 
However, it is imperative to recall
that experimental data by no means supports Bjorken's hypothesis of rapidity independence,
as is shown in Fig.~\ref{fig:gauss}. Rather, the data suggests that the rapidity shape
of produced particles is approximately Gaussian, independent of the collision energy.
This clearly limits the applicability of the boost-invariance assumption to
the central rapidity region (close to the peak of the Gaussian in Fig.~\ref{fig:gauss}).

\subsection{Initial conditions for a hydrodynamic description of heavy-ion collisions}
\label{sec:initcond}

While pure Bjorken flow assumes matter to be homogeneous and static in the transverse
(${\bf x_\perp}=(x,y)$) directions, a more realistic model of a heavy-ion collision will
have to include the dynamics in the transverse plane. This means one has to
specify the initial values for the hydrodynamic degrees of freedom as a function
of ${\bf x_{\perp}}$. While it is customarily assumed that the fluid velocities
initially vanish, $u^x(\tau_0,{\bf x_\perp})=u^y(\tau_0,{\bf x_\perp})=0$, there are two main models
for the initial energy density profile $\epsilon(\tau_0,{\bf x_\perp})$: the
Glauber and Color-Glass-Condensate (CGC) model, respectively.

The main building block for both models is the charge density of nuclei
which can be parameterized by the Woods-Saxon potential,
\beq
\rho_A(\vec{x})=\frac{\rho_0}{1+\exp[(|{\vec x}|-R_0)/a_0]},
\eeq
where $R_0,a_0$ are the nuclear radius and skin thickness parameter,
which for a gold nucleus take values of $R_0\sim 6.4$ fm and $a_0\sim 0.54$ fm.
$\rho_0$ is an overall constant that is determined by requiring $\int d^3 x \rho_A({\vec x})=A$,
where $A$ is the atomic weight of the nucleus ($A\sim 197$ for gold). In
a relativistic nuclear collision, the nuclei appear highly Lorentz contracted
in the laboratory frame, so it is useful to define the ``thickness function''
\beq
T_A({\bf x_\perp})=\int_{-\infty}^{\infty} dz \rho_A({\vec x}),
\eeq
which corresponds to ``squeezing'' the nucleus charge density into a 
thin sheet.

\begin{figure}[t]
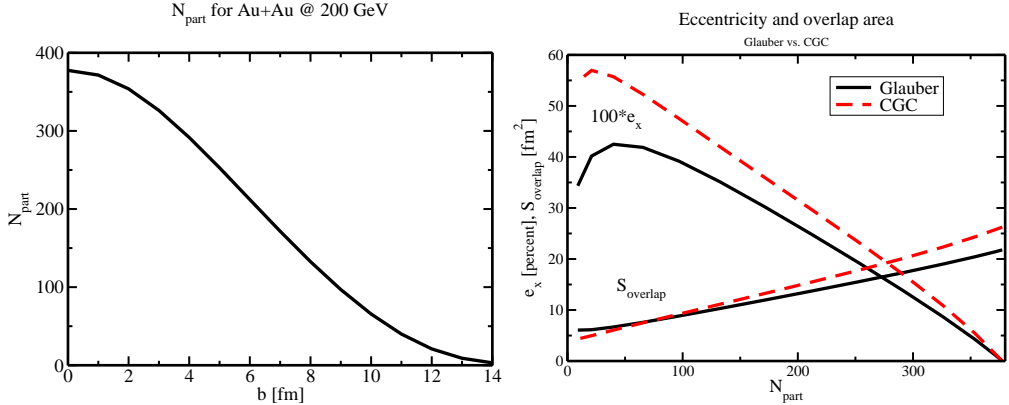

\includegraphics[width=.4\linewidth]{Fig5a.eps}
\includegraphics[width=.4\linewidth]{Fig5b.eps}
\caption{Left: Number of participants $N_{\rm part}$ in the Glauber model as a function of impact parameter $b$.
Right: Spatial eccentricity and area of the overlap region for the Glauber and CGC model, as a function
of $N_{\rm part}$.}
\label{fig:Npart}
\end{figure}

In its simplest version, the Glauber model for the initial 
energy density profile following the collision of two nuclei
at an energy $\sqrt{s}$ with impact parameter $b$ is then given by 
\beq
\epsilon({\bf x_\perp},b)={\rm const}\times T_A(x+\frac{b}{2},y) \times T_A(x-\frac{b}{2},y) \times \sigma_{NN}(\sqrt{s})\, ,
\label{epsilonGlauber}
\eeq
where $\sigma_{NN}(\sqrt{s})$ is the nucleon-nucleon cross section and 
the constant is freely adjustable (see \cite{Kolb:2001qz} for more complicated 
versions of the Glauber model). Eq.~(\ref{epsilonGlauber})
has the geometric interpretation that the energy deposited at position
${\bf x_\perp}$ is proportional to the number of binary collisions,
given by the number of charges at ${\bf x_\perp}$ 
in one nucleus times the number of charges at this position in the other nucleus,
times the probability that these charges hit each other at energy $\sqrt{s}$.
Another concept often used in heavy-ion collision literature is the
number of participants $N_{\rm part}(b)=\int d^2 x_\perp n_{\rm Part}({\bf x_\perp},b)$,
where
\bqa
n_{\rm Part}({\bf x_\perp},b)&=&n_{\rm Part}^{A}({\bf x_\perp},b)+n_{\rm Part}^{A}({\bf x_\perp},-b)\,\nonumber\\
n_{\rm Part}^{A}({\bf x_\perp},b)&=&T_A\left(x+\frac{b}{2},y\right)
\left[1-\left(1-\frac{\sigma_{NN} T_A\left(x-\frac{b}{2},y\right)}{A}\right)^A
\right]\,.
\eqa
Experiments are able to determine the number of participants, but 
cannot access the impact parameter of a heavy-ion collision directly, 
so the Glauber model $N_{\rm part}$ rather than $b$ is customarily
used to characterize the centrality of a collision (see Figure \ref{fig:Npart}).

\begin{figure}[t]
\includegraphics[width=.4\linewidth]{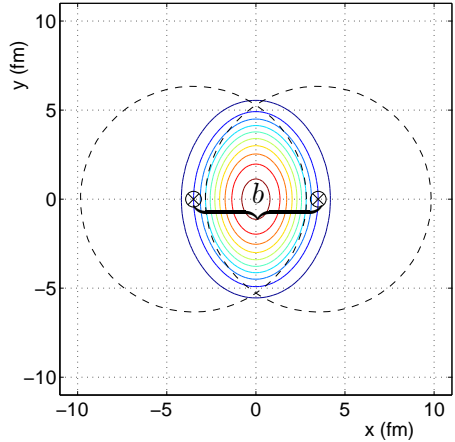}
\begin{picture}(0,2)
\put(0,0){\makebox(0,2){\vspace*{6.3cm}\hspace*{-6.7cm}$\underbrace{\qquad b\qquad}$}}
\end{picture}
\caption{Schematic view of a heavy-ion collision at impact parameter $b$ in the transverse plane (Figure from \cite{Heinz:2009xj}).}
\label{fig:schematic}
\end{figure}

The CGC model is based on the fact that a nucleus consists
of quarks and gluons which will interact according to the laws
of QCD. Accordingly, one expects corrections to the geometric
Glauber model due to the non-linear nature
of the QCD interactions. Heuristically, one can understand this
as follows \cite{Kharzeev:2000ph}: At relativistic energies, 
the nucleus in the laboratory frame is contracted into a sheet, 
so all the discussion focuses on the dynamics in the 
transverse plane. There, the area $\pi r_{gl}^2$ of a gluon
is related to its momentum $Q$ via the uncertainty principle,
$|r_{gl}|\times |Q| \sim \hbar = 1$, and the cross-section of 
gluon-gluon scattering in QCD is therefore 
\beq
\sigma\sim \alpha_s(Q^2) \pi r_{gl}^2\sim \alpha_s(Q^2) \frac{\pi}{Q^2}\, ,
\eeq
where $\alpha_s$ is the strong coupling constant.
The total number of gluons can be taken to be roughly proportional
to the number of partons in a nucleus, and hence also to its atomic weight $A$.
Therefore, the density of gluons in the transverse plane is approximately
$A/(\pi R_0^2)$, where $R_0$ is again the nuclear radius.
Gluons will start to interact with each other if the scattering
probability becomes of order unity,
\beq
1\sim \frac{A}{\pi R_0^2} \sigma = \alpha_s(Q^2) \frac{A}{R_0^2 Q^2}\, .
\eeq
Therefore, one finds that there is a typical momentum scale
$Q_s^2=\alpha_s \frac{A}{R_0^2}$ which separates perturbative
phenomena ($Q^2\gg Q_s^2$) from non-perturbative physics
at $Q^2\ll Q_s^2$ (sometimes referred to as ``saturation'').
The Color-Glass-Condensate was invented \cite{McLerran:1993ni,McLerran:1993ka} to include
the saturation physics at low momenta $Q^2$ in high energy nuclear collisions.
Due to the high occupation number at low momenta, this physics
turns out to be well approximated by classical chromodynamics.
Despite encouraging progress \cite{Lappi:2006xc}, the problem of calculating
the energy density distribution in the transverse plane at $\tau=\tau_0$
using the Color-Glass-Condensate has not been solved,
the main obstacle being the presence of non-abelian 
plasma instabilities \cite{Romatschke:2005pm,Fukushima:2006ax}.
As a consequence, there only exist phenomenological models
for the transverse energy distribution in the CGC 
(which are quite successful in describing experimental data, cf. \cite{Kharzeev:2002ei}), 
in particular the model by Ref.~\cite{Dumitru:2007qr}, which will be referred to as CGC model
in the following.

In the CGC model, the transverse energy profile at $\tau=\tau_0$ is given by
\beq
\epsilon({\bf x_\perp},b)={\rm const}\times \left[ \frac{dN_g}{d^2 {\bf x}_{T}dY}({\bf x}_T,b)\right] ^{4/3}\,
\label{edCGC}
\eeq
where $N_g$ is the number of gluons produced in the collision,
\bqa
\frac{dN_g}{d^2 {\bf x}_{T}dY} &\sim&
   \int \frac{d^2{\bf p}_T}{p^2_T}
  \int^{p_T} d^2 {\bf k}_T \;\alpha_s(k_T) \;
  \phi_+\left(\frac{({\bf p}_T+{\bf k}_T)^2}{4};{\bf x}_T\right)\;
              \phi_-\left(\frac{({\bf p}_T - {\bf k}_T)^2}{4};{\bf x}_T\right)\nonumber\\
\phi_{\pm} (k^2_{T}; {\bf x}_{T}) &=&
\frac{1}{\alpha_s (Q^2_s)} \frac{Q^2_s}{\textrm{max}(Q^2_s,k^2_{T})}\,
\left(\frac{n_{\rm part}^A({\bf x_\perp},\pm b)}{T_A(x\pm b/2,y)}\right)(1-x)^4\,\nonumber\\
Q_s^2(x,{\bf x_\perp})&=&
  \frac{2\,T_A^2(x\pm b/2,y)\,{\rm GeV}^2}{n_{\rm part}^A({\bf x_\perp},\pm b)}\left(\frac{{\rm fm}^2}{1.53}\right)
  \left(\frac{0.01}{x}\right)^{0.288}\,, \qquad x=\frac{p_T}{\sqrt{s}}.
\eqa

In order to see the difference between the Glauber and CGC model,
one defines the spatial eccentricity
\beq
e_x(b)=\frac{\langle y^2-x^2\rangle_{\epsilon}}{\langle y^2+x^2\rangle_{\epsilon}},
\label{xaniso}
\eeq
and overlap area
\beq
S_{\rm overlap}(b)=\pi \sqrt{\langle x^2\rangle_{\epsilon} \langle y^2\rangle_{\epsilon} }\,
\eeq
where $\langle\rangle_{\epsilon}$ denote integration over the transverse plane with
weight $\epsilon({\bf x_\perp},b)$. These quantities characterize the shape of the energy density profile
in the transverse plane (cf.~Fig.~\ref{fig:schematic}) and are shown in Fig.~\ref{fig:Npart}. One finds that the CGC model
typically has a larger eccentricity than the Glauber model, which will turn out
to have consequences for the subsequent hydrodynamic evolution. To see this, note that if $e_x>0$, the
energy density drops more quickly in the x-direction than in the y-direction because the 
overlap region is shaped elliptically. Using an equation of state $p=p(\epsilon)$ 
this implies that the mean pressure gradients are unequal, $\partial_x p > \partial_y p$, 
and according to the hydrodynamic equations (\ref{visceq}) one expects a larger fluid velocity to build up
in the x-direction than in the y-direction. Since the CGC model has a larger $e_x$ than the Glauber model,
this anisotropy in the fluid velocities should be larger for the CGC model, 
as will be verified below.

\subsection{Numerical solution of hydrodynamic equations}

The hydrodynamic equations are a set of coupled partial 
differential equations with known initial conditions.
Typically, it is not known how to find analytic solutions to
these set of equations, so it is necessary to come up with
numerical algorithms capable of solving the hydrodynamic equations.
As a toy problem, it is useful to study cases where the equations
simplify, e.g. the assumption of Bjorken flow discussed in section \ref{sec:BF}
where the hydrodynamic equations become a set of ordinary differential equations (\ref{0+1dsystem}).
A standard algorithm to solve Eqns.~(\ref{0+1dsystem}) numerically
is to discretize time, $\tau=\tau_0+n\times\Delta \tau$, where $\tau_0$ is the starting value,
$n$ is an integer, and $\Delta \tau$ is the step-size that has to be chosen small enough
for the algorithm to be accurate, but large enough for the
overall computing time to be reasonable. With this discretization,
derivatives become finite differences, e.g.
\beq
\partial_\tau \epsilon(\tau)=\frac{\epsilon_{n+1}-\epsilon_{n}}{\Delta \tau}\,,
\label{fd}
\eeq
and (\ref{0+1dsystem}) becomes
\bqa
\epsilon_{n+1}&=&\epsilon_n+\Delta \tau \left(
-\frac{\epsilon_n+p_n}{\tau_0+n \Delta \tau}+\frac{\pi^\xi_{\xi,n}}{\tau_0+n \Delta \tau}\right)\,,\\
\pi^\xi_{\xi,n+1} &=& \pi^\xi_{\xi,n}+\Delta \tau \left(-\frac{\pi^\xi_{\xi,n}}{\tau_\pi}
+\frac{4 \eta}{3 \tau_\pi (\tau_0+n \Delta \tau)}-\frac{4}{3 (\tau_0+n \Delta \tau)} \pi^\xi_{\xi,n}
-\frac{\lambda_1}{2\tau_\pi\eta^2} \left(\pi^\xi_{\xi,n}\right)^2\right)\,,\nonumber
\label{difeq}
\eqa
where for simplicity $\eta,\tau_\pi,\lambda_1$ were assumed to be independent of time.
Knowing $\epsilon,p,\pi^\xi_\xi$ at step $n$, the r.h.s.\ of the above equations
are explicitly known (the reason for this was the choice of ``forward-differencing'' (\ref{fd}))
and hence one can calculate $\epsilon,p,\pi^\xi_\xi$ at step $(n+1)$.
Repetition of this process gives a numerical solution for given stepsize $\Delta \tau$. 
Since the physical solution should be independent from the step size, it is 
highly recommended to create several numerical solutions for different $\Delta \tau$
and observe their convergence to a ``continuum solution'' for $\Delta \tau\rightarrow 0$.

Unfortunately, the above strategy of simple discretization does not always
lead to a well-behaved continuum solution. To see this, consider as another toy
problem the numerical solution $f(t,x)$ to the partial differential equation 
\beq
\partial_t f(t,x)=-a_0 \partial_x f(t,x),
\label{toymod}
\eeq
where $a_0$ is assumed to be constant. Again discretizing time and space
as $t=t_0+n \Delta t, x=m \Delta x$, the derivatives can be approximated
by the finite differences
\beq
\partial_t f(t,x)=\frac{f_{n+1,m}-f_{n,m}}{\Delta t}\,,\qquad 
\partial_x f(t,x)=\frac{f_{n,m+1}-f_{n,m-1}}{2 \Delta x}\,,
\label{FTCSderivs}
\eeq
which gives rise to the ``forward-time, centered-space'' or 
``FTCS'' algorithm \cite{NR} \S19. This algorithm is simple,
allows explicit integration of the differential equations 
as in Eq.~(\ref{difeq}), and usually does not work because 
it is numerically unstable. The instability can be easily
identified by making a Fourier-mode ansatz for $f(t,x)=e^{i \omega n \Delta t-i k m \Delta x}$
and calculating the dispersion relation $\omega=\omega(k)$ from the 
FTCS-discretized Eq.~(\ref{toymod})
\beq
\frac{f_{n+1,m}-f_{n,m}}{\Delta t}=f_{n,m} \frac{e^{i \omega \Delta t}-1}{\Delta t}=
-a _0 \frac{f_{n,m+1}-f_{n,m-1}}{2 \Delta x} = f_{n,m} \frac{i a_0}{\Delta x}  \sin k\Delta x\, .
\eeq
One finds
\beq
{\rm Im}\ \omega(k) = -\frac{1}{2 \Delta t} \ln \left[1+\left(a_0 \frac{\Delta t}{\Delta x}\right)^2 \sin^2 k \Delta x\right]<0\,
\label{imaom}
\eeq
which signals exponential growth in $f(t,x)$ for all modes $k$. As a consequence,
any numerical solution to Eq.~(\ref{toymod}) using the FTCS algorithm
will become unstable after a finite simulation 
time set by the inverse of Eq.~(\ref{imaom}).

However, this instability can be cured by choosing a slightly different
way of calculating the time derivative, namely replacing $f_{n,m}$ in
Eq.~(\ref{FTCSderivs}) by its space average $\frac{1}{2}(f_{n,m+1}+f_{n,m-1})$,
\beq
\partial_t f(t,x)=\frac{f_{n+1,m}-f_{n,m}}{\Delta t}-\frac{f_{n,m+1}-2 f_{n,m}+f_{n,m-1}}{2 \Delta t}\,.
\label{LAX}
\eeq 
This algorithm, known as the ``LAX'' scheme \cite{NR} \S19, has a dispersion relation
with 
\beq
{\rm Im}\ \omega(k) = -\frac{1}{2 \Delta t} \ln \left[\cos^2k \Delta x+\left(a_0 \frac{\Delta t}{\Delta x}\right)^2 \sin^2 k \Delta x\right]\,
\eeq
and hence is numerically stable for $a_0 \frac{\Delta t}{\Delta x}<1$, e.g. for sufficiently
small time steps $\Delta t$. The stability of the LAX scheme comes from the presence
of the last term in Eq.~(\ref{LAX}), which in ``continuum-form'' is a second derivative
that leads to 
\beq
\partial_t f(t,x)=-a_0 \partial_x f(t,x)+\frac{(\Delta x)^2}{2 \Delta t} \partial_x^2 f(t,x)
\label{LAXcnt}
\eeq
instead of Eq.~(\ref{toymod}). For sufficiently small $\Delta x$, this equation reduces 
to the original equation, so the LAX algorithm indeed converges to the physically interesting
solution. But the presence of this extra term, which is crucial for the numerical
stability, also has a physical interpretation: comparing Eq.~(\ref{LAXcnt}) to 
the diffusion equation (\ref{diffusioneq}) one is led to interpret the coefficient
$\frac{(\Delta x)^2}{2 \Delta t}$ as ``numerical viscosity''.
The LAX scheme works where the FTCS scheme fails because the viscous term
dampens the instabilities, in much the same way that the turbulent instability 
in fluids is damped by the viscous terms \cite{LL} \S26. Indeed, for ideal fluid
dynamics numerical viscosity is essential for stabilizing the numerical
algorithms. On the other hand, viscous fluid dynamics comes with \emph{real}
viscosity inbuilt, so it is tempting to conjecture that as long as $\eta$ or $\zeta$ are
finite and $\Delta t$ is sufficiently small, numerical viscosity is not
needed to stabilize the numerical algorithm for solving the hydrodynamic equations,
and the simple FTCS scheme can be used. Indeed, at least for the problem
of heavy-ion collision, this strategy leads to a stable algorithm 
\cite{Baier:2006gy,Romatschke:2007jx,Romatschke:2007mq,codedown}.

\begin{figure}[t]
\includegraphics[width=.3\linewidth]{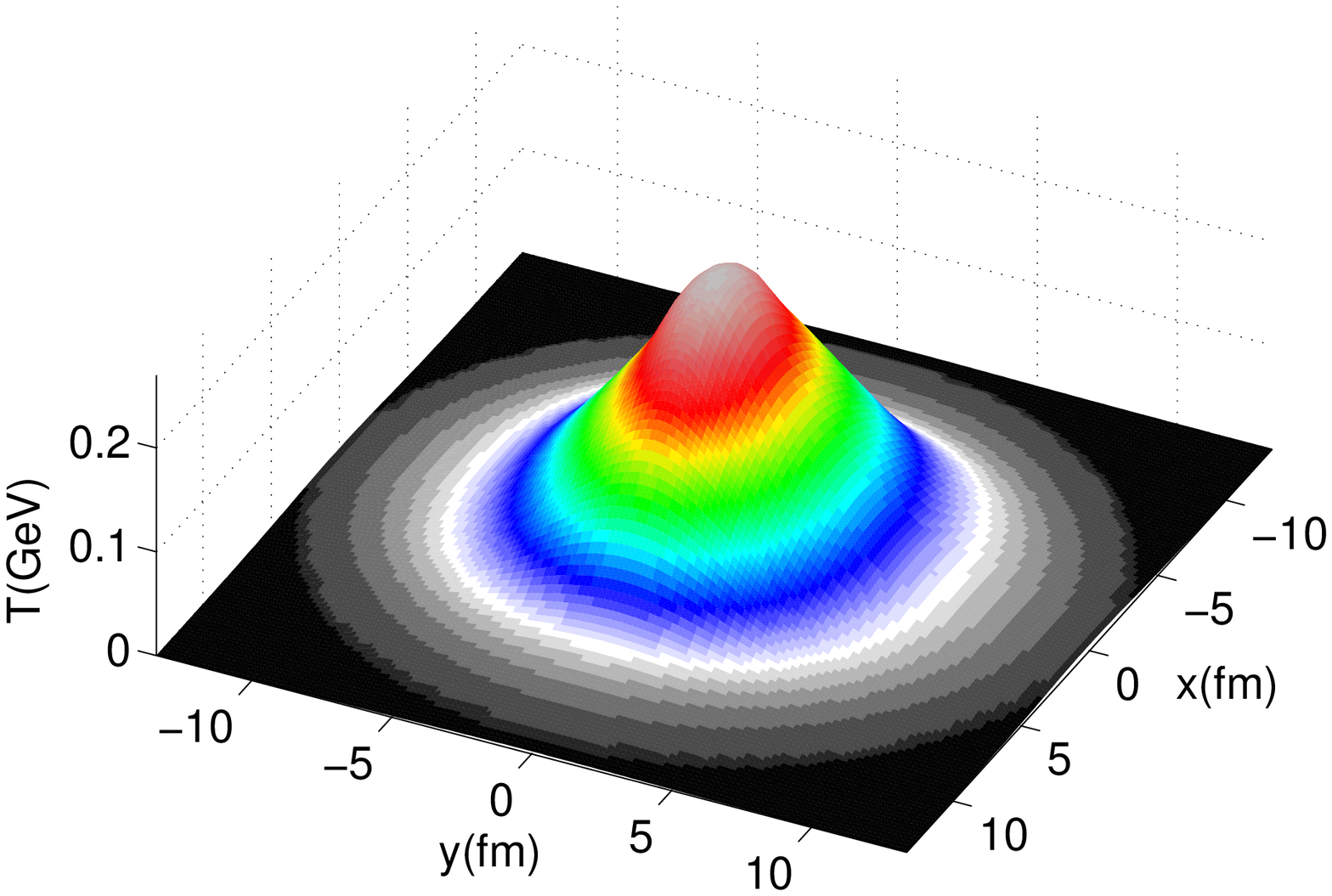}
\includegraphics[width=.3\linewidth]{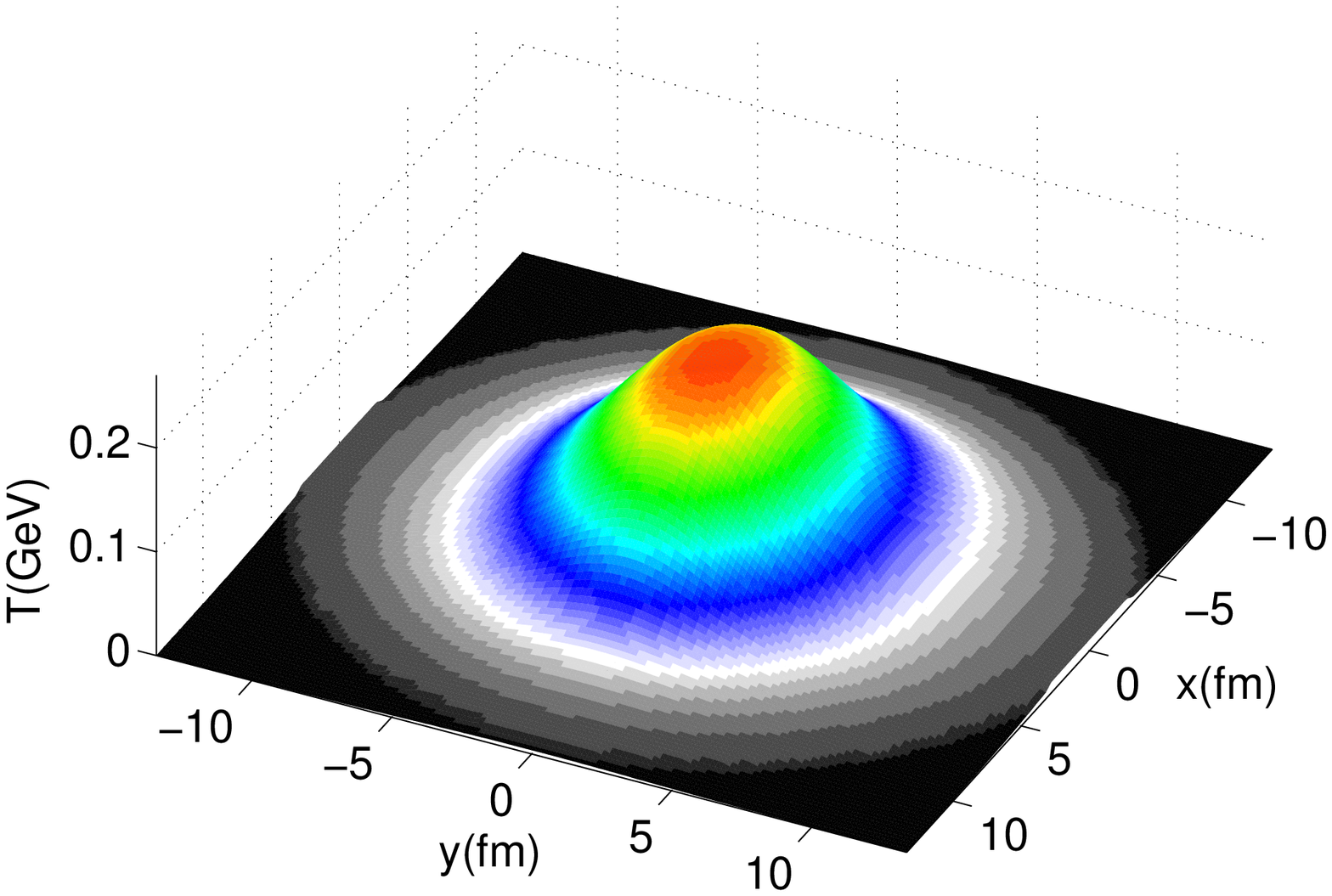}
\includegraphics[width=.3\linewidth]{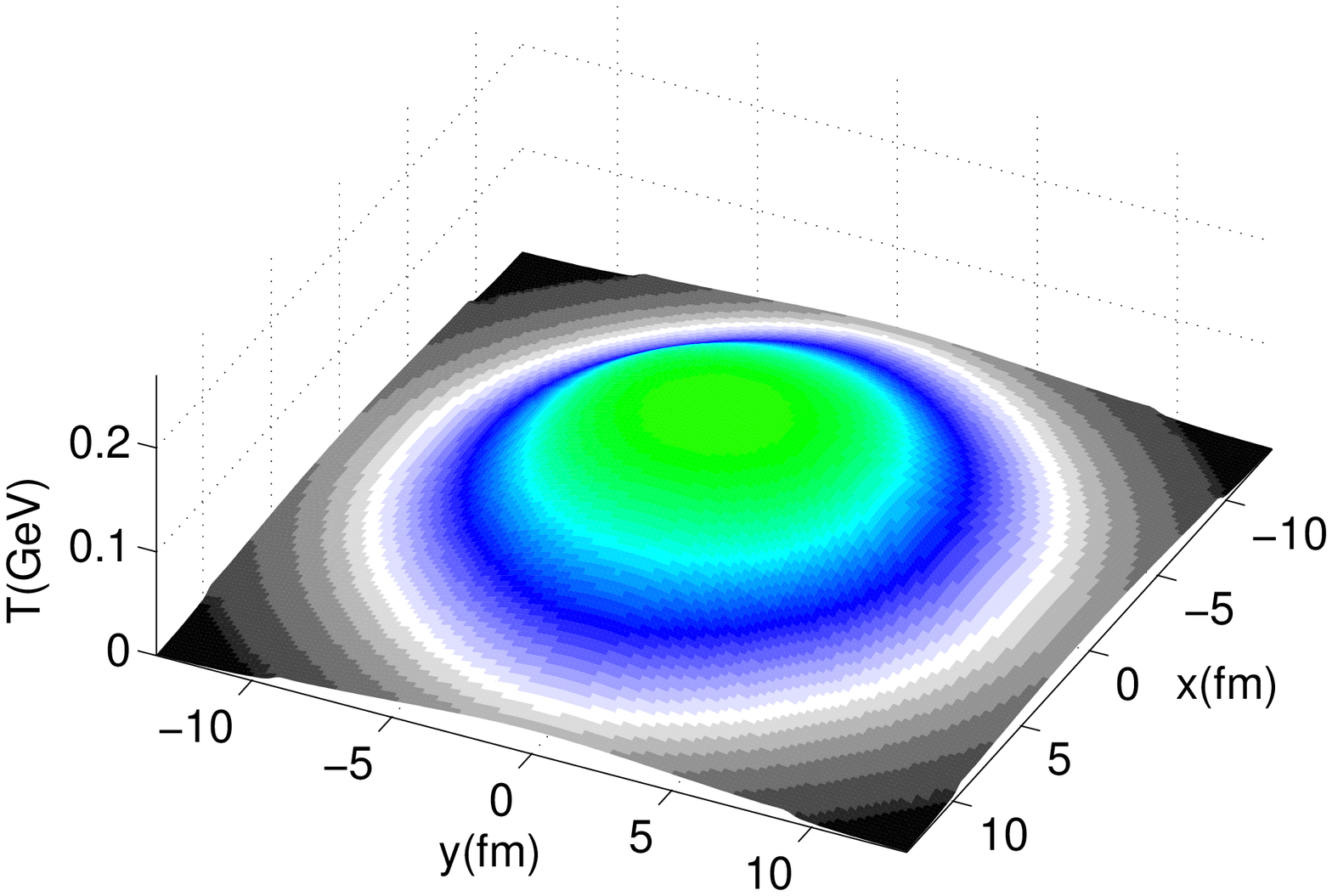}
\caption{Left to right: snapshots at $\tau=1,3$ and $7$ fm/c 
of the temperature profile $T(x,y)$ for a hydrodynamic
simulation of a $\sqrt{s}=200$ GeV Au+Au collision at $b=10$ fm. The initial
spatial eccentricity is gradually converted into momentum eccentricity and 
almost disappears at late times.}
\label{fig:snapshots}
\end{figure}

Aiming to solve the hydrodynamic equations in the transverse plane (assuming
boost-invariance in the longitudinal direction), one first has to 
choose a set of independent hydrodynamic degrees of freedom, e.g.,
$\epsilon,u^x,u^y,\pi^{xx},\pi^{xy},\pi^{yy}$ for which initial
conditions are provided along the lines of section \ref{sec:initcond}.
Only time derivatives to first order of these six quantities enter
the coupled partial differential equations (\ref{visceq}),(\ref{maineq}),
so that formally one can write the hydrodynamic equations in matrix form
\beq
{\bf A}
\cdot
\left(\begin{array}{c}
\partial_\tau \epsilon\\
\partial_\tau u^x\\
\ldots\\
\partial_\tau \Pi^{yy}
\end{array}\right)
={\bf b}\, ,
\label{mateq}
\eeq
where ${\bf A},{\bf b}$ are a matrix and vector with coefficients
that do not involve time derivatives. Using the FTCS scheme to
discretize derivatives, and matrix inversion to solve (\ref{mateq}), the
value of the independent hydrodynamic degrees of freedom at the next time step
are explicitly given in terms of known quantities (once the equation of state
and hydrodynamic transport coefficients are specified). 
Reconstructing all hydrodynamic fields from the independent components and repetition of the
above procedure then leads to a numerical solution for the hydrodynamic
evolution of a heavy-ion collision for given $\Delta \tau,\Delta x$ as long
as $\eta>0$ (in practice, values as low as $\eta/s\sim10^{-4}$ are stable with
reasonable $\Delta \tau$).
The convergence of these numerical solutions to the continuum limit
is explicitly observed when choosing a series of sufficiently small 
step sizes $\Delta \tau,\Delta x$. Snapshots of the temperature
profile in a typical simulation are shown in Fig.~\ref{fig:snapshots}.

\begin{figure}[t]
\includegraphics[width=.5\linewidth]{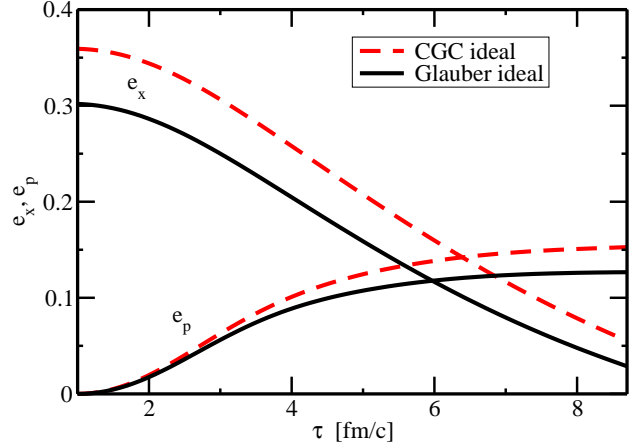}
\caption{Time evolution of the spatial and momentum anisotropies, Eq.~(\ref{xaniso}) and
Eq.~(\ref{paniso}), respectively, in the Glauber and CGC model for 
a $\sqrt{s}=200$ GeV Au+Au collision at $b=7$ fm (Figure from \cite{Luzum:2008cw}).}
\label{fig:spatialmomaniso}
\end{figure}

Fig.~\ref{fig:snapshots} also displays the gradual reduction of the eccentricity 
(the shape of the temperature profile
in the transverse plane becomes more and more circular as time progresses). 
The eccentricity corresponds to a spatial anisotropy in the pressure gradients which is converted
by hydrodynamics into a momentum anisotropy (fluid velocities $u^x\neq u^y$).
In analogy to the definition of the spatial eccentricity (\ref{xaniso}), it is therefore
useful to introduce the concept of momentum anisotropy
\beq
e_p(b)=\frac{\int d^2 {\bf x_\perp} T^{xx}-T^{yy}}{\int d^2 {\bf x_\perp} T^{xx}+T^{yy}}\, .
\label{paniso}
\eeq
The time evolution of the eccentricity and momentum anisotropy in the
Glauber and CGC model are shown in Fig.~\ref{fig:spatialmomaniso}. As 
discussed in section \ref{sec:initcond}, the higher initial
eccentricity in the CGC model gets converted in a larger momentum
anisotropy.

\subsection{Freeze-out}

Experiments in relativistic nuclear collisions measure 
momentum distributions of particles (pions, kaons, protons, etc.),
whereas hydrodynamics deals with energy density, pressure and fluid velocities.
Clearly, in order to make contact with experiment, the hydrodynamic
degrees of freedom need to be converted into measurable quantities,
which is often called the ``freeze-out''. The connection between
hydrodynamics and particle degrees of freedom is provided by
kinetic theory, which was discussed in section \ref{sec:kt}.
In particular, one requires the hydrodynamic and kinetic theory energy momentum
tensor at freeze-out to be the same,
\beq
T^{\mu \nu}_{\rm kinetic\ theory}=\int d\chi p^\mu p^\nu f(\vec{p},t,\vec{x}) = T^{\mu \nu}_{\rm hydro}\,,
\eeq
where for small departures from equilibrium the explicit form of $f$ in terms of
hydrodynamic degrees of freedom is provided by Eq.~(\ref{foutofeq}). Once
$f(\vec{p},t,\vec{x})$ is known, one can construct the particle current
from kinetic theory
\beq
n^\mu = \int d\chi p^\mu f(\vec{p},t,\vec{x})\, ,
\label{nmudf}
\eeq
which will be used to construct particle spectra that can be compared to
experimental measurements.

Freeze-out from hydrodynamic to particle degrees of freedom is expected
to occur when the interactions are no longer strong enough to keep
the system close to thermal equilibrium. Below the QCD phase transition,
this happens, e.g., when the system cools and viscosity increases 
\cite{Prakash:1993bt} until the viscous corrections 
in (\ref{foutofeq}) become too large and a fluid dynamic description is 
no longer valid. In practice, this is hard to implement, so simplified approaches such 
as isochronous and isothermal freeze-out are often used (see, however, \cite{Hung:1997du,Dusling:2007gi}). 
All of these have in common that they define a 
three dimensional hypersurface $\Sigma$ with a normal vector $d\Sigma^\mu$
such that the total number of particles after freeze-out 
is given by the particle current (\ref{nmudf}) leaving the hypersurface $\Sigma^\mu$,
\beq
N=\int n^\mu d\Sigma_\mu=\int d\chi f(\vec{p},t,\vec{x}) p^\mu d\Sigma_\mu\,.
\eeq
For energy densities sufficiently below the QCD phase transition,
the energy momentum tensor is well approximated by a non-interacting
hadron resonance gas \cite{Karsch:2003vd}. This translates to
\beq
d \chi = \sum_{i} (2s_i+1)(2g_i+1) \frac{d^4 p}{(2\pi)^3} \delta(p^\mu p_\mu - m_i^2) 2 \theta(p^0),
\eeq
where the sum is over all known hadron resonances \cite{Amsler:2008zzb} and $s_i,g_i$ are the spin and isospin of 
a resonance with mass $m_i$. As a consequence,
\beq
N=\sum_i \int d^3p \frac{1}{\sqrt{m_i^2+\vec{p}^{\, 2}}} \left(p^0 \frac{dN}{d^3 p}\right)_i\, ,
\eeq
where 
\beq
\left(p^0 \frac{dN}{d^3 p}\right)_i = \frac{d_i}{(2 \pi)^3} \int d\Sigma_\mu p^\mu f(\vec{p},t,\vec{x})\,,\quad
d_i=(2s_i+1)(2g_i+1)\,,
\label{singpspec}
\eeq
is the single-particle spectrum for the resonance $i$. Eq.~(\ref{singpspec})
is the generalization of the ``Cooper-Frye freeze-out prescription'' \cite{CooperFrye} 
to viscous fluids with $f$ given by Eq.~(\ref{dfrel}).

Arguably the simplest model is isochronous freeze-out,
where the system is assumed to convert to particles at a given constant 
time (or proper time). While fairly unrealistic, it allows
a rather intuitive introduction of the general freeze-out formalism:
constant time defines $\Sigma^\mu(t,x,y,z)$
in the hydrodynamic evolution which is parametrized by $t={\rm const}$. 
The normal vector $d\Sigma^\mu$ on this hypersurface is given by \cite{Ruuskanen:1986py,Rischke:1996em}
\beq
d \Sigma_\mu = \epsilon_{\mu \alpha \beta \gamma} 
\frac{\partial \Sigma^\alpha}{\partial x}
\frac{\partial \Sigma^\beta}{\partial y}
\frac{\partial \Sigma^\gamma}{\partial z} dx dy dz\,,
\label{dsdef}
\eeq
where $\epsilon_{\mu \alpha \beta \gamma}$ is the totally antisymmetric
tensor in four dimensions with $\epsilon_{0123}=+1$. 
A simple calculation gives $d\Sigma^\mu=(1,\vec{0})d^3 x$
and therefore the momentum particle spectra are easily obtained by integration
of the distribution function,
\beq
p^\mu d\Sigma^\mu=p^0 d^3x\,,\qquad\left(\frac{dN}{d^3 p}\right)_i = \frac{d_i}{(2 \pi)^3} 
\int d^3x f(\vec{p},t={\rm const},\vec{x})\, .
\eeq

Slightly more realistic is isochronous freeze-out in proper time, $\tau={\rm const}$, where
the freeze-out surface $\Sigma^\mu=(\tau \cosh{\xi},x,y,\tau \sinh{\xi})$ is
parametrized by $x,y,\xi$, because this incorporates Bjorken flow. 
Introducing rapidity $Y={\rm arctanh}{(p^z/p^0)}$ and 
$m_\perp=\sqrt{m^2+p_\perp^2}$ for convenience,
a short calculation for the normal vector $d\Sigma^\mu$ gives \cite{Luzum:2008cw}
\beq
d\Sigma_\mu p^\mu=\tau m_\perp \cosh{(Y-\xi)} dx dy d\xi\,.
\eeq
Considering for illustration a Boltzmann gas in equilibrium with $f_{\rm eq}=\exp{[-p^\mu u_\mu/T]}$,
for vanishing spatial fluid velocities one has
\bqa
p^0\left(\frac{dN}{d^3 p}\right)_i &=& \frac{d_i}{(2 \pi)^3} 
\tau m_\perp \int dx dy d\xi \cosh{(Y-\xi)} \exp{[-m_\perp \cosh{(Y-\xi)}/T]}\nonumber\\
&=&\frac{2 d_i}{(2 \pi)^2} \tau m_\perp \int dr r K_1\left(\frac{m_\perp}{T}\right)\, ,
\label{fo1}
\eqa
while for non-vanishing fluid velocities with azimuthal symmetry ($u^x(r)=u^y(r)=u^r(r)/\sqrt{2}$)
a short calculation gives \cite{Baier:2006gy}
\beq
p^0\left(\frac{dN}{d^3 p}\right)_i =\frac{2 d_i}{(2 \pi)^2} \tau m_\perp 
\int dr r K_1\left(\frac{m_\perp u^\tau}{T}\right) I_0 \left(\frac{|{\bf p_\perp}| u^r}{T}\right)\,,
\label{fo2}
\eeq
where $K(z),I(z)$ are modified Bessel functions and the transverse radius $r=\sqrt{x^2+y^2}$
has been introduced for convenience. Comparing the integrands in
Eq.~(\ref{fo1}),(\ref{fo2}) when transverse momenta $p_\perp=|{\bf p_\perp}|$ are much larger than
the temperature $T$ or mass $m$, one finds
\beq
\frac{K_1(m_\perp/T)}{
K_1(m_\perp u^\tau/T) I_0(p_\perp u^r/T)}
\sim \frac{\exp{[(m_\perp(1-u^\tau)+p_\perp u^r)/T]}}{u^\tau u^r p_\perp/T}\gg1
\eeq
if $u^r \sim {\cal O}(1)$. This means that the presence of $u^r>0$, or ``radial flow'', 
leads to particle spectra which are ``flatter'' at large $p_T$.
This is confirmed by numerical simulations \cite{Huovinen:2006jp}.

For a Boltzmann gas out of equilibrium and Bjorken flow only ($u^\tau=1,u^i=0$)
the viscous correction to the distribution function (\ref{dfrel}) is
\beq
\frac{\pi^{\mu \nu} p_\mu p_\nu}{2 (\epsilon+p) T^2}=
\frac{(\pi^{xx} +\pi^{yy}) p_\perp^2+2 \pi^{\xi \xi} m_\perp^2/\tau^2 \sinh^2{(Y-\xi)}}{4 (\epsilon+p) T^2}
=\frac{\pi^\xi_\xi\left(p_\perp^2-2 m_\perp^2 \sinh^2{(Y-\xi)}\right)}{4 (\epsilon+p) T^2}\,,
\eeq
so that the single particle spectrum becomes \cite{Baier:2006um}
\beq
p^0\left(\frac{dN}{d^3 p}\right)_i = \frac{2 d_i}{(2 \pi)^2} 
\tau m_\perp \int dr r \left[K_1\left(\frac{m_\perp}{T}\right)+
\pi^\xi_\xi\, \frac{p_\perp^2 K_1\left(\frac{m_\perp}{T}\right)-
2 m_\perp T  K_2\left(\frac{m_\perp}{T}\right)}{4 (\epsilon+p) T^2}\right]\,.
\eeq
Since for Bjorken flow typically $\pi^\xi_\xi>0$, 
this implies that viscous corrections tend to have the same effect of making
particle spectra flatter at large $p_T$, which hints at
the difficulty of extracting viscosity and radial flow from 
experimental data \cite{Romatschke:2007jx}. More information is needed
to disentangle these effects, so one decomposes the particle spectra
into a Fourier series with respect to the azimuthal angle in momentum 
$\phi_p=\arctan\left(p^y/p^x\right)$ \cite{Kolb:2003zi},
\beq
\left(p^0 \frac{dN}{d^3 p}\right)_i=v_0(|{\bf p_\perp}|,b)\left[1+2 
v_2 (|{\bf p_\perp}|,b)\cos{2 \phi_p}+2 v_4 (|{\bf p_\perp}|,b)\cos{4 \phi_p}+\ldots\right]\, ,
\eeq
where the coefficients $v_2,v_4$ are referred to as ``elliptic'' and ``hexadecupole'' flow \cite{Ollitrault:1992bk},
respectively. $v_2,v_4$ and even higher harmonics were measured experimentally 
for $\sqrt{s}=200$ GeV Au+Au collisions \cite{Adams:2003zg} and may be useful
to distinguish between flow and viscous effects. 

\begin{figure}[t]
\includegraphics[width=.5\linewidth]{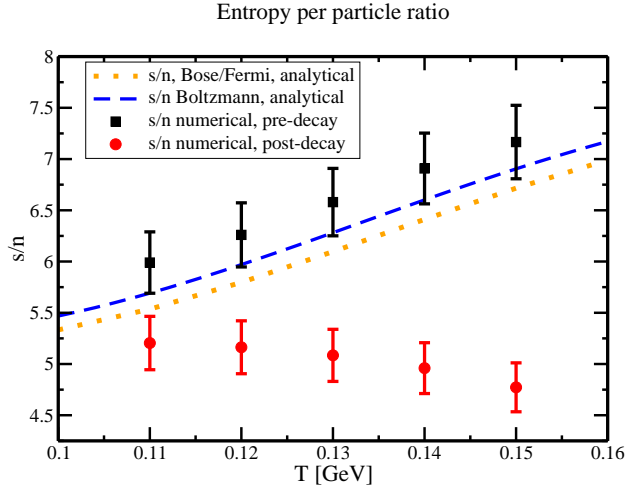}
\caption{Entropy per particle as a function of temperature for a gas of
a realistic set of massive resonances with Bose-Einstein/Fermi-Dirac
statistics (dotted line). Also shown are results when approximating by 
Boltzmann statistics (dashed line) and total initial entropy over final multiplicity 
from a numerical hydrodynamics simulation with isothermal freeze-out
using Boltzmann statistics \cite{Luzum:2008cw} (squares). The numerical results
prior to the decay of unstable resonances is in fair agreement with
the analytic prediction. The decay of unstable resonances produces additional particles, 
leading to a smaller ratio of initial entropy to final multiplicity
(circles).}
\label{fig:dsdn}
\end{figure}

A more realistic criterion than isochronous freeze-out is to assume decoupling at a
predefined temperature (isothermal freeze-out). In this case
the hypersurface $\Sigma$ can be parametrized by $\xi$, 
$\phi={\rm arctan}{\frac{y}{x}}$ and a time-like coordinate $0\le c\le 1$ with $c=0$ 
corresponding to the center $x=y=0$ of the transverse plane.
Assuming boost-invariance in the longitudinal direction, this leads to
$\Sigma^\mu=\Sigma^\mu(\tau(c) \cosh\xi,x(c,\phi),y(c,\phi),\tau(c) \sinh\xi)$,
and the normal vector is evaluated analogous to Eq.~(\ref{dsdef}) 
(cf.\cite{Luzum:2008cw}). The resulting single particle spectra are
then given by Eq.~(\ref{singpspec}), where it may be convenient
to change variables $c=\frac{\tau_{fo}-\tau}{\tau_{fo}-\tau_0}$ 
in the integral
\beq
\int_0^1 dc =-\int_{\tau_0}^{\tau_{fo}} \frac{d\tau}{\tau_{fo}-\tau_0}\,,
\eeq
where $\tau_0,\tau_{fo}$ correspond to the start and end of the hydrodynamic
evolution.
For isothermal freeze-out at a temperature $T_{fo}$, kinetic theory
specifies the entropy density $s=\frac{\epsilon+p}{T_{\rm fo}}$ and  
the number density $n=n^\mu u_\mu$ of particles. In particular, for a
massive Boltzmann gas in equilibrium Eq.~(\ref{ktemt0}),(\ref{nmudf}) lead to
\beq
s=\sum_i \frac{(2s_i+1)(2g_i+1)}{2\pi^2} m_i^3 K_3\left(\frac{m_i}{T_{\rm fo}}\right),
\quad n=\sum_i\frac{(2s_i+1)(2g_i+1)}{2\pi^2} m_i^2 T_{\rm fo} K_2\left(\frac{m_i}{T_{\rm fo}}\right)\,
\eeq
which can be used to quantify $s/n$, the amount of entropy each resonance
degree of freedom is carrying. For extremely high temperatures $s/n\rightarrow 4$,
which is the known limit for a gas of massless relativistic particles \cite{Gyulassy:1983ub},
but for temperatures below the QCD phase transition and a realistic
set of hadron resonances \cite{Amsler:2008zzb}, $s/n$ depends on $T_{\rm fo}$ 
(see Fig.~\ref{fig:dsdn}). For ideal hydrodynamics the total entropy $S$ in the fluid
is conserved (\ref{expli2ndlaw}), and hence the total number of particles $N$
created by an isothermal freeze-out should be given by $N=\frac{n}{s} S$,
which provides a non-trivial check on numerical codes.

After freeze-out, the hadron gas dynamics may be described by a
hadron cascade code such as \cite{Bass:2000ib}. A more simplistic
approach is to assume that particles stop interacting after freeze-out,
but unstable particles are allowed to decay, which changes
the spectra of stable particles \cite{Sollfrank:1990qz,Sollfrank:1991xm}. The decay of unstable resonances
can be simulated using public codes such as \cite{OSCAR} and leads to 
particle production, as can be seen in Fig.~\ref{fig:dsdn}.

\subsection{Viscous effects and open problems}

Ideal hydrodynamic simulations have been used 
quite successfully in the past to describe the properties of the 
particle spectra produced in relativistic heavy-ion collisions 
\cite{Back:2004je,Adcox:2004mh,Arsene:2004fa,Adams:2005dq}
(for reviews, see e.g. \cite{Huovinen:2006jp,Kolb:2003dz}).
Viscous effects have only been studied more recently:
the presence of viscosity leads to viscous entropy production
given by Eq.~(\ref{expli2ndlaw}), which increases the
total multiplicity for fixed initial entropy. The amount
of viscous entropy production depends on the 
hydrodynamic initialization time $\tau_0$ \cite{Dumitru:2007qr},
and for $\tau_0\sim1$ fm/c is on the order of $10$ percent for $\eta/s=0.08$
\cite{Romatschke:2007jx,Song:2008si}.

\begin{figure}[t]
\includegraphics[width=.47\linewidth]{Fig10a.eps}\hfill
\includegraphics[width=.5\linewidth]{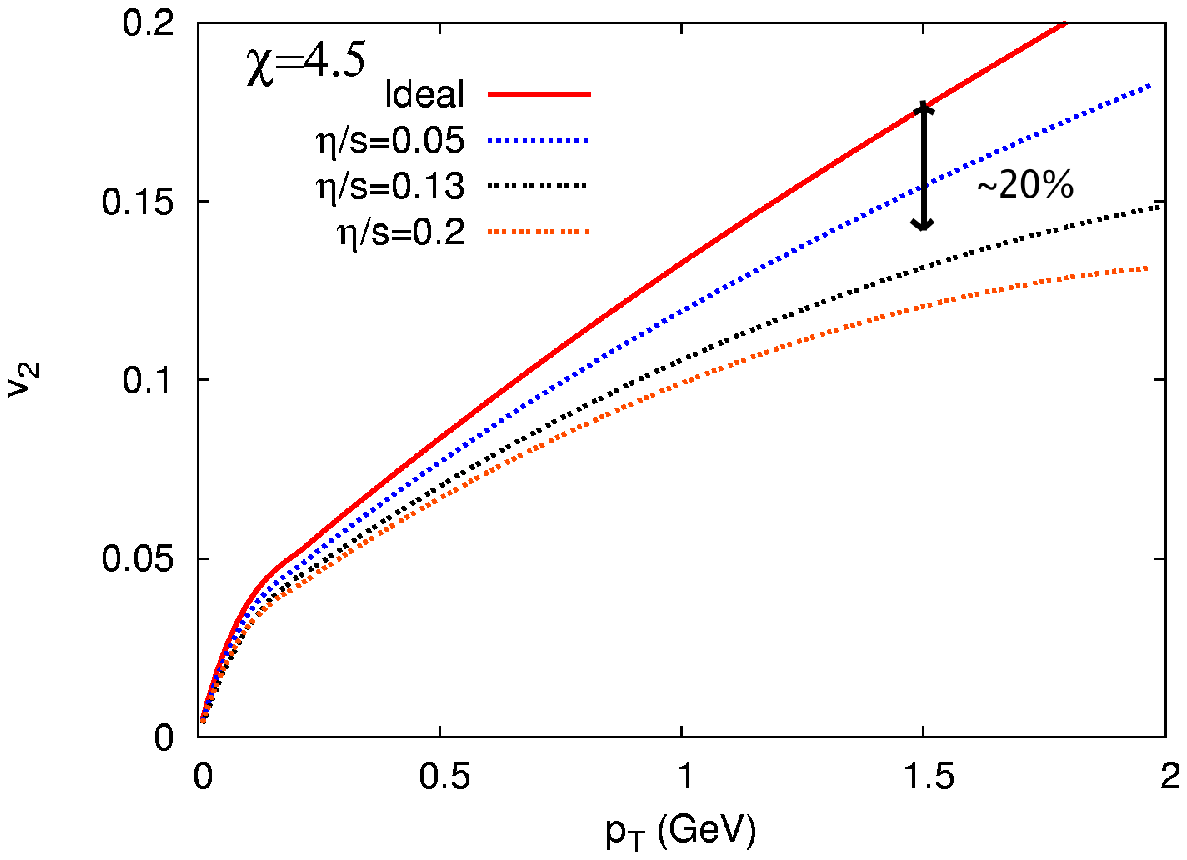}\hfill
\quad\\
\quad\\
\includegraphics[width=.49\linewidth]{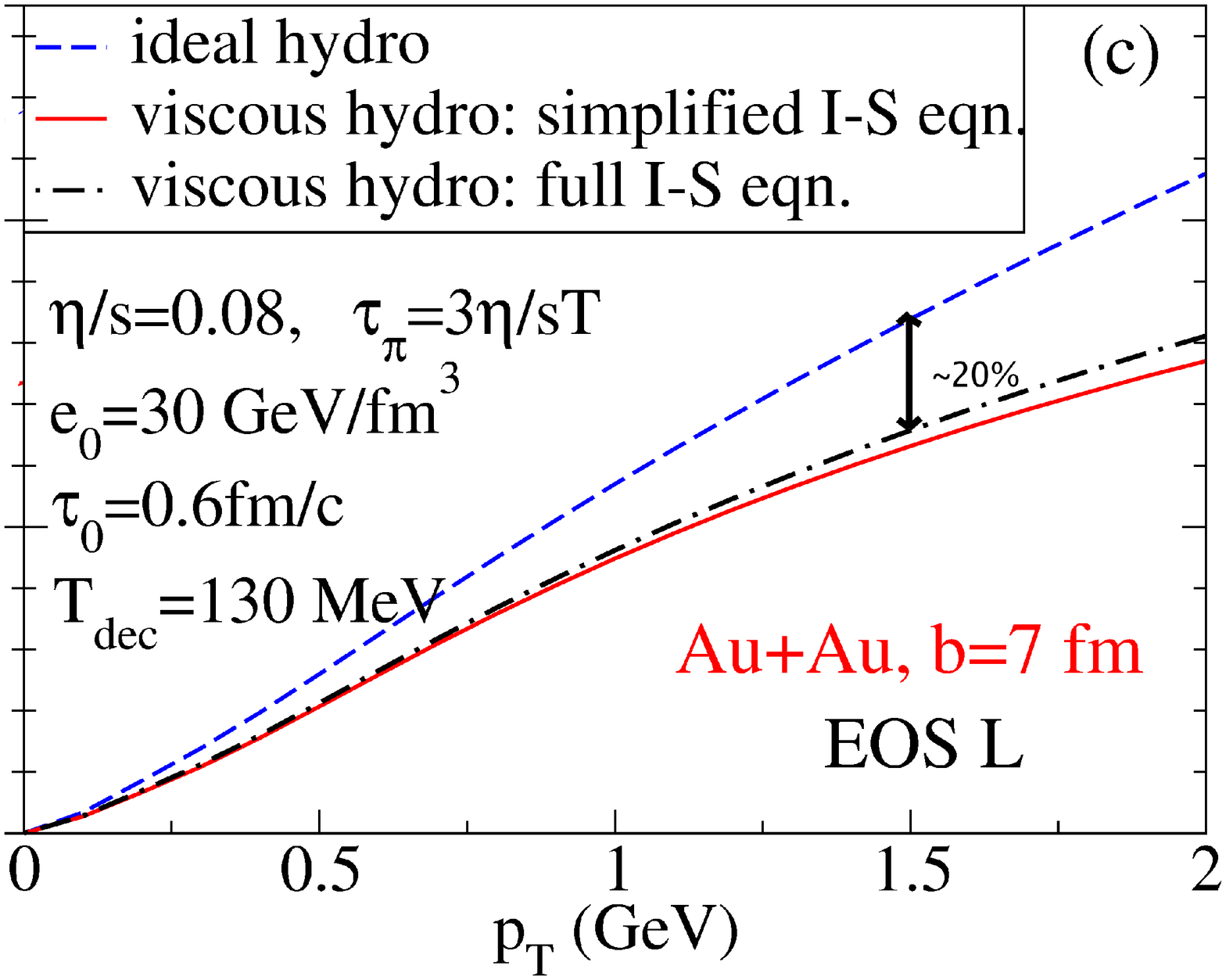}\hfill
\includegraphics[width=.44\linewidth]{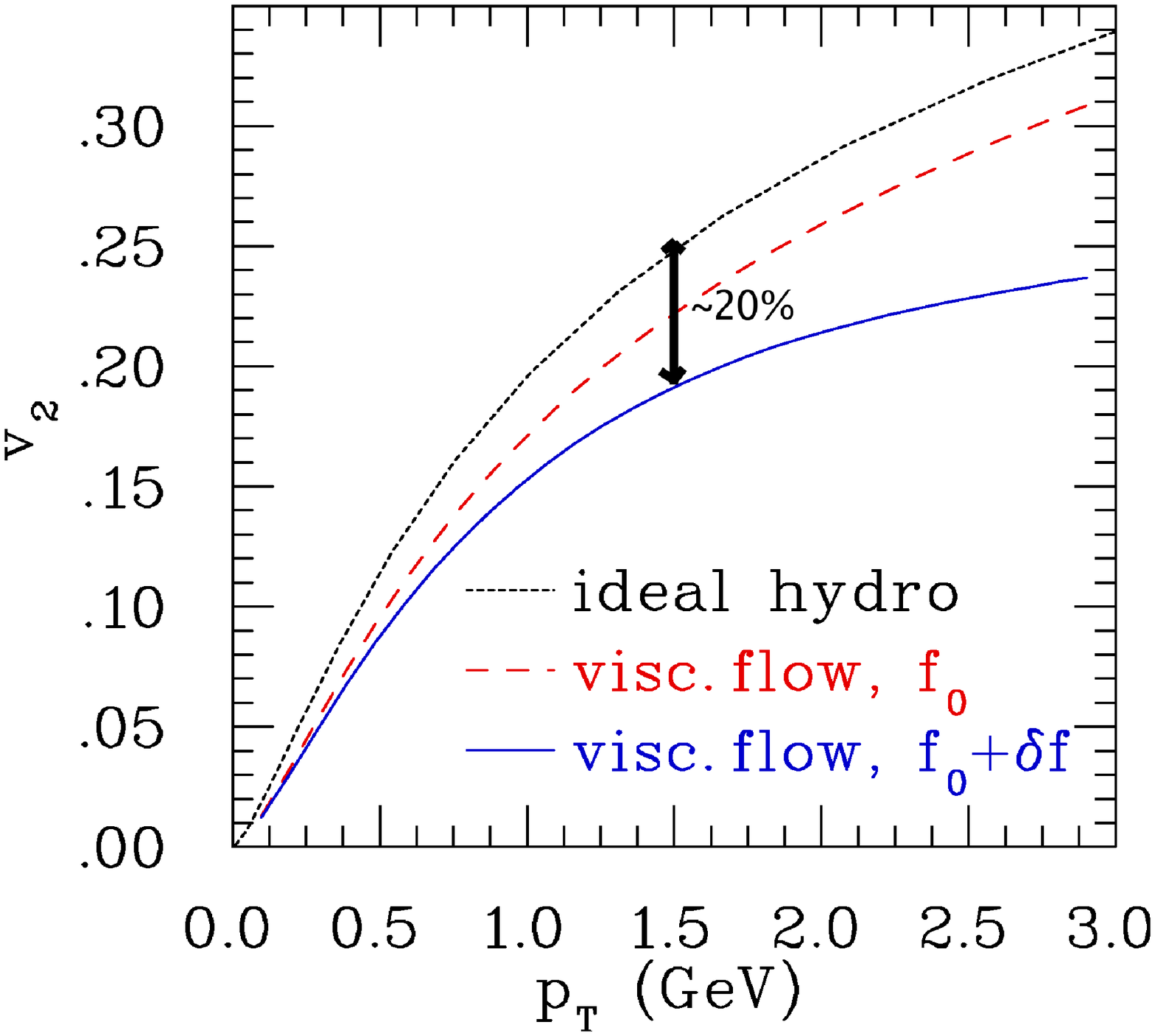}
\hspace*{0.5cm}
\caption{Reduction of elliptic flow coefficient due to shear viscosity: different groups
agree that at $p_\perp=1.5$ GeV, there is a $\sim20$ percent reduction of $v_2$
for $\eta/s=0.08$ (Figures from \cite{Romatschke:2007mq,Dusling:2007gi,Molnar:2008xj,Song:2008si}, 
clockwise from upper left.)}
\label{fig:visccomp}
\end{figure}

Viscosity also leads to stronger radial flow, which increases
the mean transverse momentum of particles \cite{Teaney:2003kp,Baier:2006gy,Romatschke:2007jx}.
Maybe more importantly, the presence of shear viscosity strongly decreases
the elliptic flow coefficient $v_2$. After some initial disagreement,
several different groups now agree on the quantitative
suppression of elliptic flow by shear viscosity, as is demonstrated
in Fig.~\ref{fig:visccomp}. Unfortunately, this does not directly
constrain the $\eta/s$ of hot QCD matter because the overall 
size of elliptic flow (which is proportional to the final momentum anisotropy
$e_p$) is dictated by the initial spatial eccentricity,
which is unknown (cf. Fig.~\ref{fig:spatialmomaniso}, see also Ref.~\cite{Luzum:2008cw}).

\begin{figure}[t]
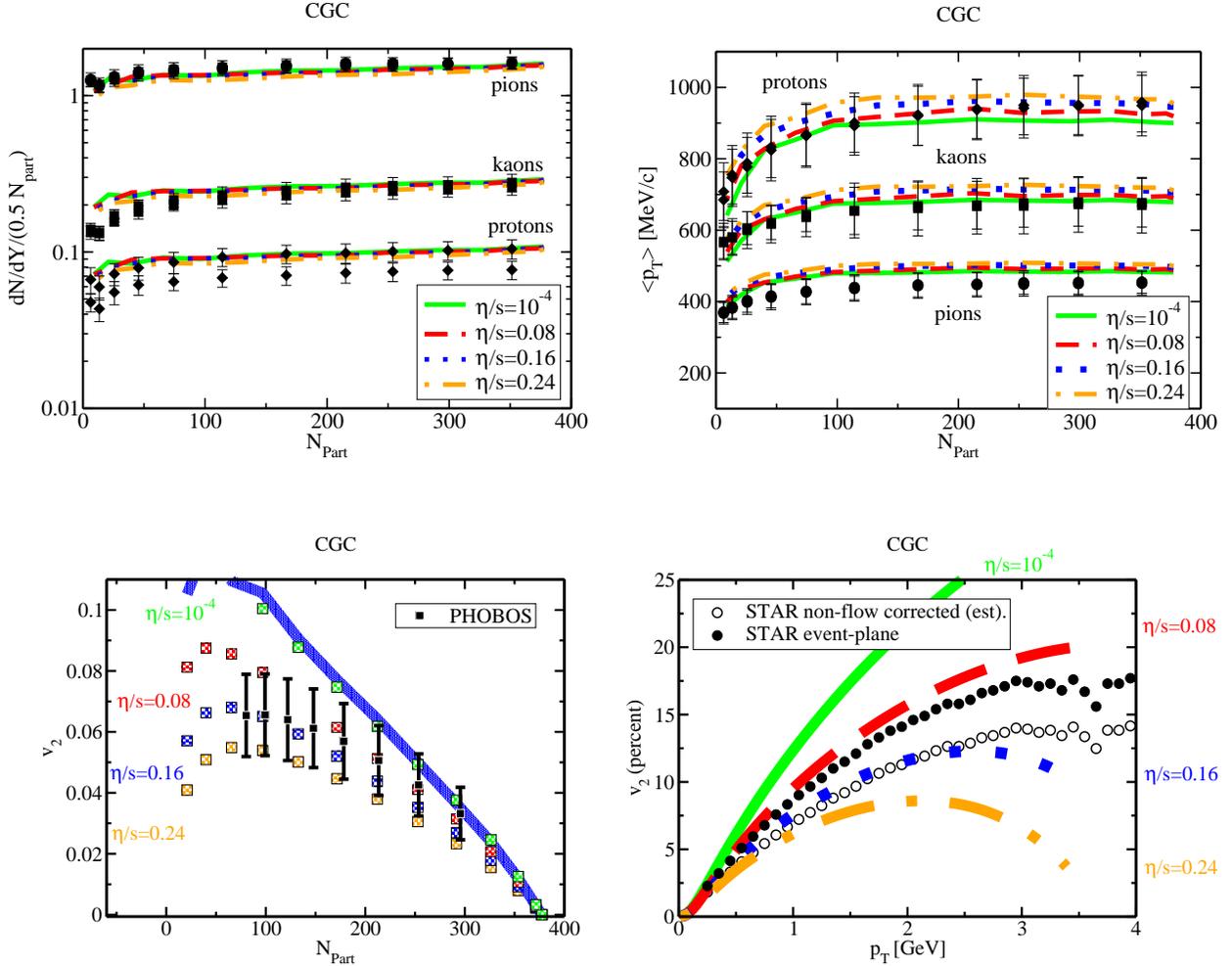

\includegraphics[width=.48\linewidth]{Fig11a.eps}
\hfill
\includegraphics[width=.48\linewidth]{Fig11b.eps}
\\
\vspace*{1cm}
\hfill
\includegraphics[width=.445\linewidth]{Fig11c.eps}
\hfill
\includegraphics[width=.49\linewidth]{Fig11d.eps}
\caption{Centrality dependence of particle spectra (multiplicity, mean momentum
and elliptic flow) as well as momentum dependence of charge hadron elliptic
flow, for a viscous hydrodynamic simulation of a $\sqrt{s}=200$ GeV Au+Au collision
using the CGC model compared to experimental data \cite{Adler:2003cb,Alver:2007qw,Abelev:2008ed} 
(Figures from \cite{Luzum:2008cw}).}
\label{fig:data}
\end{figure}

Many open problems remain, such as 
\begin{itemize}
\item
Exploring the effects of bulk viscosity. One study \cite{Karsch:2007jc} suggests 
that bulk viscosity may become large close to the QCD phase transition
(however, see \cite{Moore:2008ws,Romatschke:2009ng,CaronHuot:2009ns}),
which would have important consequences for the hydrodynamic evolution.
First phenomenological steps in this direction have been taken in \cite{Song:2008hj},
but it would be worthwile to have a classification of all the
non-conformal terms (including an estimate of their importance/size)
in the hydrodynamic equations.
\item
Implementing finite baryon chemical potential in the viscous
hydrodynamic simulations. On the one hand needed to describe the asymmetry
in the baryon/anti-baryon multiplicities, the viscous
hydrodynamic evolution in the vicinity of a possible
QCD critical point could on the other hand help to guide
experimental searches for this critical point (cf.\cite{Asakawa:2008ti}).
\item
Combining viscous hydrodynamics with a hadron cascade code
to more realistically describe the freeze-out process.
Ideally such a hybrid code would make the choice of a 
freeze-out temperature superfluous, eliminating one 
model parameter (cf.\cite{Teaney:2000cw,Hirano:2005xf,Petersen:2008dd}).
\end{itemize}
These (and other) problems are straightforward to solve in the sense
that they do not require fundamentally new ideas, but
``only'' hard work. However, there are also at least two problems
outside the framework of hydrodynamics, which would have to be
solved in order to claim a complete understanding of high
energy nuclear collisions:
\begin{itemize}
\item
What is the value of the initial spatial eccentricity $e_x$ in
high energy nuclear collision? Can it be calculated or measured
without having to rely on models (Glauber/CGC)?
Since the spatial eccentricity controls the amount of 
elliptic flow generated in a relativistic
nuclear collision, knowing $e_x$ seems necessary to 
quantify the viscosity of hot nuclear matter.
\item
How and when does the system equilibrate? An answer to this
question would give a well defined value to the hydrodynamic
starting time $\tau_0$ as well as the eccentricity at this time.
Currently both $\tau_0$ and $e_x(\tau_0)$ are ``guessed'',
with no solid arguments for any particular value.
\end{itemize}

Nevertheless, the striking ability of viscous hydrodynamics to
describe the momentum spectra of the majority of particles,
including the elliptic flow coefficient, in the highest 
energy Au+Au collisions at RHIC (see Fig.~\ref{fig:data}) make relativistic nuclear
collisions an ideal application for the old, new, and future developments
in hydrodynamic theory.

\section{Conclusions}

Relativistic viscous hydrodynamics is an effective theory 
for the long-wavelength behavior of matter. The relativistic
Navier-Stokes equation would do justice to this long-wavelength behavior,
but does not lend itself easily to direct numerical simulations
because of its ultraviolet behavior. Generalizations of the
Navier-Stokes equation including second-order gradients
have been proposed to cure this difficulty, and indeed
provide a phenomenological regularization of the Navier-Stokes
equation that can be simulated numerically unless the
regularization parameter is too small.

Interestingly, performing a complete gradient expansion to 
second order reproduces this attractive feature 
of regularizing the Navier-Stokes equation,
besides having the benefit of constituting 
an improved approximation of the underlying quantum field theory.
For all theories where the regularization parameter
obtained from this gradient expansion is known,
its value is such that the ultraviolet behavior
of the second-order hydrodynamic equations is
benign. It is not known whether this is a coincidence.

The second order hydrodynamic equations have been
applied to the problem of high energy nuclear collisions,
offering a good description of the experimentally
measured particle spectra at low momenta. Further work
is needed to extract material parameters of hot nuclear matter,
such as the shear viscosity coefficient, from experimental
data.

Many other applications of second order hydrodynamics
are possible, e.g. in astrophysics (viscous damping
of neutron star r-modes \cite{Lindblom:2001hd}) 
or cosmology (effects of bulk viscosity \cite{Zimdahl:1996ka}).

Whether in the formulation I have described in these
pages, or not, one thing is certain: relativistic
viscous hydrodynamics is here to stay.

\acknowledgments

I would like to thank the organizers of the 18th Jyv\"askyl\"a Summer School
and the 4th Torino Winter School ``Quark-Gluon Plasma and Heavy-Ion Collisions'' 
(where I presented these lectures) for their effort in preparing and conducting these
great meetings, and in particular K.~Eskola for his generous hospitality.
Also, I would like to thank M.~Luzum for a critical reading of this manuscript, 
G.A.~Miller and G.D.~Moore for constructive comments, 
D.T.~Son for numerous fruitful discussions,
and U.~Romatschke for providing Fig.~\ref{fig:snapshots}.
Finally, I would like to thank A.~Beraudo for pointing out an error in 
Eq.~(\ref{MISineq}).
This work was supported by the US Department of Energy, grant 
number DE-FG02-00ER41132.


\appendix

\section{Proof of Causality of Maxwell-Cattaneo type equations}

Let us first establish that the diffusion-type equation
\beq
\partial_t \delta u^y-\frac{\eta_0}{\epsilon_0+p_0} \partial_x^2 \delta u^y=f(t,x)\,,
\label{appde}
\eeq
which was discussed in section \ref{sec:acprob} for the homogeneous case $f(t,x)=0$, violates causality. To this end,
let us calculate the retarded Green's function $G({\bf x},{\bf x^\prime})$, ${\bf x}=(t,x)$, 
of the differential operator
(\ref{appde}),
\beq
\left[\partial_{t}-\nu \partial_{x}^2\right] G({\bf x},{\bf x^\prime})=\delta^2({\bf x}-{\bf x^\prime}),\quad \nu=\frac{\eta_0}{\epsilon_0+p_0}
\eeq
Doing a Fourier-transform of $G$ one finds
\beq
G({\bf x},{\bf x^\prime})=\int \frac{d^2{\bf k}}{(2\pi)^2} 
\frac{e^{-i \omega (t-t^\prime)+i k (x-x^\prime)}}{-i \omega+ \nu k^2}
\eeq
which can be solved by the usual contour integration methods and Gaussian integration,
$$
G({\bf x},{\bf x^\prime})=
\frac{\theta(t-t^\prime)}{\sqrt{4 \pi \nu (t-t^\prime)}}
\exp{\left[-\frac{(x-x^\prime)^2}{4 \nu (t-t^\prime)}\right]}\ .
$$
The solution to Eq.~(\ref{appde}) is then
\beq
\delta u^y(t,x)=\int d^2{\bf x^\prime} G({\bf x},{\bf x^\prime}) f({\bf x^\prime})=
 \frac{\theta(t)}{\sqrt{4 \pi \nu t}}
\exp{\left[-\frac{x^2}{4 \nu t}\right]}\,,
\eeq
where the system was started with an initial ``kick'', $f(t,x)=\delta(t)\delta(x)$.
One can see that for any finite time $t>0$, the perturbation is non-vanishing for all values of $x$,
not only for those $x<t$. This obviously violates causality.

Considering instead of the diffusion-type equation (\ref{appde}) the Maxwell-Cattaneo law 
\beq
\partial_t \delta u^y+\frac{1}{\epsilon_0+p_0} \partial_x \pi^{xy}=0, \quad
\tau_\pi \partial_t \pi^{xy}+\pi^{xy}=-\eta_0 \partial_x \delta u^y\,,
\label{appMC}
\eeq
the Green's function has to fulfill
\beq
\left[\partial_t^2+\frac{\partial_t}{\tau_\pi}-\frac{\nu}{\tau_\pi} \partial_x^2\right]
G({\bf x},{\bf x^\prime})=\frac{1}{\tau_\pi}\delta^2({\bf x}-{\bf x^\prime})
\eeq
and hence is given by
\beq
G({\bf x},{\bf x^\prime})=\int \frac{d^2{\bf k}}{(2\pi)^2} \frac{e^{-i \omega (t-t^\prime)+i k (x-x^\prime)}}
{-\omega^2 \tau_\pi-i \omega+\nu k^2}\,.
\eeq
The frequency integration proceeds as before, and one finds
\beq
G({\bf x},{\bf x^\prime})=\theta(t-t^\prime) \int_{-\infty+i \epsilon}^{\infty+i \epsilon} \frac{dk}{2\pi}
\frac{i}{\tau_\pi} 
e^{i k (x-x^\prime)} 
\left[\frac{e^{-i \omega^+ (t-t^\prime)}-e^{-i \omega^- (t-t^\prime)}}{\omega^+-\omega^-}\right] 
\,,
\label{anotherint}
\eeq
where $2 \tau_\pi \omega^{\pm}=-i\pm\sqrt{4 \tau_\pi \nu k^2 -1}$. 
The integral over $k$ is chosen in the upper half-plane
and the branch cut of the square-root is chosen to run from 
$k=-(4 \tau_\pi \nu)^{-1/2}$ to $k=(4 \tau_\pi \nu)^{-1/2}$
along the real axis \cite{MF1}\S7.4.
To evaluate the integral
$$
I_+\equiv
e^{-\frac{t-t^\prime}{2 \tau_\pi}}\int_{-\infty+i \epsilon}^{\infty+i \epsilon}\frac{dk}{2\pi}
\frac{i\exp{\left[i k (x-y)-i \frac{(t-t^\prime)}{2 \tau_\pi}
\sqrt{4 \tau_\pi \nu k^2-1}\right]}}
{\sqrt{4 \tau_\pi \nu k^2-1}}
 \, ,
$$
note that if $x>x^\prime+\frac{(t-t^\prime)}{2\tau_\pi}\sqrt{4 \tau_\pi\nu}$, then the contour can be closed by a semicircle
in the upper half plane, giving a vanishing contribution since there are no singularities in that halfplane.
For $x<x^\prime+\frac{(t-t^\prime)}{2\tau_\pi}\sqrt{4 \tau_\pi\nu}$ on the other hand, the contribution will not
vanish. It can be calculated by using a table of Laplace transforms \cite{MF1}\S7.4, \cite{MF2}\S11, giving 
\beq
I_+=\theta\left((t-t^\prime)\sqrt{\frac{\nu}{\tau_\pi}}-(x-x^\prime)\right) \frac{e^{-\frac{t-t^\prime}{2 \tau_\pi}}}
{\sqrt{4 \nu \tau_\pi}} I_0\left(\sqrt{\frac{(t-t^\prime)^2}{4 \tau_\pi^2}-\frac{(x-x^\prime)^2}{4 \nu \tau_\pi}}
\right)\,,
\eeq
where $I_0(x)$ is a modified Bessel function.
Similarly, one can calculate the other component of Eq.~(\ref{anotherint}), so that
one finds 
\beq
G({\bf x},{\bf x^\prime})=\theta(t-t^\prime)\,
\theta\left(\frac{(t-t^\prime)^2\nu}{\tau_\pi}-(x-x^\prime)^2\right)
\frac{e^{-\frac{t-t^\prime}{2 \tau_\pi}}}
{\sqrt{4 \nu \tau_\pi}}
I_0\left(\sqrt{\frac{(t-t^\prime)^2}{4 \tau_\pi^2}-\frac{(x-x^\prime)^2}{4 \nu \tau_\pi}}\right)\, .
\label{appGxyresult}
\eeq

\begin{figure}[t]
\includegraphics[width=.47\linewidth]{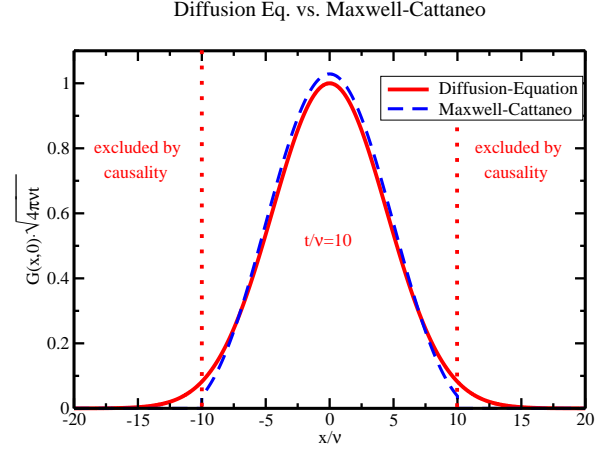}
\caption{Green's function for the diffusion equation and Maxwell-Cattaneo law
for $\tau_\pi=\nu$. See text for details}
\label{fig:diffeq}
\end{figure}

From the step-function in Eq.~(\ref{appGxyresult}), one
can easily convince oneself that the solution $\delta u^y(t,x)$ to Eq.~(\ref{appMC})
is confined to $|x|<t\, v_T^{\rm max}$, where $v_T^{\rm max}=\sqrt{\frac{\eta_0}{\tau_\pi (\epsilon_0+p_0)}}$
coincides with the limit found in Eq.~(\ref{vT}). 
The difference between the Maxwell-Cattaneo solution and the 
diffusion equation is highlighted in Fig.~\ref{fig:diffeq}, where $\sqrt{4 \pi \nu t} G({\bf x},{\bf 0})$
is plotted for $t=10 \nu$ as a function of $x$. One can see that $G({\bf x},0)$
has non-vanishing support in the region excluded by causality for the diffusion equation, while
this does not happen for the Maxwell-Cattaneo law.

\section{Notations and Conventions}

This appendix is a collection of notations and conventions used in the
main part of the article.

\begin{itemize}
\item
The metric sign convention is $(+,-,-,-)$
\item
Projectors:
\beq
\Delta^{\mu \nu}=g^{\mu \nu}-u^\mu u^\nu\,,\quad
P^{\mu \nu}_{\alpha \beta}=\Delta^{\mu}_\alpha \Delta^\nu_\beta+\Delta^{\mu}_\beta \Delta^\nu_\alpha
-\frac{2}{3} \Delta^{\mu \nu} \Delta_{\alpha \beta}\,,
\eeq
with properties $u_\mu\Delta^{\mu \nu}=u_\mu P^{\mu \nu}_{\alpha \beta}=0$, 
$g_{\mu \nu} P^{\mu \nu}_{\alpha \beta}=0$.
\item
Derivatives:
\beq
D=u^\mu D_\mu\,,\quad
\nabla_\mu= \Delta^\alpha_\mu D_\mu\,,\quad
D_\mu = u_\mu D+\nabla_\mu\,,
\eeq
where $D_\mu$ is the geometric covariant derivative that reduces to $D_\mu \rightarrow \partial_\mu$
for flat space. In the non-relativistic, flat-space limit,
\beq
D=\partial_t+\vec{v}\cdot\vec{\partial}+{\cal O}\left(|\vec{v}|^2\right)\,,\quad
\vec{\nabla}=-\vec{\partial}+{\cal O}\left(|\vec{v}|\right)\,,
\eeq
which supports the interpretation of time-, and space-like derivatives for
$D$ and $\nabla$, respectively.
\item
Brackets:
\bqa
&A^{(\alpha}B^{\beta)}=\frac{1}{2}\left(A^\alpha B^\beta+A^\beta B^\alpha\right)\,,\quad
A^{[\alpha}B^{\beta]}=\frac{1}{2}\left(A^\alpha B^\beta-A^\beta B^\alpha\right)\,,&\nonumber\\
&A^{<\alpha}B^{\beta>}=P^{\alpha \beta}_{\mu \nu} A^\mu B^\nu\,,&
\eqa
which are used to define e.g. the vorticity, $\Omega_{\alpha \beta}=\nabla_{[\alpha} u_{\beta]}$.
Note that the above definition of $A^{<\alpha}B^{\beta>}$ differs from others (e.g.
\cite{Baier:2007ix}) by a factor of $2$.

\end{itemize}

\end{document}